\documentclass[10pt,journal]{IEEEtran}
\IEEEoverridecommandlockouts
\usepackage{cite}
\usepackage{amsmath,amssymb,amsfonts}
\usepackage{algorithmic}
\usepackage{graphicx}
\usepackage{textcomp}
\usepackage[table, dvipsnames]{xcolor}

\usepackage{mathtools}
\usepackage[english]{babel}
\begin{document}
	\title{Optimal transport meets speech: a tutorial review}
	
	\author{\IEEEauthorblockN{Xugang Lu$^{1}$, Yu Tsao$^{2}$}\\
		\IEEEauthorblockA{\textit{1. National Institute of Information and Communications Technology, Japan}}\\
		\IEEEauthorblockA{\textit{2. Research Center for Information Technology Innovation, Academia Sinica, Taiwan} }\\
	}
	
	\maketitle
	
	\begin{abstract}
		Optimal Transport (OT) provides a principled framework for comparing and transforming probability distributions while preserving geometric structure. Recently, OT has gained significant attention in machine learning due to its ability to measure discrepancies between distributions, even when their supports do not overlap, making it effective for tasks such as generative modeling, domain adaptation, and transfer learning. Despite its success in fields such as computer vision and natural language processing, OT remains relatively underexplored in speech research. Speech signals present unique challenges, including temporal dynamics, speaker variability, noise, reverberation, and heterogeneous multimodal representations involving audio, text, and visual information. These factors often lead to distribution mismatches, where OT offers a natural framework for alignment and interpretation. This work aims to promote broader adoption of OT in speech processing by: (1) reviewing OT foundations through intuitive physical interpretations and highlighting connections to modern generative models; (2) presenting computational algorithms suitable for deep learning frameworks; and (3) demonstrating OT applications in cross-domain and cross-modal speech tasks, including speech enhancement, automatic speech recognition, language and speaker recognition, and audio spoof detection. We highlight OT’s strong potential for addressing distributional variations in real-world speech applications.      
		
	\end{abstract}
	
	\begin{IEEEkeywords}
		Optimal transport, Cross-domain adaptation, Cross-modal knowledge transfer learning, Alignment and matching.
	\end{IEEEkeywords}
	
	\section{Introduction}
	Optimal transport (OT), also known as mass transport, can be simply described as the search for the most efficient way to transform from one mass distribution to another, considering a specified transform cost \cite{VillaniBook2003,Kolouri2016}. This theory which has been extensively studied by many researchers has found extensive applications in operational research, mathematics, economics, and physics \cite{VillaniBook2003,Kolouri2016}. In particular, it has garnered widespread attention in statistical machine learning (ML) in recent years \cite{KolouriSPM2017, MontesumaPAMI2025, Gabriel2025}. 
	\subsection{Optimal transport in machine learning}
	In statistical ML, whether dealing with pattern classification or regression problems, numerous research tasks involve comparing or manipulating the probability distributions of the underlying features or objects. The role of applying OT in ML lies in its ability to furnish an effective metric measure (e.g., discrepancy or distance measure) between two probability distributions, establishing correspondences between sets of samples with a geometric-aware distance between distributions. This metric measure can then function as a learning objective or metric when handling feature vectors within a probability space. OT-based learning has been shown to be effective in measuring the distribution discrepancy even when the compared distributions have disjoint supports \cite{PeyreBook2019}. Furthermore, this learning can incorporate the geometric structures of datasets, demonstrating a form of geometric-aware learning \cite{Courty2014}. In addition to the increasing popularity of ML in recent years, especially in computational ML theories (such as generative model learning, domain adaptation, or transfer learning) \cite{Courty2014}, OT theory has been extensively applied in real-world scenarios, particularly in the fields of computational graph, image classification \cite{ZhangEMD2020}, and computer vision \cite{XingTOG2022, Bonneel2023}. It has also gained recognition for a range of natural language processing (NLP) tasks \cite{ChenNLP2019,Yokoi2020, Bhardwaj2022}, including document retrieval, semantic matching \cite{Bhardwaj2022}, machine translation \cite{ZhouACL2023}, and word embedding \cite{KusnerICML2015}. In contrast, the speech science community has only recently begun to explore applications of OT. The inherently sequential and temporal nature of speech signals poses significant challenges in integrating OT into speech processing frameworks, which has limited its exploration in this domain.
	
	\subsection{Optimal transport in speech}
	In speech tasks, various challenges need consideration, including feature representation learning, feature and model fusion, domain adaptation, knowledge transfer learning, etc. These learning algorithms primarily manipulate probability distributions. A crucial aspect for improving performance is the thoughtful selection of a meaningful measurement metric, such as discrepancy, similarity, or distance. The use of OT extends to estimating, comparing, and aligning the distributions for feature aggregation, model fusion, cross-domain and cross-modal adaptation, and transfer learning. Hence, the primary goal of this review paper is to review fundamental knowledge about OT and its computational algorithms, demonstrating their adaptability to deep learning model frameworks with ease. Through this review, a clear understanding emerges that OT can offer meaningful explanations in model learning. 
	
	In real-world applications of speech technologies, the performance of speech applications often experiences notable degradation due to various factors. On the one hand, speech signals are susceptible to external or extrinsic degradations such as noise, reverberation, and noise suppression processing, significantly impacting speech quality and intelligibility \cite{YuBook2014, LiBook2015, DeepRobustASR2018}. On the other hand, intrinsic variations pose challenges as speech exhibits considerable diversity in articulation styles among speakers and co-articulation effects influenced by different contexts and domains. A co-occurring pattern involves representing acoustic speech in multi-views, including text content and visual information during speaking. Addressing these challenges, learning representations that encompass all these multi-view information sources can potentially mitigate uncertainty. Robust representations learned through cross-modal feature learning have been shown to enhance performance and adaptability in diverse speech scenarios. The OT serves as a powerful mathematical tool for quantifying and manipulating distribution variations. Thus, the second objective of this review paper is to introduce ideas and techniques to seamlessly integrate OT to describe the underlying distribution variations of speech, with the aim of enhancing overall performance.
	
	In summary, the purpose of this review is to introduce OT in speech signal processing and to share our findings, offering our insights to speech researchers seeking entry points into this promising area. In this review paper, our aim is to: (1) Provide a historical review of the OT problem, mathematical foundations, connections in transport modeling to generative diffusion models, and computational algorithms, particularly focusing on those integrated into deep learning model frameworks. (2) Introduce potential applications of OT in ML and related fields, particularly in domain adaptation tasks; (3) Explain how OT can be seamlessly and effectively integrated into speech-related application tasks. We will present research topics that integrate OT for speech application problems, including speech translation, spoken document retrieval, semantic analysis in NLP, and speech acoustic modeling. Specifically, we will address cross-domain problems related to speech enhancement (SE) \cite{LinNeurIPS2021, HsiehIS2021}, speaker recognition or verification (SR/SVR) \cite{ZhangICASSP2023, YangTIFS2026}, spoken language identification/recognition (SLID/SLR) \cite{LuICASSP2021}, and multi-modal or cross-modal knowledge transfer and adaptation for automatic speech recognition (ASR) \cite{LuASRU2023, LuICASSP2024, LuIS2025, LuICASSP2026}; (4) Share our review and perspective on the potential usage of OT in speech. 
	
	\subsection{Notations and symbols}
	For a self-contained introduction of the main concepts of OT, we begin by giving several fundamental mathematical notations, symbols and operators that will be frequently used throughout this paper. 
	\subsubsection{Gradient}
	The gradient of a scalar function \( f: \mathbb{R}^n \to \mathbb{R} \) is the vector consisting of its partial derivatives with respect to each coordinate direction. It points in the direction of the steepest increase in the function and encodes how the function changes in space,  a vector field denoted by \( \nabla f \) formulated as:
	\begin{equation}
		{\rm grad}(f):  \nabla f \mathop  = \limits^\Delta  \left( {\frac{{\partial f}}{{\partial x_1 }},\frac{{\partial f}}{{\partial x_2 }},...,\frac{{\partial f}}{{\partial x_n }}} \right).
	\end{equation}
	\subsubsection{Divergence}
	The divergence of a vector field is the scalar quantity obtained by summing the partial derivatives of each vector component. It measures the net rate at which ``mass" or ``flow" is expanding or contracting at a given point. The divergence of a vector field \( \mathbf{F}: \mathbb{R}^n \to \mathbb{R}^n \) with \( \mathbf{F} = (F_1, F_2, \ldots, F_n) \) is a scalar field denoted by \( \nabla \cdot \mathbf{F} \) and is defined as:
	\[
	{\rm div}({\bf F}):  \nabla  \cdot {\bf F} \mathop  = \limits^\Delta \sum\limits_i {\frac{{\partial F_i }}{{\partial x_i }}}.  
	\]
	\subsubsection{Laplacian}
	The Laplacian operator is defined as the divergence of the gradient. Applied to a scalar function, it produces a scalar quantity that describes how the value of the function at a point compares to its average value in a small neighborhood. The Laplacian plays a central role in diffusion processes, heat equations, and many partial differential equations (PDEs) relevant to OT.
	The Laplacian of a scalar field \( f: \mathbb{R}^n \to \mathbb{R} \) is a scalar field denoted by \( \Delta f \) or \( \nabla^2 f \) and is defined as the divergence of the gradient of \( f \):
	\[
	\Delta f = \nabla^2 f = \nabla \cdot (\nabla f)=\rm{div}(\rm{grad}(f)).
	\]
	Explicitly, in \( \mathbb{R}^n \), it is given by:
	\[
	\Delta f \mathop  = \limits^\Delta  \frac{\partial^2 f}{\partial x_1^2} + \frac{\partial^2 f}{\partial x_2^2} + \cdots + \frac{\partial^2 f}{\partial x_n^2}.
	\]
	\subsubsection{Probability space and Push-Forward Measures}
	The notion of a probability space and the concept of a push-forward measure \cite{Lasota1994} are fundamental in OT. A probability space (or a measure space) is formally described as a triple $(X, \mathcal{A}, \mu)$, where
	\begin{itemize}
		\item $X$ is the sample space, representing all possible outcomes,
		\item $\mathcal{A}$ is a $\sigma$-algebra of measurable subsets of $X$, representing the collection of admissible events,
		\item $\mu$ is a probability measure that assigns a probability value to each event in $\mathcal{A}$.
	\end{itemize}
	
\begin{figure}[tbp]
		\centering	
		\includegraphics[width=6cm, height=4cm]{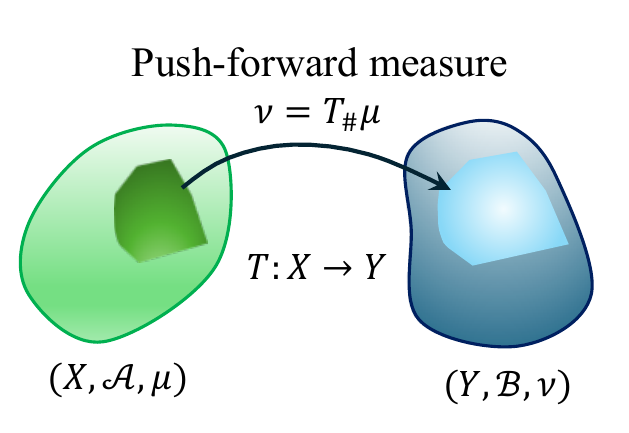}
		\caption{Push-forward measure in probability spaces.}
		\label{fig:fig55}
	\end{figure}

	Although these definitions arise from abstract measure theory (e.g., $X$ is often assumed to be a Polish space), our purpose here is practical. Therefore, it suffices to consider a probability space as a structured way to assign probabilities to events. Now consider a measurable transformation $T : X \to Y$ that maps elements from one probability space to another. This map naturally induces a new measure in $Y$, called the push-forward measure, as shown in Fig. \ref{fig:fig55}. 
	Intuitively, the push-forward describes how the probability mass is transported from $X$ to $Y$ through $T$. Formally, given a measure space $(X, \mathcal{A}, \mu)$ and a measurable space $(Y, \mathcal{B})$, the push-forward measure $\nu = T_{\#}\mu$ in $(Y, \mathcal{B})$ is defined by:
	\begin{equation}
		T_{\#}\mu(B) = \mu\!\left(T^{-1}(B)\right), \qquad \forall\, B \in \mathcal{B},
	\end{equation}
	where $T^{-1}(B) = \{ x \in X \mid T(x) \in B \}$.
	
	The push-forward measure satisfies the usual change-of-variables formula. For any measurable function $
	\varphi : Y \to \mathbb{R}$, we have:
	\begin{equation}
		\int_{Y} \varphi \, d(T_{\#}\mu)
		= \int_{X} \varphi \circ T \, d\mu.
	\end{equation}
	This identity is used extensively in OT theory, particularly when describing how probability distributions transform under transport maps.
	
	\section{Historic review of Optimal transport}
	This section provides a comprehensive and historical review of the fundamental concept of OT and its connections to concepts used in other disciplines.
	\subsection{Problems of OT}
	OT problems have been described through various intuitive examples in different fields, for example, delivering baked goods to coffee shops, moving sand or soil for construction, or relocating rubble for site preparation \cite{VillaniBook2003}. In simple terms, OT seeks the most efficient way to achieve a given goal, typically by minimizing the effort or cost required to move the mass from one distribution to another. Despite the simplicity of its underlying idea, solving OT problems is often highly nontrivial. The field has a history spanning more than two centuries and has been marked by extensive and ongoing research activities. Fig. \ref{fig:fig1} describes the important characters related to OT along the time line. 
	\subsection{Historical figures in the development of OT theory}
	Over the past two centuries, many influential scholars have contributed to the development of OT theory, which continues to thrive with vigorous research activities, especially in the 21st century.
	\begin{figure}[tbp]
		\centering
		\includegraphics[width=8.5cm, height=4cm]{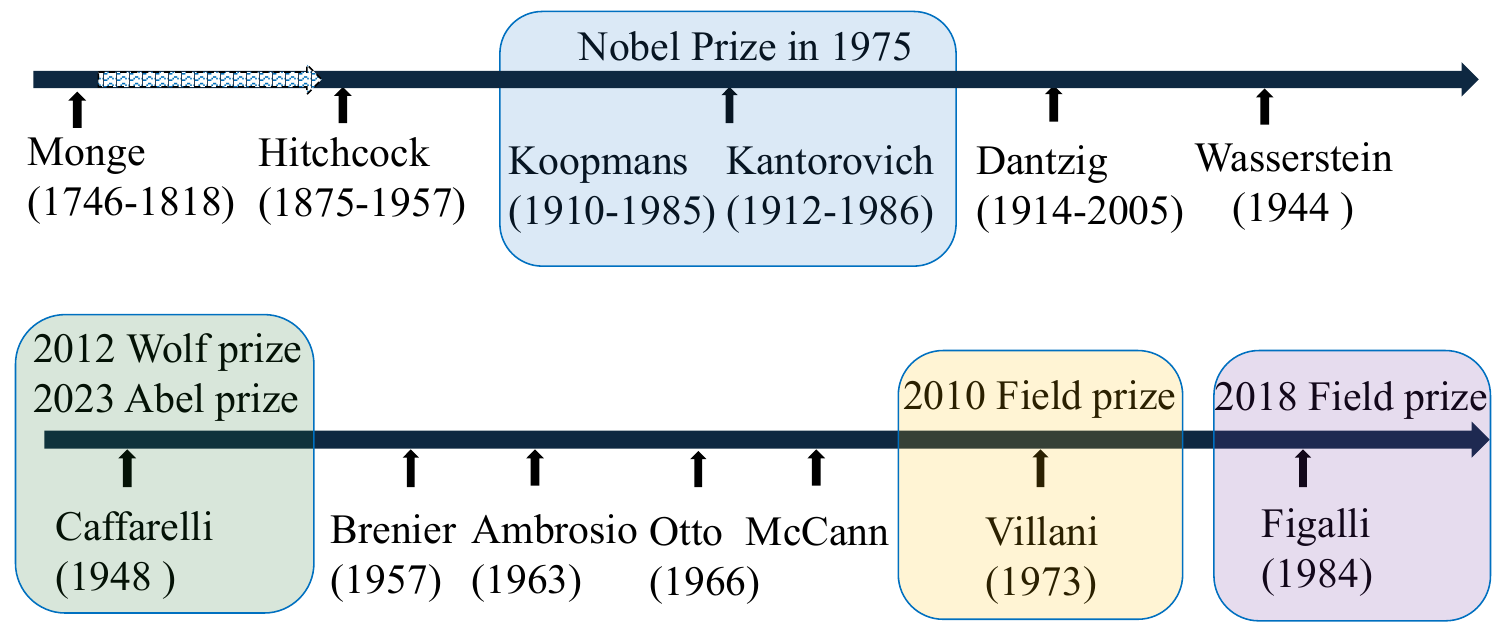}
		\caption{Historical figures in the development of OT theory.}
		\label{fig:fig1}
	\end{figure} 
	As shown in Fig. \ref{fig:fig1}, OT theory traces its origins back to 1781, when Monge first posed the problem of determining the most efficient way to transport mass between two locations. The development of this theory progressed slowly until 1942, when Kantorovich introduced a relaxed formulation of Monge’s problem by incorporating duality and linear programming, thus establishing OT as a powerful tool for resource allocation \cite{VillaniBook2003, Galichon2016}. In parallel, Dantzig-known as the ``father of the simplex algorithm'' proposed efficient algorithms that further advanced the solution of optimization problems in OT. Wasserstein later formalized the Kantorovich problem through what is now known as the Wasserstein distance \cite{AmbrosioBook2008}. This formulation opened the door to new insights and methodologies that have influenced a wide range of disciplines, including physics, probability theory, financial mathematics, and economics. Toward the new century, significant theoretical progress was made by mathematicians such as Brenier \cite{Brenier1991}, who addressed the existence of OT maps, and Otto and McCann, who investigated the geometric structure underlying OT. In recent decades, Villani has made profound contributions to OT theory, and his influential books on OT are widely known \cite{VillaniBook2003,VillaniBook2008}. Figalli has also advanced the field, particularly through applications to PDEs. The historical trajectory of these contributions highlights not only the brilliance of the researchers involved but also the recognition the field has received through numerous prestigious awards. Together, these developments underscore the importance and enduring vitality of OT as a research area. 
	\subsection{Landmark research activities}
\begin{figure}[tbp]
		\centering	
		\includegraphics[width=9cm, height=3.2cm]{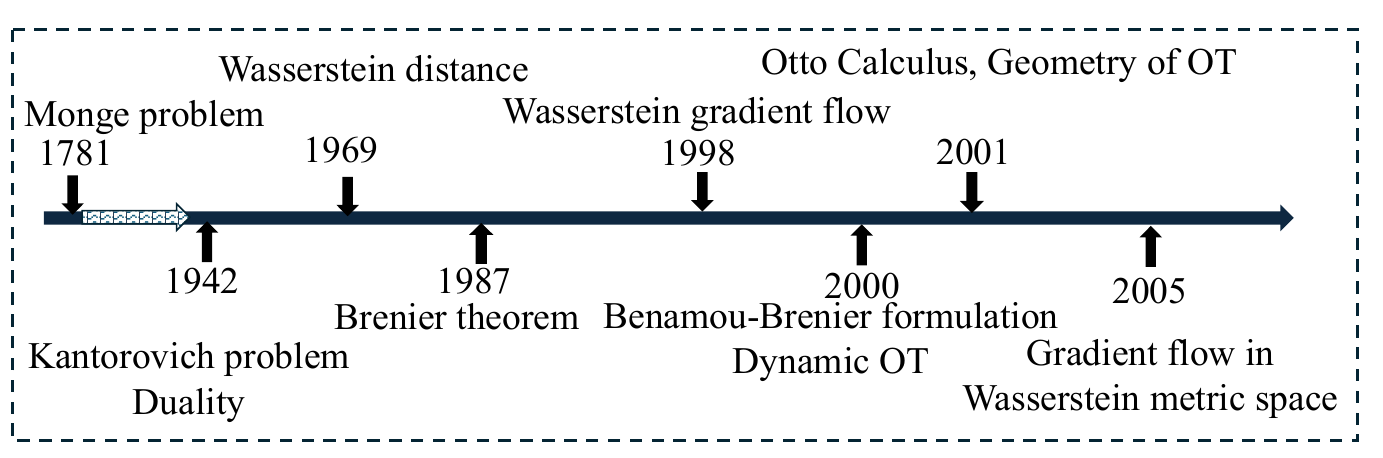}
		\caption{Landmark research activities.}
		\label{fig:fig2}
	\end{figure}

	During the development of OT theory, several historical moments or landmark activities have been recorded. A rough figure is shown in Fig. \ref{fig:fig2}. 
	The figure summarizes the major milestones in the development of OT theory. The concept originated in 1781, when Monge introduced what is now known as the Monge OT problem: finding an OT map that moves mass from one distribution to another while minimizing a prescribed cost. This formulation is elegant but difficult to solve, and, moreover, an optimal map does not always exist. Substantial progress was made in 1942 when Kantorovich proposed a relaxed version of Monge’s formulation by introducing the notion of a transport plan or coupling. This relaxation transformed the OT problem into a LP framework and allowed solutions to exist under far more general conditions. In the late 1960s, the concept of the Wasserstein distance was articulated, revealing that the Wasserstein metric is equivalent to the Kantorovich formulation. This insight provided a geometric interpretation of OT and connected it to probability theory, analysis, and geometry. However, for many years, the existence of OT maps remained unclear. A breakthrough occurred around 1987, when Brenier established his celebrated theorem, proving the existence and uniqueness of OT maps under suitable conditions, and showing that such maps arise as gradients of convex functions \cite{Brenier1991,MaggiBook2023}. This result has become foundational for modern computational OT, for example, many neural network-based OT methods such as an neural OT models and algorithms based on the Input Convex Neural Network (ICNN) model \cite{AmosICML2017, Makkuva2020}, directly build on the structure revealed by Brenier’s theorem. Further significant advances took place in the late 1990s. Around 1998-1999, researchers demonstrated that the evolution from one probability distribution to another can be interpreted as a Wasserstein gradient flow, governed by a PDE \cite{FigalliBook2023}. Shortly thereafter, around 2000, OT was linked to fluid dynamic formulations, giving rise to a dynamic viewpoint in which OT corresponds to minimizing kinetic energy over time \cite{Benamou2000}. This dynamic formulation inspired numerous extensions, including time-dependent OT models and applications in continuum mechanics \cite{Benamou2000}. The study of gradient flows in the Wasserstein space also led to the discovery of rich geometric properties of OT, most notably the development of Otto’s Calculus \cite{Jordan1998, Otto2001}. These geometric insights have had far-reaching consequences: they connected OT to the analysis of PDEs, informed variational approaches to diffusion equations, and motivated modern interpretations of stochastic differential equations (SDEs) through the lens of Wasserstein geometry \cite{Gangbo1996}. Together, these theoretical developments form the backbone of both classical and modern OT theory, laying the foundation for its widespread influence in mathematics \cite{Jordan1998}, physics, engineering, data science, and machine learning \cite{VillaniBook2003}. 
	\subsection{Wide concept connections to OT}
\begin{figure}[tbp]
		\centering
		\includegraphics[width=9cm, height=6cm]{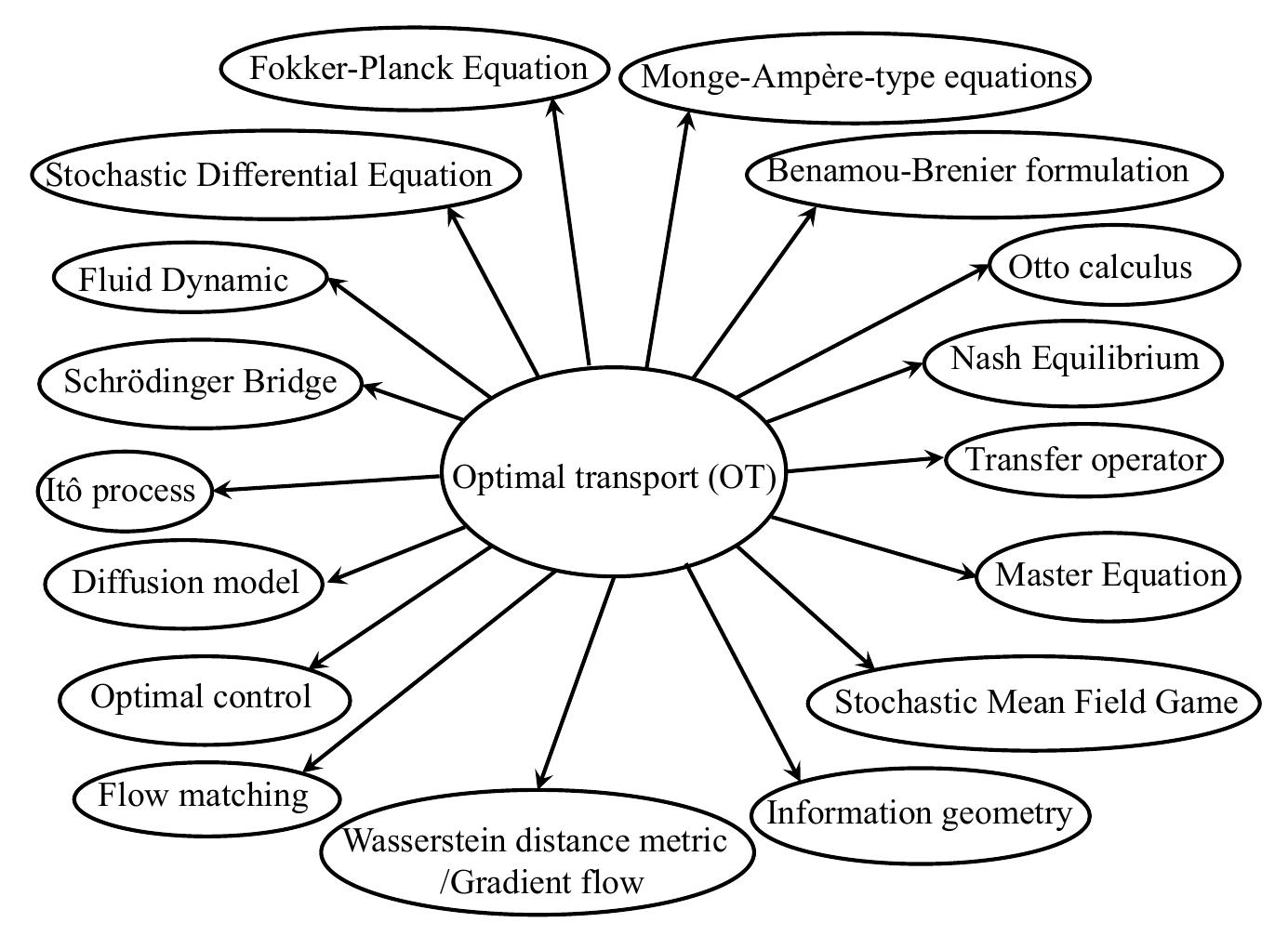}
		\caption{Wide concept connections to OT.}
		\label{fig:fig3}
	\end{figure}

	Based on OT theory, many concepts and methods could be connected to OT. Fig. \ref{fig:fig3} shows several concepts and methods we have ever met in different research fields which have a close relationship to OT. 
	For example, fluid dynamics, diffusion model, flow matching, etc., particularly their connections with recent generative model based machine learning algorithms. These concepts are original from physics and mathematics, and there is a trend in the machine learning society trying to apply these concepts to learning algorithms \cite{PeyreBook2019}. In applications, a lot of tools have been derived from these research fields and concepts. In the following sections, we will introduce the mathematical formulations and properties related to OT.
	
	\section{Monge's Optimal Transport Problem}
	\label{sect:MOT}
	Monge's OT problem is one of the classical formulations in OT theory. Given a probability measure \( \mu \) in a measurable space \( (X, \mathcal{A}) \), a probability measure \( \nu \) in a measurable space \( (Y, \mathcal{B}) \), and a cost function \( c : X \times Y \to [0, \infty) \), the goal is to find a measurable map \( T : X \to Y \) that minimizes the total transportation cost. The problem is formulated as \cite{VillaniBook2003, AmbrosioBook2021}:
	\begin{equation}
		\inf_{T} \int_{X} c(x, T(x)) \, d\mu(x),
		\label{eq:mot}
	\end{equation}
	subject to the constraint that the map \(T\) pushes the measure \(\mu\) onto \(\nu\):
	\begin{equation}
		T_{\#}\mu = \nu.
	\end{equation}
	
	Several key components appear in this formulation:
	\begin{itemize}
		\item \textbf{Transport map \(T\)}:  
		A measurable map that transfers the mass from the source measure \(\mu\) to the target measure \(\nu\). 
		The push-forward operation ensures that \(T\) redistributes \(\mu\)-mass to match \(\nu\).
		
		\item \textbf{Cost function \(c(x,y)\)}:  
		A nonnegative function defining the transportation cost between \(x \in X\) and \(y \in Y\). 
		Common choices include the Euclidean distance, its powers, or more general Minkowski-type distances.
		
		\item \textbf{Total cost of transport}:  
		The integral \(\int_X c(x, T(x)) \, d\mu(x)\), representing the total cost of transporting the mass 
		according to the map \(T\).
		
		\item \textbf{Infimum}:  
		The problem seeks the minimum possible value of the total cost over all admissible transport maps (indicated as symbol ``inf'' in Eq. (\ref{eq:mot}) ). 
		Informally, this may be regarded simply as a minimization problem.
	\end{itemize}
	
	In summary, the Monge formulation asks for an OT map \(T\) that moves the mass of \(\mu\) to 
	match \(\nu\) while achieving the smallest possible transport cost.
	
	\section{Kantorovich Relaxation of the Optimal Transport Problem}
	\label{sect:KOT}
	The formulation of the OT problem in Monge's sense suffers from several fundamental limitations. Monge's formulation requires a transport map $T$ that sends each point in the source domain to exactly one point in the target domain. This prohibits any splitting of the mass and implicitly assumes that $\mu$ and \(\nu\) have the same total mass and compatible structures. Consequently, the Monge problem may be ill-posed: a minimizer may not exist, and even when a solution exists, it is difficult to compute because the problem is inherently combinatorial \cite{PeyreBook2019}. A major breakthrough occurred in 1942 due to Kantorovich, who proposed a relaxed formulation of the transport problem. In Kantorovich's version, mass may be split and redistributed across multiple destinations, making the transportation process stochastic rather than deterministic. This relaxation replaces transport maps with transport plans (also called couplings). With this modification, the existence of minimizers is guaranteed under very mild conditions, and the theory becomes substantially more tractable. This relaxation marks the beginning of modern OT theory \cite{VillaniBook2003, AmbrosioBook2021}.
	
	\subsection{Kantorovich OT Problem}
	Kantorovich's formulation generalizes Monge's problem by replacing measurable maps with joint measures. Given probability measures \( \mu \) on \( (X, \mathcal{A}) \) and \( \nu \) on \( (Y, \mathcal{B}) \), and a cost function $c$ as defined in Monge's problem, the objective is to find a transport plan \( \gamma \) that minimizes the total transportation cost. A transport plan is a probability measure \( \gamma \) in \( X \times Y \) whose marginals are \( \mu \) and \( \nu \).
	The Kantorovich OT problem is then defined as:
	\begin{equation}
		\inf_{\gamma \in \Pi(\mu,\nu)} 
		\int_{X \times Y} c(x,y)\, d\gamma(x,y),
		\label{eq:kot}
	\end{equation}
	where the admissible set of couplings is
	\begin{equation}
		\begin{array}{l}
			\begin{aligned}
				\gamma(A \times Y) &= \mu(A), \qquad \forall A \in \mathcal{A}, \\
				\gamma(X \times B) &= \nu(B), \qquad \forall B \in \mathcal{B}. \\
			\end{aligned}
		\end{array}
		\label{eq:kotc1}
	\end{equation}
	
	In this formulation, a transport plan \( \gamma \) is a joint distribution $\prod {\left( {\mu ,\nu } \right)} $ on \( X \times Y \) that describes how mass is transferred from \( X \) to \( Y \). Its marginals satisfy:
	\begin{equation}
		\int_{Y} \gamma(x, dy) = \mu(dx),
		\qquad 
		\int_{X} \gamma(dx, y) = \nu(dy).
		\label{eq:kotc2}
	\end{equation}
	Unlike Monge's transport map, a transport plan allows the mass at a single location \( x \in X \) to be split and sent to multiple destinations.
	
	\subsection{Discrete Kantorovich OT}
\begin{figure}[tbp]
		\centering	
		\includegraphics[width=7cm, height=4cm]{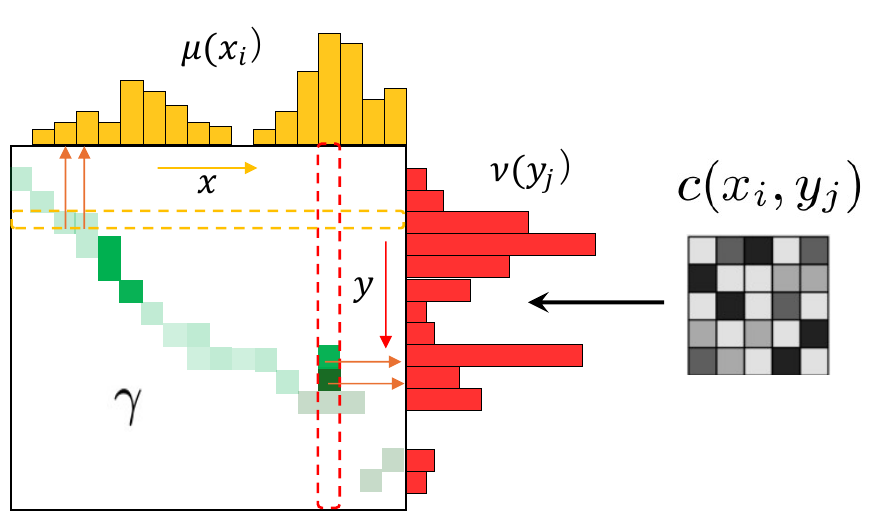}
		\caption{Mass splitting in discrete Kantorovich OT.}
		\label{fig:DKOTF}
	\end{figure}

	In the discrete setting, let  
	\( X = \{x_1, \ldots, x_m\} \) and  
	\( Y = \{y_1, \ldots, y_n\} \)  
	with corresponding probability distributions \( \mu \) and \( \nu \). Let \( c(x_i, y_j) \) denote the transportation cost from \( x_i \) to \( y_j \).  
	A discrete transport plan is represented by a nonnegative matrix $\gamma = (\gamma_{ij}) \in \mathbb{R}_{\ge 0}^{m \times n}$, where \( \gamma_{ij} \) specifies the fraction of mass moved from \( x_i \) to \( y_j \). The discrete Kantorovich OT problem becomes \cite{PeyreBook2019}:
	\begin{equation}
		\min_{\gamma \in \Pi(\mu,\nu)} \sum_{i=1}^{m} \sum_{j=1}^{n} \gamma_{ij}\, c(x_i, y_j), 
		\label{eq:dkot}
	\end{equation}
	subject to:
	\begin{equation}
		\begin{array}{l}	
			\begin{aligned}		
				& \sum_{j=1}^{n} \gamma_{ij} = \mu(x_i), \qquad \forall i, \\
				& \sum_{i=1}^{m} \gamma_{ij} = \nu(y_j), \qquad \forall j, \\
				& \gamma_{ij} \ge 0, \qquad\qquad\quad \forall i,j. \\
			\end{aligned}	
		\end{array}
		\label{eq:dkotc1}
	\end{equation}
	This discrete transport process is explained in Fig. \ref{fig:DKOTF}.
	As shown in this figure, Kantorovich's relaxation allows the mass at each \(x_i\) to be divided among multiple target points \(y_j\), making the coupling matrix \( \gamma \) a probabilistic representation of all permissible transport assignments. This relaxation not only guaranties the existence of minimizers but also provides the foundation for modern numerical OT algorithms and theoretical advancements.
	
	\section{Wasserstein Distance Metric}
	Based on the concepts and definitions explained in sections \ref{sect:MOT} and \ref{sect:KOT}, we now introduce an important notion that appears frequently in our research, i.e., the Wasserstein distance. This distance is defined between two probability measures and quantifies how far one distribution \( \mu \) is from another distribution \( \nu \). The terminology ``Wasserstein'' originates from the mathematician Leonid Vaseršteĭn, whose name appeared in the literature in this transliterated form. The \( p \)-Wasserstein distance is defined using the cost of transporting the mass with respect to a ground distance metric \( d(x,y) \). This construction leads to a mathematically well-behaved notion of distance with strong geometric and analytic properties. In particular, the Wasserstein distance is a metric, which allows us to study probability measures within the framework of metric geometry and functional analysis. For \( p \ge 1 \), the \( p \)-Wasserstein distance between probability measures \( \mu \) and \( \nu \) on a metric space \( (X, d) \) is defined as \cite{PeyreBook2019, VillaniBook2003}:
	\begin{equation}
		W_p(\mu, \nu)
		\mathop  = \limits^\Delta  \left(
		\inf_{\gamma \in \prod(\mu, \nu)}
		\int_{X \times X} d(x,y)^p \, d\gamma(x,y)
		\right)^{1/p},
	\end{equation}
	where \( \prod(\mu, \nu) \) denotes the set of all couplings of \( \mu \) and \( \nu \), i.e., all joint probability measures in \( X \times X \) whose marginals coincide with \( \mu \) and \( \nu \), respectively. Several properties are related to this Wasserstein distance:
	\begin{itemize}
		\item \textbf{Non-negativity:} \( W_p(\mu,\nu) \ge 0 \) and \( W_p(\mu,\nu)=0 \) if and only if \( \mu = \nu \).
		\item \textbf{Metric property:} \( W_p \) defines a metric in the space of probability measures with finite \( p \)-th moment.
		\item \textbf{Invariance:} \( W_p(\mu,\nu) \) is invariant under isometries of the underlying metric space.
	\end{itemize}
	Based on this definition, several special cases could be obtained:
	\begin{itemize}
		\item When \( p = 1 \), the Wasserstein distance \( W_1 \) is known as the Earth Mover’s Distance or the Kantorovich-Rubinstein metric. It is widely used in computer vision and machine learning, including the Wasserstein GAN framework \cite{ArjovskyICML2017}.
		\item When \( p = 2 \), the Wasserstein distance \( W_2 \) corresponds to the quadratic Wasserstein metric, central in OT theory and geometric interpretations of PDEs.
	\end{itemize}
	
	Because the Wasserstein distance is a genuine metric, the induced space:
	\begin{equation}
		\mathcal{W}_p(X)
		:\mathop  = \limits^\Delta  \left( \mathcal{P}_p(X),\, W_p \right),
	\end{equation}
	which is known as the Wasserstein space. This geometric structure enables various operations on probability distributions, including interpolation, computation of geodesics, and analysis of distributional flows. Furthermore, the geometric viewpoint has deep connections to PDEs. Many evolution equations, such as the continuity equation and several SDEs, can be interpreted as gradient flows in the Wasserstein space \cite{AmbrosioBook2008}. This perspective has become foundational in modern generative modeling, for example in diffusion-based generative AI, where stochastic diffusion processes admit gradient-flow formulations in Wasserstein geometry. The broader theoretical development of Wasserstein spaces is extensive and influential, although its details are beyond the scope of this review paper \cite{AmbrosioBook2008, SantambrogioBook2015, MaggiBook2023}.
	\section{Dual Problem of Optimal Transport}
	\label{sect:DualOT}
	OT theory, originating from Monge and later rigorously reformulated by Kantorovich, aims to determine the most cost-efficient way to transport the mass from one probability distribution to another. Kantorovich’s relaxation naturally leads to a linear programming (LP) formulation, which in turn provides access to its dual problem through LP duality. In this section, we introduce the primal and dual problems and explain their connections through the $c$-transform and Kantorovich potentials \cite{PeyreBook2019, VillaniBook2003}.
	
	\subsection{Primal Problem}
	Given two probability measures \( \mu \) and \( \nu \), and a measurable cost function  $c$, the Kantorovich OT problem seeks a coupling \( \gamma \in \Pi(\mu,\nu) \) minimizing the total transportation cost as defined in eqs. (\ref{eq:kot}), (\ref{eq:kotc2}) and (\ref{eq:dkot}), (\ref{eq:dkotc1}). Looking at their structures, they are formulated the same as used in LP which structure implies the existence of a corresponding dual problem. Intuitively, while the primal minimizes transportation cost, the dual maximizes the ``benefit'' associated with assigning potential values to the source and target points, subject to feasibility constraints.
	
	\subsection{Dual Problem}
	The dual of the Kantorovich OT problem introduces functions:
	\begin{equation}
		\phi : X \to \mathbb{R}, \qquad
		\psi : Y \to \mathbb{R},
	\end{equation}  
	called the Kantorovich potentials, satisfying the inequality constraint:
	\begin{equation}
		\phi(x) + \psi(y) \le c(x,y),
		\qquad \forall (x,y) \in X \times Y.
	\end{equation}
	The dual objective is then
	\begin{equation}
		\sup_{\phi,\psi}
		\left\{
		\int_{X} \phi(x)\, d\mu(x)
		+
		\int_{Y} \psi(y)\, d\nu(y)
		\right\}.
	\end{equation}
	
	\subsubsection*{$c$-transform and $c$-concavity}
	A key concept for understanding the dual problem is the $c$-transform \cite{VillaniBook2008}. For a function \( \phi : X \to \mathbb{R} \), define:
	\begin{equation}
		\phi^c(y)
		:\mathop  = \limits^\Delta  \inf_{x \in X}
		\left( c(x,y) - \phi(x) \right),
	\end{equation}
	and similarly,
	\begin{equation}
		\phi^{cc}(x)
		:\mathop  = \limits^\Delta  \inf_{y \in Y}
		\left( c(x,y) - \phi^c(y) \right).
	\end{equation}
	When \( \phi^{cc} = \phi \), the function \( \phi \) is called $c$-concave. The dual potentials satisfy relations of the form \cite{VillaniBook2008}:
	\begin{equation}
		\begin{aligned}
			\psi(y) &= \phi^c(y), \\
			\phi(x) &= \psi^{\bar c}(x)
			:\mathop  = \limits^\Delta  \inf_{y \in Y}
			\left( c(x,y) - \psi(y) \right).
		\end{aligned}
	\end{equation}
	Thus, dual optimization may be recast as optimizing a single $c$-concave function. This formulation is sometimes referred to as the semi-dual problem.
	
	\subsubsection*{Interpretation of Dual Variables}
	The dual functions \( \phi \) and \( \psi \) represent the marginal prices or potentials associated with the transport of mass. The inequality constraint ensures that no transport between \( x \) and \( y \) costs less than the assigned potential difference. Geometrically, these potentials describe the supporting hyperplanes of the cost function and play an essential role in understanding optimal couplings \cite{VillaniBook2008}.
	
	\subsection{Relationship Between Primal and Dual Problems}
	A fundamental result in OT is strong duality:
	\begin{equation}
		\begin{array}{l}
			{\rm inf}_{\gamma  \in \prod {\left( {\mu , \nu } \right)} } \int_{X \times Y} {c\left( {x,y} \right)d\gamma \left( {x,y} \right)}  \\ 
			\mathop  = \limits^\Delta  {\rm sup}_{\phi ,\psi } \left\{ {\int_X {\phi \left( x \right)d\mu \left( x \right)}  + \int_Y {\psi \left( y \right)d\nu \left( y \right)} } \right\}, \\ 
		\end{array}	
	\end{equation}
	subject to:
	\begin{equation}
		\phi(x) + \psi(y) \le c(x,y).
	\end{equation}
	Special cases include:
	\begin{itemize}
		\item \( c(x,y) = -\langle x,y \rangle \), in which case the dual takes a Legendre-transform-like form:
		\begin{equation}
			\psi(y)
			\mathop  = \limits^\Delta  \sup_{x \in X}
			\left( \langle x,y \rangle - \phi(x) \right).
		\end{equation}	
		\item \( c(x,y) = \|x - y\|^2 \), important in quadratic OT and Brenier’s theorem (will be introduced in section \ref{sect:Brenier}).
	\end{itemize}
	
	\subsection{Kantorovich-Rubinstein Distance from Duality}
	For the special case \( p = 1 \), duality yields the classical Kantorovich-Rubinstein formula:
	\begin{equation}
		W_1(\mu,\nu)
		\mathop  = \limits^\Delta  \sup_{\phi \in \mathrm{Lip}_1}
		\left(
		\int_{X} \phi(x)\, d\mu(x)
		-
		\int_{Y} \phi(y)\, d\nu(y)
		\right),
	\end{equation}
	where \( \mathrm{Lip}_1 \) is the set of 1-Lipschitz functions. This dual form is widely used in machine learning, particularly in the Wasserstein GAN (WGAN) framework \cite{ArjovskyICML2017}.
	\subsection{Applications based on the dual properties}
	The dual formulation plays a central role in economics, physics, PDE theory, and modern machine learning. Entropic regularization, for example, modifies the primal LP to yield smooth and computationally efficient dual objectives, forming the basis of Sinkhorn algorithms. When the ground cost \( c(x,y) \) is metric or convex in its arguments, alternating $c$- and $\bar c$-transforms naturally produces a pair of $c$-concave potentials. Hence, the dual problem can often be reduced to maximizing over a single $c$-concave potential, significantly simplifying numerical optimization \cite{Makkuva2020}. In summary, the dual problem provides an elegant and powerful perspective on OT, revealing the structure of optimal couplings through potential functions and enabling efficient computational methods across diverse applications.
	\section{Brenier's Theorem}
	\label{sect:Brenier}
	Brenier's theorem is a cornerstone of OT theory, particularly for the quadratic cost function. It provides powerful conditions that guaranty the existence and uniqueness of an OT map, and further shows that this map is the gradient of a convex function \cite{Brenier1991,VillaniBook2008}. Up to this point, we have introduced the basic concepts of Monge and Kantorovich formulations and the associated dual problems. A natural and fundamental question in OT is whether an OT map exists. If so, is it unique, and how is it related to the OT plan? Historically, these questions have been highly challenging to answer. For many decades it was unclear whether solutions to Monge’s original formulation even existed, except in special cases. In the late 20th century, Brenier resolved these issues in Euclidean spaces for the quadratic cost, establishing the existence, uniqueness, and structural form of the optimal map. This achievement is one of the landmark results in modern OT theory \cite{VillaniBook2003,PeyreBook2019}. Later, Gangbo and McCann extended Brenier's result to Riemannian manifolds, greatly broadening its applicability \cite{Gangbo1996}. In this section, we highlight the conclusions of Brenier's theorem that are most relevant for practical applications including machine learning, where the convex potential is often approximated using ICNNs \cite{AmosICML2017, Makkuva2020}.
	
	\subsection{Theorem Statement}
	Let \( \mu \) and \( \nu \) be two probability measures in \( \mathbb{R}^n \). Assume that \( \mu \) is absolutely continuous with respect to the Lebesgue measure. Consider the quadratic cost function
	$c(x,y) = |x - y|^2$, then there exists a convex function \( \varphi : \mathbb{R}^n \to \mathbb{R} \) such that the OT map \( T : \mathbb{R}^n \to \mathbb{R}^n \) is given by:
	\begin{equation}
		T(x) = \nabla \varphi(x).	
	\end{equation}
	Furthermore, this map advances \( \mu \) to \( \nu \) with meaning $T_\# \mu = \nu$, or equivalently for every measurable set \( B \subseteq \mathbb{R}^n \),
	$\nu(B) = \mu(T^{-1}(B))$.
	
	\subsection{OT Map}
	The OT map \( T = \nabla \varphi \) minimizes the cost of quadratic transportation:
	\begin{equation}
		\int_{\mathbb{R}^n} |x - T(x)|^2 \, d\mu(x).
	\end{equation}
	Because the cost is strictly convex, any mass at point \( x \) is optimally transported to a unique point \( T(x) \).
	
	\subsection{Existence and Uniqueness}
	The convex potential \( \varphi \) is unique up to an additive constant. Consequently, the OT map $T = \nabla \varphi$ is unique almost everywhere with respect to \( \mu \). This result gives a complete solution to Monge's formulation under the stated conditions. Based on the above explanations, Brenier’s theorem gives the following implications:
	
	\begin{itemize}
		\item \textbf{Convex Potential Representation.}  
		For quadratic cost, the optimal map is always the gradient of a convex function. This greatly simplifies analysis and computation.
		
		\item \textbf{Regularity.}  
		If \( \mu \) and \( \nu \) have smooth, strictly positive densities and satisfy the appropriate convexity conditions, then \( \varphi \) is smooth, and so is \( T \).
		
		\item \textbf{Machine Learning.}  
		In modern applications, the convex potential \( \varphi \) is often parameterized using ICNNs \cite{AmosICML2017, Makkuva2020}, making Brenier's theorem a foundation of neural OT methods.
	\end{itemize}
	
	The theorem answers two classical questions at once:  
	(1) the existence of an optimal map, and  
	(2) its structural relationship with the optimal plan.  
	Moreover, whenever a Monge map exists, it always induces the corresponding Kantorovich optimal plan.
	
	\subsection{Relationship Between OT Map and Plan}
	OT theory distinguishes between two objects:
	\begin{itemize}
		\item The OT map \( T : X \to Y \), which gives a deterministic assignment.  
		\item The OT plan \( \gamma \in \Pi(\mu,\nu) \), which is a joint probability distribution describing how mass is transported.
	\end{itemize}
	The two concepts are closely related but not always identical. Brenier's theorem clarifies this relationship in the quadratic-cost case. For a strictly convex cost of the form:
	$c(x,y) = h(x-y)$ with \( h(\cdot) \) convex, the optimal map has the form:
	\begin{equation}
		T(x) = x - \nabla h^{-1}(\nabla \varphi(x)),	
	\end{equation}
	where \( \varphi \) is a convex potential associated with the dual problem. If a unique optimal map \( T \) exists, the map will induce a plan that is deterministic and given by: 
	\begin{equation}
		\gamma = (\mathrm{Id} \times T)_\# \mu ,
	\end{equation}
	where $\mathrm{Id}$ is an identity matrix. In most general cases, if \( \mu \) is not absolutely continuous, or if the cost is not strictly convex, then an optimal map may not exist. In such cases, the mass may split: the optimal plan \( \gamma \) encodes probabilistic couplings rather than deterministic assignments. For example, in the deterministic case, if both \( \mu \) and \( \nu \) have continuous densities and the cost is strictly convex (e.g., \( |x - y|^2 \)), the optimal plan is induced by a unique map \( T = \nabla \varphi \). For discrete distributions, the OT plan is typically a transport matrix \( \gamma \). There is generally no single map unless the problem reduces to a perfect matching. In summary, the OT plan is the most general object, always guaranteed to exist. The OT map is a special but highly valuable structure: it exists under regularity and convexity assumptions, and when it exists, it fully determines the plan. Brenier’s theorem provides the precise conditions under which this powerful simplification holds.
	
	\section{Optimal Transport in Non-Comparable Spaces}
\begin{figure}[tbp]
		\centering
		\includegraphics[width=8cm, height=3.5cm]{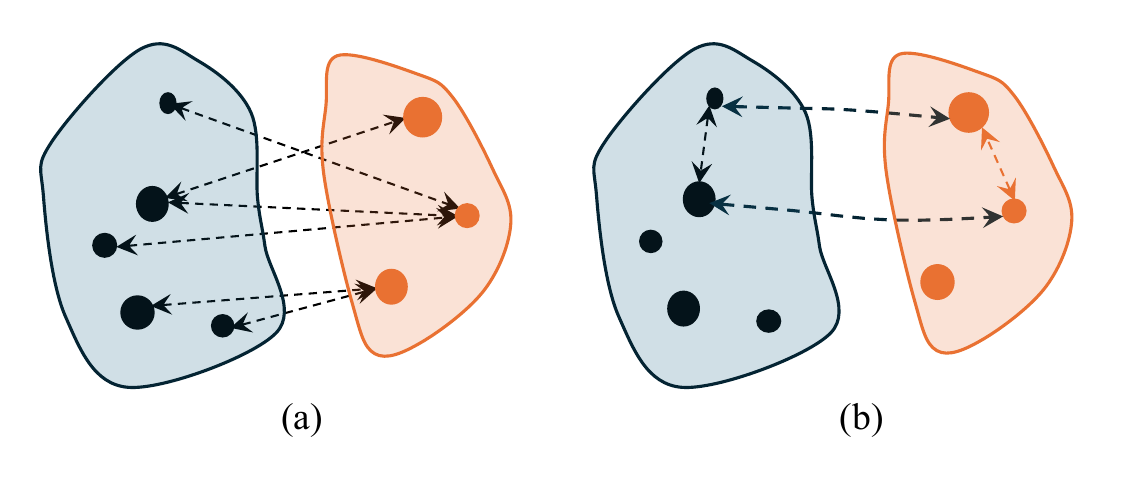}
		\caption{OT as graph matching on nodes (a) and edges (b) as conventional definition of OT and GWOT, respectively.}
		\label{fig:fig6}
	\end{figure}

	In previous sections, we discussed OT in settings where the cost function or distance metric is defined on a shared space, i.e., both probability measures live on the same underlying domain, allowing distances between samples to be computed directly. However, many real-world applications involve probability distributions defined on different, and often non-comparable feature spaces. In such situations, the classical Wasserstein distance is no longer applicable because the ground cost $d(x,y)$ (instead of using $c(x,y)$ as a distance representation) cannot be defined meaningfully. This motivates the use of Gromov-Wasserstein optimal transport (GWOT), which compares distributions through the intrinsic structure of their own metric spaces \cite{Peyre2016,Vayer2020}. Rather than comparing samples directly, GWOT aligns the pairwise distances within each space. This concept is illustrated in Fig.~\ref{fig:fig6}, where matching can be operated on either nodes (in comparable spaces) or edges (in non-comparable spaces). 
	GWOT is particularly useful for comparing heterogeneous modalities. For instance, acoustic features, text embeddings, and video descriptors, each lies in distinct feature spaces where cross-modal distances are undefined. Yet, intra-modal distances (within each modality) remain meaningful. GWOT constructs an optimal coupling by aligning these intra-modal geometries across modalities. This is conceptually similar to graph matching, where correspondences between nodes must also respect the similarity between edges. Moreover, when temporal structure is present (e.g., in speech or video), GWOT naturally relates to directed graph matching, because temporal order introduces directionality into the pairwise relationships. 
	
	In this section, we introduce the Gromov-Wasserstein distance (GWD) and show how it leads to a more general hybrid formulation known as the fused Gromov-Wasserstein distance (FGWD), which incorporates both the intrinsic geometric discrepancy between spaces and the conventional Wasserstein discrepancy between samples. These tools will later be used for cross-modal graph matching and knowledge transfer in ASR applications \cite{LuIS2025}.
	
	\subsection{Gromov-Wasserstein OT}
	GWOT aligns two metric measurement spaces \( (X, d_X, \mu) \) and \( (Y, d_Y, \nu) \) by minimizing the distortion between their respective pairwise distances. In the discrete setting, let \( \mathbf{D}^X \in \mathbb{R}^{n \times n} \) and \( \mathbf{D}^Y \in \mathbb{R}^{m \times m} \) denote the intra-space distance matrices. GWOT seeks a coupling \( \gamma \in \Pi(\mu,\nu) \) such that the relational structures (i.e., the edges) of the two spaces are best aligned. The GWOT problem is defined as \cite{Peyre2016,Vayer2020}:
	\begin{equation}
		\min_{\gamma \in \Pi(\mu,\nu)} 
		\left\langle 
		\mathcal{D}\!\left(\mathbf{D}^X , \mathbf{D}^Y \right) \otimes \gamma, \gamma 
		\right\rangle,
	\end{equation}
	where $\mathcal{D}(\cdot)$ is a distance function. In fully discrete form as:
	\begin{equation}	
		\min_{\gamma \in \Pi(\mu,\nu)} 
		\sum_{i,j,k,l} 
		d_{i,j,k,l}\,\gamma_{i,k}\gamma_{j,l}, 
	\end{equation}
	where the distortion tensor evaluates pairwise discrepancies:
	\begin{equation}
		d_{i,j,k,l} :\mathop  = \limits^\Delta \mathcal{D}\!\left( {\mathbf{D}}^X_{i,j},\, {\mathbf{D}}^Y_{k,l} \right).	
	\end{equation}
	
	This formulation emphasizes that GWOT compares structures (edges) rather than samples (nodes) as a graph matching schematic point of view conceptually illustrated in Fig.~\ref{fig:fig6}. It is therefore well suited for matching distributions whose supports are not directly comparable.
	
	\subsection{Fused Gromov-Wasserstein OT}
	In many applications, both node features and edge structures provide complementary information. For example, in multimodal learning, acoustic and text embeddings (node features) can be compared via a learned cross-modal cost, while their temporal or relational structures (edges) must also be considered. To capture both aspects simultaneously, the fused Gromov-Wasserstein OT (FGWOT) combines the classical Wasserstein alignment of node features with GW alignment of pairwise structures. The fused objective is defined as:
	\begin{equation}
		\min_{\gamma \in \Pi(\mu,\nu)}
		(1 - \alpha)\,
		\left\langle \mathbf{D}^{X,Y},\, \gamma \right\rangle
		+
		\alpha\,
		\left\langle 
		\mathcal{D}\!\left( \mathbf{D}^X, \mathbf{D}^Y \right) \otimes \gamma,\gamma
		\right\rangle,
		\label{eq:fgwot}
	\end{equation}
	where $\mathbf{D}^{X,Y}$ is the ground cost comparing node features across spaces (used in classical OT as $\mathbf{C}^{X,Y}$), $\alpha \in [0,1] $ balances node-level vs. edge-level alignment, the second term enforces relational (edge) consistency. When \( \alpha = 0 \), FGWOT reduces to classical OT; when \( \alpha = 1 \), FGW reduces to pure GWOT. Based on this formulation of the fused objective in Eq. (\ref{eq:fgwot}), we can observe the following.
	\begin{itemize}
		\item The first term enforces point-wise similarity, aligning samples across modalities.
		\item The second term enforces structural similarity, encouraging pairs of points with similar relationships in one space to correspond to pairs of similar relationships in the other.
		\item The FGW distance thus behaves like a graph-matching distance that incorporates both node attributes and edge structures.
	\end{itemize}
	
	This makes FGWOT particularly suitable for:
	\begin{itemize}
		\item Cross-modal learning (acoustic-text, text-vision, etc.),
		\item Temporal sequence alignment (speech frames, video frames),
		\item Knowledge transfer across heterogeneous architectures,
		\item Graph matching and graph-based signal processing.
	\end{itemize}
	In a later section \ref{sec:gmotASR}, we will show the application of GWOT and FGWOT formulations to cross-modality knowledge transfer in ASR tasks, where acoustic and text distributions lie in different feature spaces but share structural relationships that can be exploited through OT.
	
	\section{Unbalanced Optimal Transport}
	In the original definition of OT, marginal distribution constraints are required to be kept (Eqs. (\ref{eq:kotc1}), (\ref{eq:kotc2}), (\ref{eq:dkotc1})), i.e., the masses of source and target should be equal in transport. These constraints can be relaxed to increase the potential of OT applications. For example, in cross-modal alignment between acoustic and linguistic distributions, alignment for matching requires to identify meaningful and reliable correspondences between acoustic and linguistic spaces, acoustic frames with no corresponding linguistic transcriptions (e.g., noisy backgrounds) could be ignored or discarded in matching to linguistic transcriptions (tokens) \cite{LuICASSP2026}. Also in semantic interactions between sentences through word alignment \cite{Arase2023}, it is natural that information inequality exists between sentences. This requirement fits well to the mathematical theory of un-balanced OT (UOT), i.e., controlling the marginal distributions cross modalities during OT. The UOT has been investigated in \cite{ChizatUOT2016,  UOT2023, NguyenUOT2023}. Suppose that the two discrete probability distributions with weight coefficients are as follows:
	\begin{equation}
		\begin{array}{l}
			{\mu}  = \left( {u_1 ,u_2 ,...,u_m } \right)^{\top} \in \mathbb{R}^{m}, \\
			{\nu}  = \left( {v_1 ,v_2 ,...,v_n } \right)^{\top}  \in \mathbb{R}^{n}, \\
		\end{array}	
	\end{equation} 
	with the pairwise cost matrix $\mathbf{C} \in \mathbb{R}^{m \times n} $, the UOT is defined as:
	\begin{equation}
		L_{{\rm UOT}} \mathop  = \limits^\Delta  \mathop {\min }\limits_{\gamma  \in \mathbb{R}_ + ^{m \times n} } \sum\limits_{i,j}^{m,n} {\gamma _{i,j} C_{i,j}  + \alpha _1 \mathcal{D}(\gamma {\bf 1}_n ||\mu ) + \alpha _2 \mathcal{D}(\gamma ^{\top} {\bf 1}_m ||\nu )}, 
		\label{eq:uot}		
	\end{equation}
	where $ \mathbf{1}_m \in \mathbb{R}^{m} $ and $ \mathbf{1}_n \in \mathbb{R}^{n} $ are vectors of ones, $ \gamma \mathbf{1}_n \in \mathbb{R}^{m} $ and $\gamma^\top \mathbf{1}_m \in \mathbb{R}^{n} $ are the marginals of row and column of $\gamma$, $\mathcal{D}(\cdot \| \cdot)$ is a Kullback-Leibler (KL)-divergence based distance function, $\alpha_1, \alpha_2 \geq 0 $ control the penalty on deviation from the two original marginals, when $\alpha_1, \alpha_2 \to \infty  $, equal mass constraint is guaranteed which is the original definition of OT. With setting a different threshold ratio to maintain the mass transport, partial OT (POT) shares a similar function to the partial alignment between two distributions which has also been investigated \cite{ChapelPOT2020}.
	
	\section{Dynamic Optimal Transport}
\begin{figure}[tbp]
		\centering
		\includegraphics[width=7cm, height=5cm]{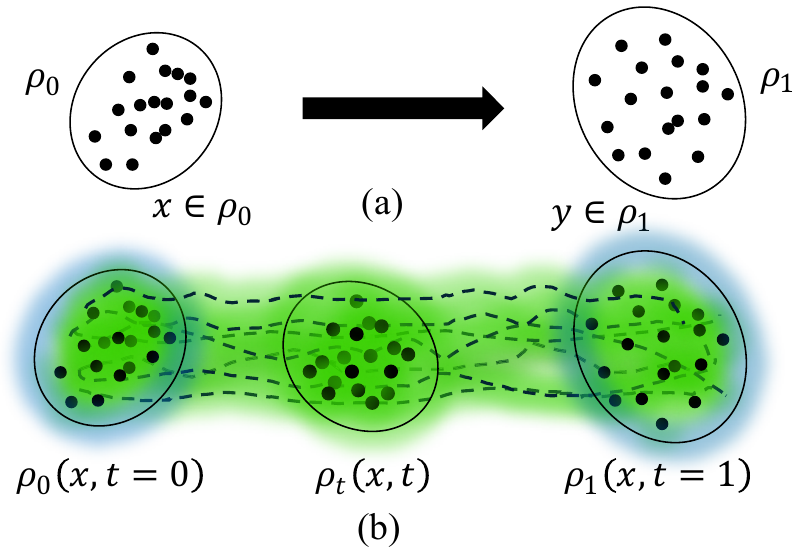}
		\caption{(a) Static OT; (b) Dynamic OT, distribution changes along an efficient path.}
		\label{fig:fig7}
	\end{figure}

	In the previous sections, OT was introduced in its static formulations, where no temporal evolution is considered. Conceptually, the classical OT problem seeks a transport map that moves an initial probability distribution $\rho _0 (x)$ to a target distribution $\rho _1 (y)$ with minimal transportation cost. A natural extension of this idea is to ask: how does a distribution evolve over time along an energetically efficient path on the space of probability measures? This perspective introduces an additional temporal dimension and leads to the framework of dynamic OT as illustrated in Fig. \ref{fig:fig7}, i.e., the distributions are changed along a trajectory with time-variant distribution $\rho _t$. 
	In this sense, OT can be regarded as a transport problem on a manifold of probability distributions, endowed with a well-defined geometry. The dynamic formulation of OT has a strong connection to fluid mechanics and optimal control theory \cite{Chen2021}. The celebrated Benamou-Brenier formulation interprets OT as a fluid flow problem, where the goal is to determine both the evolution of the density and the velocity field that together transport the fluid from the initial to the final configuration with minimal kinetic energy \cite{Benamou2000}.
	
	\subsection{Lagrangian and Eulerian Descriptions of Fluid Dynamics}
	The Benamou-Brenier formulation is based directly on the principles of fluid dynamics, incorporating both spatial and temporal structures of mass evolution \cite{Benamou2000}. In fluid mechanics, two complementary modeling viewpoints are widely used, as shown in Fig. \ref{fig:fig8}:
	\begin{figure}[tbp]
		\centering
		\includegraphics[width=6cm, height=3cm]{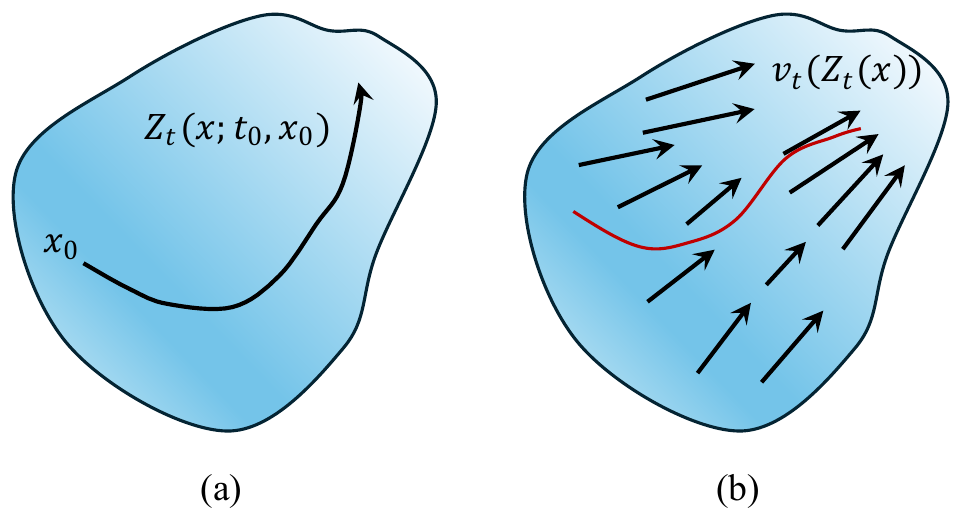}
		\caption{Lagrangian (a) and Eularian (b) point of views for fluid dynamics.}
		\label{fig:fig8}
	\end{figure} 
	\begin{itemize}
		\item \textbf{Lagrangian description}: tracks individual particles along their trajectories, where each particle is initially located at position $x_0$ at time $t_0$.
		\item \textbf{Eulerian description}: studies the evolution of physical quantities (density, velocity, etc.) at fixed spatial locations described by a velocity field $v_t$.
	\end{itemize}
	
\begin{figure}[tbp]
		\centering
		\includegraphics[width=7cm, height=3.5cm]{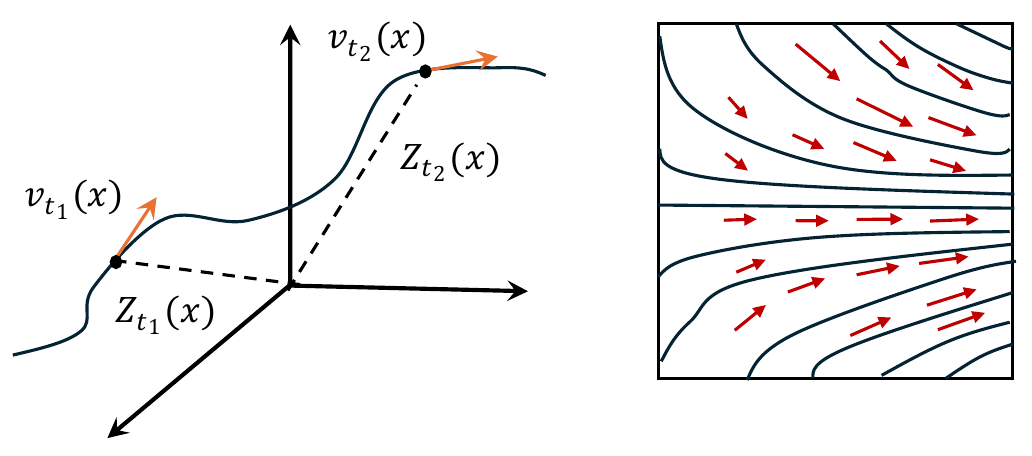}
		\caption{Velocity as tangent directions along a particle trajectory (a) and on streamlines of population of particles (b).}
		\label{fig:fig9}
	\end{figure}

	These two descriptions can be formally linked by a flow map $Z_t(x)$ that satisfies the differential equation:
	\begin{equation}
		\label{eq:Lagrangian}
		\begin{cases}
			\dfrac{dZ_t(x)}{dt} = v_t\left(Z_t(x)\right), \\
			Z_0(x) = x,
		\end{cases}
	\end{equation}
	where $v_t$ denotes the velocity field at time $t$. The flow map induces an evolution of densities through the push-forward operation:
	\begin{equation}
		\label{eq:pushforward}
		\rho_t = (Z_t)_\# \rho_0.
	\end{equation}
	Hence, a time-varying velocity field generates a corresponding temporal evolution of the density. As further illustrated in Fig. \ref{fig:fig9}, intuitively, the velocity field describes the tangent direction along each streamline of the flow, while the set of all streamlines describes the motion of a large population of particles.
	
	\subsection{Continuity Equation in Fluid Dynamics}
	In fluid dynamics, mass conservation is expressed through the continuity equation, which governs the evolution of density $\rho_t(x)$ under a velocity field $v_t(x)$ as \cite{VillaniBook2003, SDEBook2019}:
	\begin{equation}
		\label{eq:continuity}
		\frac{\partial \rho_t(x)}{\partial t} + \nabla \cdot \big( \rho_t(x)\, v_t(x) \big) = 0.
	\end{equation}
	Here, $\partial_t \rho_t$ represents the rate of change in density, while $\nabla \cdot (\rho_t v_t)$ describes the divergence of the mass flux generated by the velocity field. This equation enforces the principle of mass conservation: mass is neither created nor destroyed, but is only redistributed in space. In the context of OT, the continuity equation acts as a constraint linking the density evolution and the velocity field. It ensures that feasible density paths $\rho_t$ correspond to physically valid mass flows. This continuity equation forms the basis for understanding how densities evolve under OT processes, ensuring that the fundamental conservation laws are satisfied throughout the dynamics which is frequently mentioned in the later sections.
	
	\subsection{Benamou-Brenier Formulation of Dynamic OT}
	Based on the above introduction, the dynamic OT problem aims to find both the density path $(\rho_t)_{t \in [0,1]}$ and the velocity field $(v_t)_{t \in [0,1]}$ that transport $\rho_0$ to $\rho_1$ while minimizing the kinetic energy. The Benamou-Brenier formulation expresses the squared Wasserstein-2 distance as the optimal value of the following variational problem \cite{Benamou2000, VillaniBook2003}:
	\begin{equation}
		\label{eq:BB-functional}
		W_2^2(\rho_0, \rho_1)
		\mathop  = \limits^\Delta  \min_{\rho,\, v}
		\left\{
		\int_0^1 \int_{X}
		L\!\left( v(x,t) \right)\,
		\rho(x,t)\, dx\, dt
		\right\},
	\end{equation}
	subject to the constraints:
	\begin{equation}
		\label{eq:boundary}
		\rho(\cdot,0) = \rho_0, \qquad
		\rho(\cdot,1) = \rho_1,
	\end{equation}
	and the continuity equation \eqref{eq:continuity}. A common choice for the Lagrangian is the kinetic energy $L(v) = \frac{1}{2}\|v\|^2$. This formulation reveals that, rather than computing a static coupling, OT with a quadratic cost is equivalent to finding a time-varying flow of probability densities that smoothly interpolates between the two measures along a minimal-energy path. The optimal path $\rho_t$ serves as the geodesic between $\rho_0$ and $\rho_1$ in the Wasserstein space.
	
	\subsection{Benamou-Brenier Formula and Gradient Flows in Wasserstein Space}
	The Benamou-Brenier formulation also provides a geometric interpretation of the Wasserstein distance and plays a fundamental role in the study of gradient flows in the Wasserstein space. For two probability measures $\mu_0$ and $\mu_1$, the dynamic formulation expresses the Wasserstein distance as \cite{ SantambrogioBook2015, AmbrosioBook2021,FigalliBook2023}:
	\begin{equation}	
		W_2^2(\mu_0, \mu_1)
		\mathop  = \limits^\Delta  \int_0^1 \int_\mathcal{X} 
		\| \nabla \phi_t(x) \|^2\, d\rho_t(x)\, dt,
		\label{eq:BB-gradient-flow}
	\end{equation}
	where $\phi_t$ is the time-dependent OT potential and $(\rho_t, \nabla \phi_t)$ satisfy the continuity equation
	\[
	\frac{\partial \rho_t}{\partial t} + \nabla \cdot ( \rho_t \nabla \phi_t ) = 0.
	\]
	
	This dynamic formulation plays a central role in modern analysis and machine learning, enabling:
	
	\begin{itemize}
		\item Variational formulations of diffusion equations,
		\item Schr{\"o}dinger bridge problems,
		\item Score-based generative modeling and diffusion models,
		\item Optimal control formulations,
		\item Continuous-time interpolation between probability distributions.
	\end{itemize}
	
	The Benamou-Brenier perspective therefore provides not only a physically intuitive framework for OT but also a mathematically rich foundation for dynamic modeling on the space of probability measures which plays a crucial role in understanding the geometry of Wasserstein spaces and provides a variational characterization of the Wasserstein distance, connecting it directly to OT maps and gradient flows. 
	
	\section{Transport-Based Modeling}
	In the Benamou-Brenier formulation of OT, the transport trajectory is described by a deterministic differential equation whose boundary conditions are given by the initial and terminal distributions $\rho_0$ and $\rho_1$. A natural extension of this idea is to ask: what if the transport trajectory itself follows a stochastic process? This question leads to the diffusion-bridge view of OT, most notably expressed by the Schr{\"o}dinger Bridge Problem (SBP) \cite{Leonard2014, Chen2014, Chen2021}. 
	\begin{figure}[tbp]
		\centering
		\includegraphics[width=7cm, height=6.5cm]{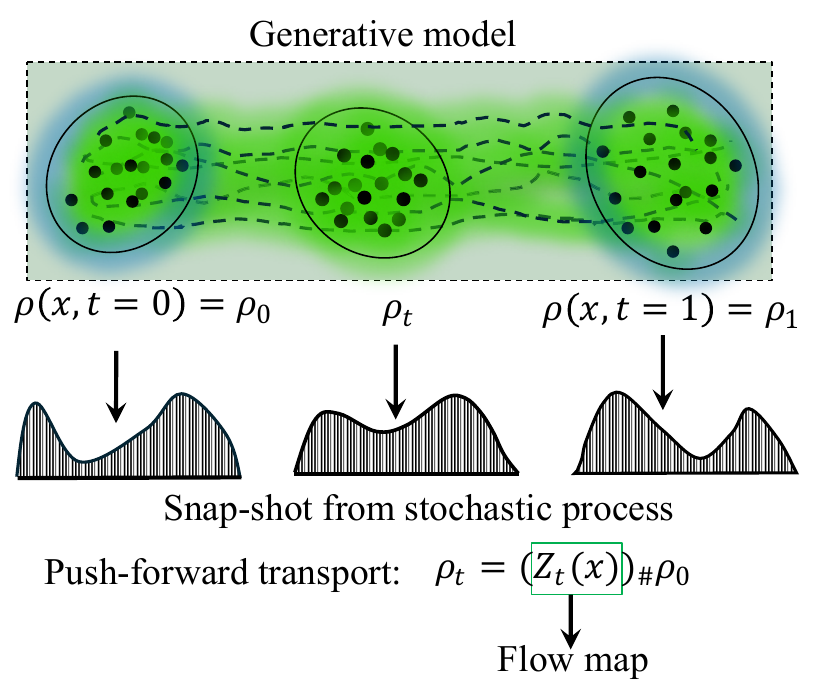}
		\caption{Transport-based modeling with a flow map.}
		\label{fig:fig10}
	\end{figure} 
	In recent years, several generative modeling frameworks have emerged-including diffusion models, score-based models, flow matching, and normalizing flows that all describe transformations between probability distributions through continuous-time dynamics. These generative modeling frameworks can be simply described as in Fig. \ref{fig:fig10}. Observations or samples are regarded as snap-shots from different distributions along time evolution $\rho_t$. Although their algorithmic motivations differ, these approaches share a unified structure: each specifies a transport mechanism (deterministic or stochastic) that maps a source distribution to a target distribution under suitable constraints. From this perspective, they can be collectively viewed as transport-based modeling under a unified stochastic differential equation (SDE) framework.
	
	In this section, we introduce the key theoretical foundations underlying these connections, beginning with the Schr{\"o}dinger Bridge and diffusion bridges, followed by SDE-based diffusion models and flow-matching methods, and concluding with a transfer-operator perspective that further unifies these viewpoints.
	
	\subsection{Schr{\"o}dinger Bridge Problem and Diffusion Bridges}
	The SBP is the most classical and influential example of a diffusion bridge. Originally studied by Schr{\"o}dinger in 1931-1932, the problem asks for the most likely evolution of a cloud of particles observed at two time points, with prescribed initial and terminal marginals. In modern terms, the SBP seeks a probability law on the space of stochastic paths that: (i) matches the two marginals, and (ii) is closest in relative entropy to a given reference process. In Schr{\"o}dinger's original setting, this reference was the standard Brownian motion. Formally, let $P_{0:T}$ be the probability law of the reference process. The SBP seeks a path measure $Q_{0:T}$ that satisfies the marginal constraints $Q_0 = \rho_0$ and $Q_T = \rho_1$ and minimizes the KL-divergence \cite{Chen2021}:
	\begin{equation}
		Q_{0:T}^\ast 
		:\mathop  = \limits^\Delta \arg\min_{Q_{0:T}:Q_0=\rho_0,\;Q_T=\rho_1}
		\mathrm{KL}(Q_{0:T}\,\|\,P_{0:T}).
		\label{eq:SBPKL}
	\end{equation}
	Intuitively, the SBP constructs a stochastic ``bridge'' between the two distributions by perturbing the reference process as little as possible while still matching the desired marginals. Thus, SBP can be interpreted as a stochastic analog of OT, where randomness is explicitly incorporated through diffusion.
	
	\subsubsection{Reference Process in Schr{\"o}dinger Bridge}
	Let the reference process be an It\^{o} diffusion:
	\begin{equation}
		dX_t = f(X_t,t)\,dt + g(X_t,t)\, dW_t,
		\label{eq:ItoSDE}
	\end{equation}
	where $f(X_t,t)$ is a drift term, $g(X_t,t)$ denotes diffusion coefficient, and $W_t$ is a Wiener process. The SBP seeks an alternative process $\mathbb{Q}$, absolutely continuous with respect to the reference law $\mathbb{P}$, that minimizes:
	\begin{equation}
		\inf_{\mathbb{Q}}
		\mathbb{E}_{\mathbb{Q}} 
		\left[
		\log \frac{d\mathbb{Q}}{d\mathbb{P}}
		\right],
	\end{equation}
	subject to
	\[
	\mathbb{Q}(X_0)=P_{0},
	\qquad
	\mathbb{Q}(X_T)=P_{T}.
	\]
	
	From the viewpoint of stochastic optimal control, this minimization induces a controlled SDE:
	\begin{equation}
		dX_t = f(X_t,t)\,dt + g(X_t,t)\,dW_t + u(X_t,t)\, dt,
	\end{equation}
	where the control $u$ is the drift adjustment required to satisfy the marginal constraints and minimize the relative entropy. This structure yields a dynamic programming equation closely related to the Monge-Amp{\`e}re equation in deterministic OT \cite{Bunne2023}.
	
	\subsubsection{Connection Between Static OT and SBP via KL-Divergence}
	Consider the entropy-regularized OT (entropic OT) problem defined as (this will be further discussed in section \ref{sect:EROTSH} for fast computational algorithms):
	\begin{equation}
		\min_{\gamma\in\Pi(\mu,\nu)}
		\left\{
		\int c(x,y)\, d\gamma(x,y)
		\;-\;
		\lambda\, H(\gamma)
		\right\},
	\end{equation}
	where $H(\gamma) = -\!\int \gamma\log\gamma$ is Shannon entropy of transport plans. With re-arrangements, the entropic OT becomes:
	\begin{equation}
		\min_{\gamma\in\Pi(\mu,\nu)}
		\lambda\, \mathrm{KL}\!\big(
		\gamma
		\;\Vert\;
		e^{-\tfrac{1}{\lambda} C}
		\big).
	\end{equation}
	Thus, entropic OT is also a KL projection onto couplings, with the Gibbs kernel $e^{-\tfrac{1}{\lambda} C}$ playing the role of a reference measure. Comparing this with the SBP's KL projection onto a path measure $Q_{0:T}$ in Eq. (\ref{eq:SBPKL}) reveals a deep equivalence: SBP is the dynamic version of entropy-regularized OT \cite{Leonard2014}. This analogy forms the basis of many modern computational algorithms that unify SBP and entropic OT, enabling tractable approximations of OT in high-dimensional spaces.
	
	\subsubsection{Connection Between Dynamic OT and SBP via SDEs}
	A general stochastic process is given by the It\^{o} SDE as in Eq. (\ref{eq:ItoSDE}), if $g= 0$, the dynamics reduce to a deterministic ODE and correspond to the Benamou-Brenier formulation of OT. If only the diffusion term is present (as in Brownian motion), the dynamics describe the original SBP. In the general controlled-SDE form, SBP can also be expressed as the minimization of drift energy \cite{Chen2014, Chen2021}:
	\begin{equation}
		\min_{f\in\mathcal{U}}
		\mathbb{E}
		\left[
		\int_0^T \frac{1}{2}\|f(X_t,t)\|^2\, dt
		\right],
	\end{equation}
	subject to the marginal constraints
	\[
	X_0\sim P_0, \qquad X_T\sim P_T.
	\]
	
	Through the Fokker-Planck equation, these Lagrangian (path-space) formulations have Eulerian (density-space) counterparts. This connection places OT, SBP, and SDEs within a unified framework of stochastic optimal control \cite{Chen2021}.
	
	\subsection{Stochastic Differential Equations and Diffusion Models}
	In physics and applied mathematics, a fundamental question is how to model interacting particles and describe their temporal evolution. Such systems often exhibit both deterministic trends and random fluctuations, motivating a probabilistic formulation of dynamics. SDEs provide a natural and expressive framework for modeling these behaviors, where particle ensembles evolve as probability distributions that change continuously over time. The SDEs describe the evolution of a stochastic process \( X_t\) with dynamics of the form formulated in Eq. (\ref{eq:ItoSDE}) with two terms \( f \) is the drift field, \( g \) is the diffusion coefficient, where \( \mathrm{d}W_t \) denotes a Wiener process. These two terms correspond respectively to deterministic flow and random perturbations. The associated evolution of probability densities is governed by the Fokker-Planck equation (FPE) \cite{SDEBook2019}:
	\begin{equation}
		\frac{\partial \rho_t(x)}{\partial t}
		= -\nabla \cdot ( f_t(x)\rho_t(x) ) 
		+ \frac{1}{2}\nabla \cdot \nabla\!\left( D_{ij}(x,t)\rho_t(x) \right),
	\end{equation}
	where \( D_{ij}(x,t) = [g_t(x)g_t^\top(x)]_{ij} \). When the diffusion term is spatially independent \( g_t(x)=\sigma(t) \), the equation reduces to:
	\begin{equation}
		\frac{\partial \rho_t(x)}{\partial t}
		+ \nabla \cdot (f_t(x)\rho_t(x))
		= \frac{\sigma^2(t)}{2}\Delta \rho_t(x).
		\label{eq:FKP}
	\end{equation}
	
	This equation describes how the probability density function evolves over time as a result of the combined effects of drift and diffusion in the stochastic process. The term $\nabla  \cdot (f_t (x)\rho _t (x))$ represents the drift-induced transport of probability density, while $\frac{{\sigma ^2 (t)}}{2}\Delta \rho _t (x)	$ accounts for the spread of the diffusion of the density. In the stochastic process, if the drift term is zero, i.e., no drift field, the FPE follows the diffusion process only. If the drift term is not zero, the probability density will shift from their original positions as well with diffusion. Therefore, in a SDE, it can manipulate complex probability density changes with controlling the drift and diffusion terms. That is why it is convenient to use SDEs for generative modeling in machine learning. 
	
	The FPE, also known as the Kolmogorov forward equation, in the context of OT theory, plays a crucial role in describing the evolution of probability densities associated with stochastic processes underlying OT problems. OT theory provides a natural framework to study how probability distributions \( \mu_t \) evolve under SDEs. The Wasserstein distance \( W_p(\mu_t, \nu_t) \) between two distributions \( \mu_t \) and \( \nu_t \) quantifies the minimum cost of transforming \( \mu_t \) into \( \nu_t \). This connection is essential because OT seeks to find the most efficient way to transport one probability distribution to another, and stochastic processes described by such equations naturally arise in this context. The FPE thus provides a mathematical framework for analyzing the evolution of probability distributions and their relation to OT problems. 
	
	Modern diffusion-based generative models leverage SDEs to construct a forward noising process and a learned reverse process. The forward SDE gradually perturbs the data \( \mathbf{x}_0 \sim p(\mathbf{x}) \) (usually the original data distribution) into a simple Gaussian distribution over time \( T \). The reverse-time SDE reconstructs the data by denoising \cite{Anderson1982, SongICLR2021}:
	\begin{equation}
		\mathrm{d}\mathbf{x}_t
		= \left[ f(\mathbf{x}_t,t) 
		- g(t)^2 \nabla_{\mathbf{x}_t}\log p_t(\mathbf{x}_t) \right]\mathrm{d}t
		+ g(t)\,\mathrm{d}\mathbf{w}_t.
	\end{equation}
	Here, \( p_t(\mathbf{x}_t) \) is the marginal distribution of \( \mathbf{x}_t \) at time \( t \), $ \nabla_{\mathbf{x}} \log p(\mathbf{x}) $ is a score function, the gradient of the log-density of data distribution. The training objective is often formulated as a variational bound on the negative log-likelihood, which can be optimized using stochastic gradient descent \cite{Song2019,SongICLR2021}. However, these methods can be computationally expensive and may struggle with high-dimensional data. Neural score matching addresses these challenges by directly learning the score function using a neural network \cite{Hyvarinen2005, Hyvarinen2007, SongICLR2021, Lipman2024, Holderrieth2026}. This approach leverages the expressive power of neural networks to model the score function across various noise levels, allowing more accurate and efficient data generation. The core idea behind neural score matching is to train a neural network \( \mathbf{s}_\theta(\mathbf{x}, t) \) to approximate the score function \( \nabla_{\mathbf{x}} \log p_t(\mathbf{x}) \) at each time step \( t \). Given a data set \( \{\mathbf{x}_i\}_{i=1}^N \), the training objective is to minimize the score matching loss defined as:
	\begin{equation}
		\mathcal{L}(\theta) =\mathbb{E}_{_{\scriptstyle t \sim \mathcal{U}(0,T) \hfill \atop 
				{\scriptstyle {\bf x}_0  \sim p_0 ({\bf x}_0 ) \hfill \atop 
					\scriptstyle {\bf x}_t  \sim p_t ({\bf x}_t |{\bf x}_0 ) \hfill}} } \left[ {w (t)|{\bf s}_\theta  ({\bf x}_t ,t) - \nabla _{{\bf x}_t } \log p_t ({\bf x}_t |{\bf x}_0 )|^2 } \right],
	\end{equation}
	where \( w(t) \) is a weighting function that can be tuned to emphasize different levels of noise during training, and \( p_t(\mathbf{x}_t \mid \mathbf{x}_0) \) denotes the distribution of noisy data at time \( t \) conditioned on the original data \( \mathbf{x}_0 \), $t$ belongs to a uniform distribution in the interval as $t \sim \mathcal{U}(0,T)$. In practice, the true score function \( \nabla_{\mathbf{x}_t} \log p_t(\mathbf{x}_t \mid \mathbf{x}_0) \) is intractable. Instead, it can be estimated by adding Gaussian noise to the data and using the score matching objective to learn a close approximation. Once trained, this network can be used to perform denoising at each time step, effectively reversing the diffusion process \cite{SongICLR2021,DDPM2020}.
	
	\subsection{Normalizing Flow and Flow Matching}
\begin{figure}[tbp]
		\centering
		\includegraphics[width=9cm, height=5cm]{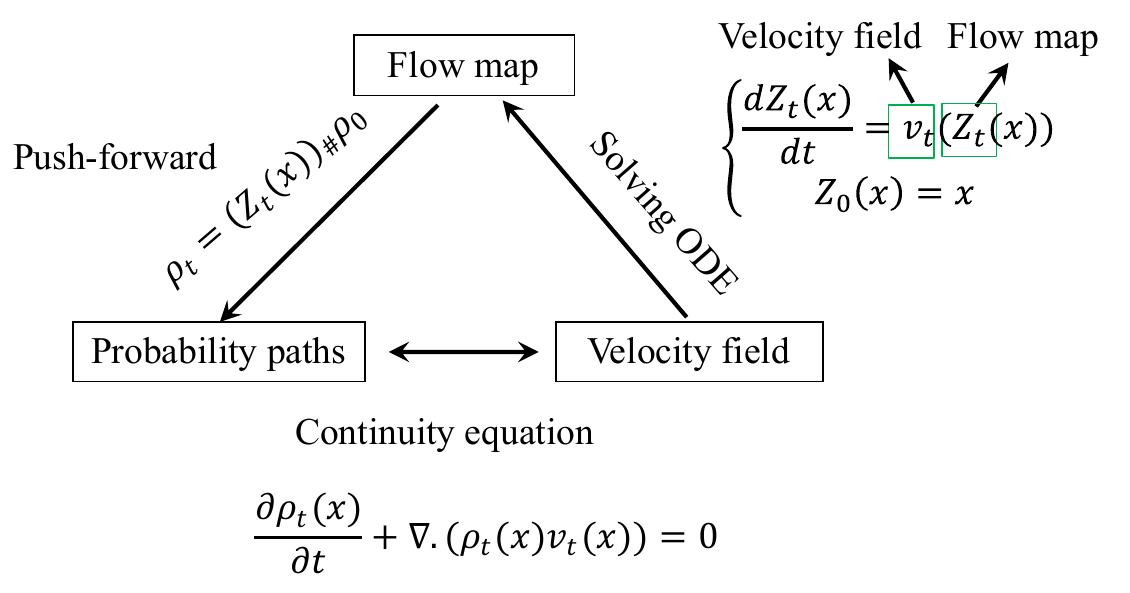}
		\caption{Unified concepts in flow mapping via ODEs.}	
		\label{fig:fig11}
	\end{figure}

\begin{figure}[tbp]
		\centering
		\includegraphics[width=8cm, height=3.5cm]{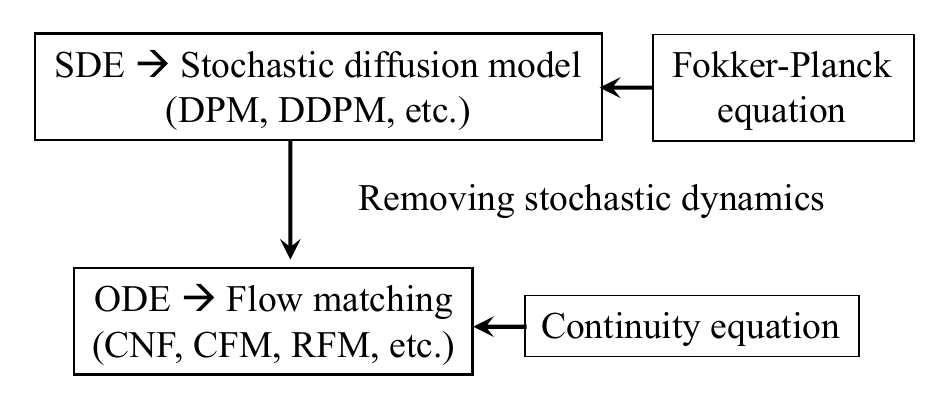}
		\caption{Unified concepts with diffusion and flow matching models via SDEs/ODEs.}	
		\label{fig:fig12}
	\end{figure}

	Flow matching provides a simulation-free alternative to SDE-based generative models developed from the idea of simplified normalizing flow. When the diffusion term in the FPE (Eq.\ref{eq:FKP}) is zero, the density evolution reduces to the continuity equation corresponding to an ordinary differential equation (ODE):
	\begin{equation}
		\frac{{dZ_t \left( x \right)}}{{dt}} = v_t \left( {Z_t \left( x \right)} \right),
	\end{equation}
	where \( Z_t \) is a transport field or flow map generated by the ODE with \( v_t(\cdot) \) as its velocity field. The evolution of the associated density is given by the change-of-variables formula as in a normalizing flow \cite{Lipman2022, Lipman2024, Holderrieth2026}:
	\begin{equation}
		\rho_t(x)
		= \rho_0\!\left( Z_t^{-1}(x) \right)
		\det\!\left(
		\frac{\partial Z_t^{-1}}{\partial x}
		\right).
		\label{eq:NMF}
	\end{equation}
	In flow matching and normalizing flow models, the goal is to learn a parameterized velocity field:
	\[
	v_t(Z_t(x)) \ \rightarrow\  v_t(Z_t(x);\theta),
	\]
	such that pushing forward a simple base distribution (e.g., a Gaussian) along the learned flow produces a complex target distribution $\rho_t(x)$, and learning model parameters with maximizing negative log-likelihood of parameterized $\rho_t(x;\theta)$. This viewpoint unifies several key concepts as shown in Fig. \ref{fig:fig11}.
	As shown in this figure, the velocity field governs the evolution of particles through an ODE, the flow map-solution of the ODE, pushing forward probability mass, continuity equation links density evolution and the velocity field. The push-forward operator maps the initial densities to the future densities. Other transport-based methods can be understood as instances of ODE/SDE-driven probabilistic transport \cite{NODE2018}, for example, Diffusion Probabilistic Model (DPM), Denoising Diffusion Probabilistic Model (DDPM) \cite{DDPM2020}, Continuous Normalizing Flow (CNF) \cite{CNF2021}, Conditional Flow Matching (CFM) \cite{Lipman2022}, Rectified Flow Matching (RFM) \cite{LiuX2023}, etc., all share common tasks or purposes, and can be explicitly unified as in Fig. \ref{fig:fig12}. From this figure, we could figure out the relationship between different generative model frameworks proposed in recent years, as well as their relationship with OT \cite{Kornilov2024,Tong2024,LiuX2023,LTFlow2025}.
	
	\subsection{Transfer Operator Perspective}
\begin{figure}[tbp]
		\centering
		\includegraphics[width=8cm, height=3.5cm]{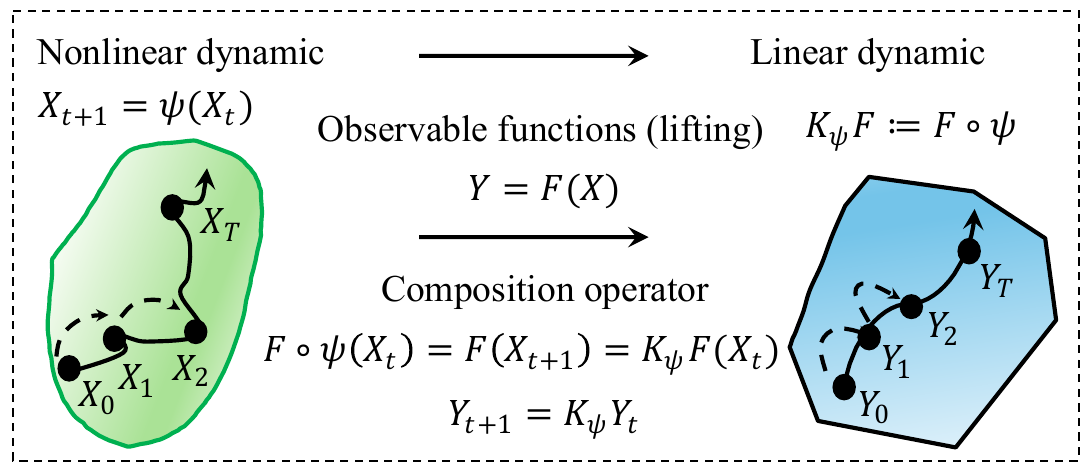}
		\caption{Transfer operator point of view for state dynamics.}	
		\label{fig:fig13}
	\end{figure}

	In the previous subsections, dynamics was described via SDEs or ODEs, which require knowledge of drift and diffusion terms. An alternative viewpoint uses transfer operators with the mathematical insight that a dynamic system intrinsically determined by a differential equation even with complex nonlinearity might correspond to a linear operator, i.e., the Koopman operator, which characterizes system evolution without explicitly referencing the underlying governing equations. The idea is illustrated in Fig. \ref{fig:fig13}.  
	The evolution of the state $X_t \in \mathbb{R}^n$ under nonlinear dynamics $\Psi: \mathbb{R}^n \to  \mathbb{R}^n$ is formulated as:
	\begin{equation}
		X_{t + 1}  = \Psi \left( {X_t } \right).
	\end{equation}
	For a space of measurement or observable functions defined as $\mathcal{G}: = \left\{ {F:\mathbb{R}^n  \to \mathbb{R}} \right\}$, where the element of the real-valued measurement function is $Y = F\left( X \right)$ taking values in the state space, the Koopman operator (composition operator) $K_\Psi  :\mathcal{G} \to \mathcal{G}$ is defined as \cite{Brunton2022}:
	\begin{equation}
		K_\Psi   F: = F\circ \Psi. 
	\end{equation}
	By this Koopman operator, the evolution of dynamics in the whole lifting space (an infinite-dimensional space of the collection of 
	observable functions) is linearly represented as:
	\begin{equation}
		Y_{t + 1}  = K_\Psi  Y_t.
	\end{equation}    	
	That is to say, the Koopman operator captures the evolution of observables under nonlinear dynamics through a linear infinite-dimensional operator, and it is defined as the composition of observable with the flow induced by the dynamic system (in practice, Koopman operator can be approximated by a finite-dimensional matrix with data-driven based algorithms, e.g., Extended Dynamic Mode Decomposition (EDMD) algorithm \cite{EDMD2015}). A much more clear explanation is shown in the bottom panel of Fig. \ref{fig:fig13} where the nonlinear dynamics could be analyzed in a linear space with the Koopman operator acting on scalar-valued functions driven based on the dynamics rather than on the original states. The adjoint of the Koopeman operator, the Perron-Frobenius operator, describes the evolution of probability densities with uncertainty of the initial condition through the propagation of a dynamic system \cite{Lasota1994}. It induces a measure-preserving transform of a probability density as a push-forward map. This viewpoint connects naturally to OT and SDE dynamics: the Perron-Frobenius operator corresponds to density evolution (Eulerian point of view), the Koopman operator corresponds to observable evolution (Lagrangian point of view), the FPE is a PDE representation of the same evolution, and SBP and entropic OT correspond to KL projections of path or density evolution. Thus, OT, SBP, SDE-based generative models, and transfer-operator approaches fit within a common theoretical framework for modeling the transport and evolution of probability distributions. In summary, optimal transport, diffusion processes, and transfer-operator analysis form a coherent theoretical framework that connects probability flows, stochastic control, and generative modeling via analysis of SDEs which is an exciting field for further exploration.
	
	\section{Computational Algorithms for Optimal Transport}
	The practical applications of OT in engineering, machine learning, and data science require efficient numerical algorithms. Classical OT formulations lead to large-scale linear programs, whose computational cost can be prohibitive for modern datasets. We first recall that the discrete Kantorovich OT problem is a linear program. Thus, many classical LP solvers can be directly applied, for example:
	\begin{itemize}
		\item \textbf{Network Simplex Method:} A variant of the simplex algorithm tailored for network flow, effective for medium-scale OT problems.
		\item \textbf{Interior Point Methods:} Algorithms that traverse the interior of the feasible polytope, providing strong theoretical guaranties for solving large-scale LP problems.
		\item \textbf{Classical Simplex Solvers:} Applicable but often prohibitively slow for dense OT matrices of large size.
	\end{itemize}
	Although theoretically sound, these methods scale poorly: the cost of solving a \(n \times n\) OT problem is typically \(O(n^3)\) \cite{PeyreBook2019}, which historically limited the use of OT in large-scale machine learning. During the past decade, significant progress has been made in designing fast, scalable OT solvers, which has greatly accelerated the adoption of OT in machine learning, graphics, geometry processing, domain adaptation, scalable OT, Wasserstein GANs, and statistical modeling. Many researchers and studies have contributed foundational algorithms and open-source tools that made OT widely usable in practice \cite{PeyreBook2019, KhamisPAMI2024}. Among them, a major breakthrough came in 2013 when Cuturi introduced entropy-regularized OT and the Sinkhorn algorithm \cite{Sinkhorn1967,Cuturi2013, MatrixScale2016}. This approach dramatically reduces computational cost and is often referred to as ``optimal transport at light speed'' \cite{Cuturi2013}. Since then, entropic OT has become one of the most widely used computational algorithms. First, let us introduce the concept of entropy-regularized OT. 
	
	\subsection{Sinkhorn Algorithm for Entropy-Regularized OT}
	\label{sect:EROTSH}
	Given probability measures \( \mu \) and \( \nu \) in spaces $X$ and $Y$, with transport plans \( \gamma \in \Pi(\mu,\nu) \), the OT plan based on the entropy-regularized OT problem (EOT) is defined as follows:
	\begin{equation}
		\gamma^\ast \mathop  = \limits^\Delta 
		\arg\min_{\gamma \in \Pi(\mu,\nu)} 
		\left\{
		\int c(x,y)\, d\gamma(x,y)
		- \lambda\, H(\gamma)
		\right\}.
	\end{equation}
	Here \( H(\gamma) \) is the Shannon entropy defined as:
	\begin{equation}
		H(\gamma) = -\int \gamma(x,y) \log \gamma(x,y)\, dx\,dy.
	\end{equation}
	Equivalently, it can be expressed using the KL-divergence:
	\begin{equation}
		\gamma^\ast 
		\mathop  = \limits^\Delta 
		\arg\min_{\gamma \in \Pi(\mu,\nu)}
		\left\{
		\langle \gamma, C \rangle 
		+ \lambda\, \mathrm{KL}(\gamma \mid \mu \otimes \nu)
		\right\},
	\end{equation}
	where the KL-divergence \( \mathrm{KL}(\gamma | \mu \otimes \nu) \) is defined as:
	\begin{equation}
		\mathrm{KL}(\gamma | \mu \otimes \nu) \mathop  = \limits^\Delta \int_{X \times Y} \gamma(x,y) \log \left( \frac{\gamma(x,y)}{\mu(x) \nu(y)} \right) \, dx \, dy.
	\end{equation}
	
	The Sinkhorn algorithm is a powerful method for efficiently computing approximate solutions to the entropy-regularized OT problem. This formulation introduces an entropy term to the traditional OT cost function, promoting sparsity and numerical stability in large-scale applications \cite{Cuturi2013}. For the discrete case in real implementation, the formulation is given:
	\begin{itemize}
		\item Two discrete measures \( \mu = \sum_{i=1}^{m} a_i \delta_{x_i} \) and \( \nu = \sum_{j=1}^{n} b_j \delta_{y_j} \) with positive masses \( a_i \) and \( b_j \), and locations \( x_i \) and \( y_j \).
		\item Cost matrix \( C \) where \( C_{ij} = c(x_i, y_j) \), representing pairwise transportation costs.
		\item Regularization parameter \( \lambda > 0 \) to control the regularization strength of the entropy.
	\end{itemize}
	The EOT problem becomes:
	\begin{equation}
		\mathrm{EOT}_\lambda(\mu,\nu)
		\mathop  = \limits^\Delta
		\min_{\gamma\in\Pi(\mu,\nu)}
		\left(
		\langle \gamma, C \rangle 
		- \lambda H(\gamma)
		\right),	
		\label{eq:deot}
	\end{equation}
	where \( H(\gamma) = -\sum_{ij} \gamma_{ij} (\log \gamma_{ij}-1) \). With defining the Gibbs kernel as:
	\[
	{\bf K} = e^{-C/\lambda},
	\]
	the Sinkhorn algorithm alternates between row and column scaling/normalization:
	\begin{equation}
		{\bf u}^{(k+1)} = \frac{\bf a}{{\bf K} {\bf v}^{(k)}}, 
		\qquad
		{\bf v}^{(k+1)} = \frac{\bf b}{{\bf K}^\top {\bf u}^{(k+1)}},
		\label{eq:deotiter}
	\end{equation}
	where ${\bf a} = \left[ {a_1 ,a_1 , \ldots a_m } \right]$ and ${\bf b} = \left[ {b_1 ,b_1 , \ldots b_n } \right]$ are the marginal weight coefficients defined on sample locations in $X$ and $Y$. The transport plan is then obtained as:
	\begin{equation}
		\gamma^\ast = \mathrm{diag}({\bf u}) {\bf K} \mathrm{diag}({\bf v}),
		\label{eq:deotsolver}
	\end{equation}
	where ${{\bf u} }$ and ${{\bf v} }$ are two scaling (or re-normalization) vectors. This procedure is equivalent to performing iterated Bregman projections onto the marginal constraints. Sinkhorn’s original scaling algorithm dates back to 1964, but its use for OT was popularized only recently due to its remarkable efficiency, stability, and compatibility with GPU acceleration \cite{Cuturi2013, PeyreBook2019}. The advantages of this Entropy regularization are summarized as follows:
	\begin{itemize}
		\item \textbf{Smoothness:} Entropy spreads the mass across the support, avoiding degeneracy.
		\item \textbf{Strict convexity:} Providing a unique solution.
		\item \textbf{Stability:} Prevents numerical collapse of the transport plan.
		\item \textbf{Scalability:} Transforming an OT into a diagonal matrix-scaling problem allows fast GPU parallelization.
	\end{itemize}
	
	Many computational OT algorithms are inspired by this Entropy-regularized OT algorithm with algorithmic variants and improvements, including Iterative Bregman Projections (IBP) \cite{Benamou2014}, Bregman ADMM \cite{WangNIPS2014}, IPOT(Inexact Proximal Point OT) \cite{Xie2018}, etc. IPOT is particularly useful when no entropic smoothing is desired, since it approximates the exact OT solution via proximal iterations without requiring extremely small \(\lambda\). Moreover, a more general regularization of OT beyond entropy can be formulated as:
	\begin{equation}
		\gamma^\ast
		\mathop  = \limits^\Delta
		\arg\min_{\gamma\in\Pi(\mu,\nu)}
		\left\{
		\langle C, \gamma\rangle 
		+ \lambda\, \mathrm{Reg}(\gamma)
		\right\}.	
	\end{equation}
	Depending on the prior structure, different regularizers $\mathrm{Reg}(\gamma)$ may be used, for example, Low-rank OT \cite{Scetbon2021}, Sparse OT \cite{Blondel2018, LiuICLR2023}, Laplacian-regularized OT \cite{FlamaryNIPS2014}, Group-Lasso regularized OT \cite{Blondel2018}, etc. Regularization lets practitioners embed structural priors (smoothness, sparsity, geometry) into the transport plan, enabling OT to be adapted to specific application domains. In real applications, we need to deal with the numerical instability problem. As Sinkhorn iterations involve exponentials and divisions, making numerical underflow or overflow common. Stability is often improved through log-domain Sinkhorn iterations, cost scaling, stabilization through absorbing constants (as in log-sum-exp), GPU-friendly normalization \cite{PeyreBook2019}.
	
	\subsection{Neural Optimal Transport}
	\label{sec:neural_optimal_transport}
	Neural Optimal Transport (Neural OT) integrates deep learning with OT by parameterizing potentials or transport maps using neural networks \cite{Makkuva2020, NOPT2021, NOPT2022}, particularly with neural solver algorithms by connecting deep network architecture with ODE \cite{NODE2018}. This makes OT feasible in high-dimensional spaces where classical solvers struggle. A common approach parameterizes the Kantorovich potential \( \phi(x;\theta) \) via a neural network. Based on \(c\)-transform, the dual OT problem becomes:
	\begin{equation}
		\max_{\phi}
		\left[
		\int \phi(x;\theta)\, d\mu(x)
		+
		\int \phi^c(y;\theta)\, d\nu(y)
		\right],
	\end{equation}
	where the \(c\)-transform is defined as (refer to the dual property of OT in section \ref{sect:DualOT}):
	\begin{equation}
		\phi^c(y;\theta)
		\mathop  = \limits^\Delta
		\sup_{x\in X}
		\left( c(x,y) - \phi(x;\theta) \right).	
	\end{equation}
	
	Neural networks allow flexible, high-capacity approximations of optimal potentials. For the quadratic cost \(c(x,y)=\frac{1}{2}\|x-y\|^2\), based on Brenier’s theorem: the optimal Kantorovich potential is convex, the optimal OT map is the gradient of this potential. Thus, neural OT often parameterizes \(f(x;\theta)\) as an ICNN \cite{AmosICML2017, Makkuva2020}:
	\begin{equation}
		\inf_{f \in \mathrm{cvx}(X)}
		\left[
		\int f(x;\theta)\, d\mu(x)
		+
		\int f^*(y;\theta)\, d\nu(y)
		\right],	
	\end{equation}
	where the convex conjugate $f^*(y;\theta)$ is defined as:
	\begin{equation}
		f^*(y;\theta)	\mathop  = \limits^\Delta	\sup_x \left( \langle x,y\rangle - f(x;\theta) \right).	
	\end{equation}
	The primal OT formulation is the following.
	\begin{equation}
		\inf_{\gamma\in\Pi(\mu,\nu)}
		\int \frac{1}{2}\|x-y\|^2\, d\gamma(x,y),
	\end{equation}
	and the dual formulation connects directly to the neural parameterization above. Neural OT methods therefore provide scalable tools for Wasserstein distance estimation, generative modeling, domain adaptation, barycenter computation, and gradient-based optimization in OT geometry.
	
	\section{Applications of optimal transport in speech}
	This section provides a comprehensive review of how OT has been applied in modern speech-processing tasks. We first explain the use of OT in speech signal processing with reference to its applications in machine learning, where the applications can be broadly categorized into two paradigms: (i) employing OT as a loss function to guide model training and (ii) using OT for domain adaptation or distribution alignment. Following this, we present a series of successful applications that illustrate the effectiveness of OT in real-world scenarios. These include cross-domain and multi-modal speech enhancement \cite{LinNeurIPS2021,HsiehIS2021}, cross-domain speaker and language recognition \cite{LuICASSP2021,ZhangICASSP2023,YangTIFS2026, ZhangTASLP2024}, cross-corpus or cross-language emotion recognition \cite{ZhangICASSP2024}, domain-adaptive audio spoof detection \cite{ZhangIS2023,ZhangTASLP2025}, and cross-modal knowledge transfer for ASR \cite{LuASRU2023,LuICASSP2024, LuSLT2024, LuIS2025}. Finally, we discuss emerging research trends and potential future directions, highlighting how OT may continue to advance speech processing in increasingly complex, multi-domain, and multi-modal environments.
	
	\subsection{Recall of OT for Machine Learning}
\begin{figure}[tbp]
		\centering
		\includegraphics[width=6cm, height=2.5cm]{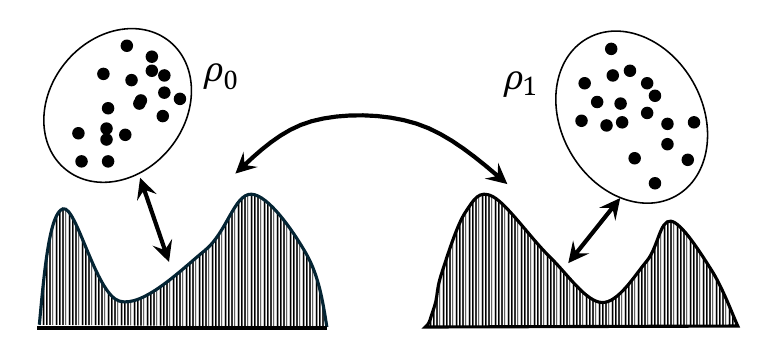}
		\caption{Distance measure and transform between distributions.}
		\label{fig:fig16}
	\end{figure}

	In machine learning, many problems can be viewed through the lens of three fundamental operations: (i) estimating distributions from observed data, (ii) sampling from these distributions, and (iii) transforming or manipulating one distribution into another. OT is particularly relevant to the third operation. For example, as illustrated in Fig. \ref{fig:fig16}, given observations from two distributions $\rho_0$ and $\rho_1$, how can we evaluate their similarity either for classification or generation tasks. 
	It provides both a principled geometric measure of discrepancy between probability distributions and a framework for constructing transport maps or interpolation paths between them. Because of this, OT shares conceptual similarity with a wide class of generative modeling techniques whose goal is to transform one distribution into another. Examples include Generative Adversarial Networks (GANs) \cite{GAN2014}, Variational Autoencoders (VAEs) \cite{VAE2013, VAE2019}, normalizing flows \cite{CNF2021,Papamakarios2021}, diffusion models \cite{DDPM2020}, diffusion Schr{\"o}dinger bridges \cite{BortoliNIPS2021, ShiNIPS2023}, and flow matching approaches \cite{Lipman2022}. OT provides a geometrically grounded alternative or complement to these methods.
	
	In speech signal processing, the use of OT typically falls into two major categories: OT as a loss function within the learning objective and OT as a discrepancy metric for cross-domain or modality knowledge transfer and adaptation. Traditional loss functions, such as sample-wise metrics (e.g., Mean Square Error (MSE), Mean Absolute Error (MAE), or task-specific measures, like Signal-to-Noise Ratio (SNR) and Signal-to-Distortion Ratio (SDR), do not explicitly consider the geometry of the underlying data distributions. In the following subsection, we highlight the advantages of incorporating OT into learning objectives.
	
	\subsection{Conventional Loss Functions vs. OT-Based Losses}
\begin{figure}[tbp]
		\centering
		\includegraphics[width=6cm, height=3cm]{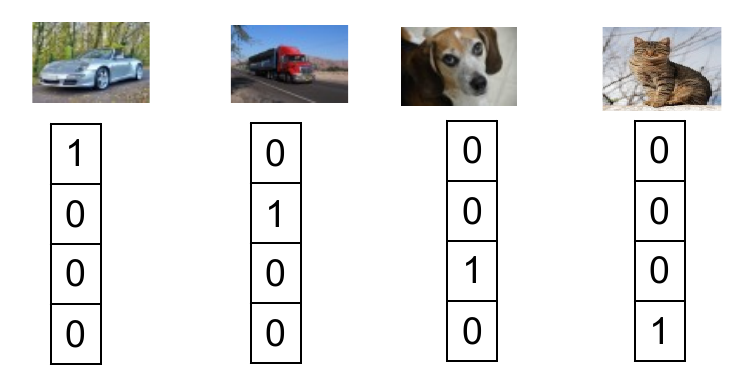}
		\caption{Equal distance between different categories (no information of category geometric relationship).}
		\label{fig:fig17}
	\end{figure}

\begin{figure}[tbp]
		\centering
		\includegraphics[width=6cm, height=3cm]{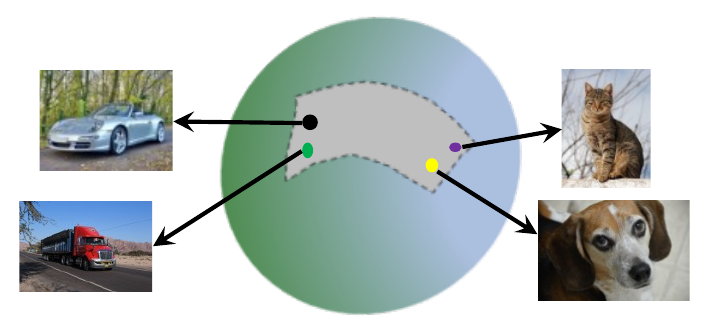}
		\caption{Geodesic distance in Wasserstein space.}
		\label{fig:fig18}
	\end{figure}

	A central objective in machine learning is to quantify the degree to which the predicted distribution deviates from the true data distribution. Conventional losses such as Cross-Entropy (CE), KL-divergence, and Jensen–Shannon (JS)-divergence are widely used, but they often fail to account for geometric relationships between target outcomes. For example, misclassifying a car as a truck incurs the same penalty as misclassifying a car as a dog (as illustrated in Fig. \ref{fig:fig17}), although the former error is semantically and geometrically much closer.
	A key advantage of OT is that it incorporates the geometry of the sample space directly into the loss function as illustrated in Fig. \ref{fig:fig18}. 
	Wasserstein distances, for example, compute the minimal cost of transporting mass from one distribution to another along geodesic paths in Wasserstein space. Unlike KL-divergence or JS-divergence, which are not true metrics and require overlapping supports, OT defines a proper distance even when the two distributions have disjoint supports. The well-known OT distances used in ML include the Earth Mover’s Distance (EMD) and the Wasserstein distance. A notable example demonstrating the value of OT in representation learning is the Word Mover’s Distance (WMD) \cite{KusnerICML2015}, which computes the similarity of the document by measuring the minimum amount of ``work" required to move embedded words from one document to another in the word2vec space. This approach effectively aligns semantically related words before measuring their distance, illustrating the geometric insight that OT introduces. In speech processing, OT can also be used as an auxiliary alignment loss. For example, instead of relying solely on the Connectionist Temporal Classification (CTC) loss in speech translation, OT has been added to enforce alignment between speech and text embeddings, yielding significant improvements \cite{LeICML2023}.
	
	\subsection{Cross-Domain Adaptation and Transfer Learning}
	The second major application of OT in machine learning is domain adaptation. Domain mismatch is a widespread challenge in many learning problems. In computer vision, for example, images collected under different lighting or camera conditions often lead to severe degradation when models trained in one domain are applied directly to another. Similar issues arise in speech processing tasks. In speech enhancement, mismatches between training and testing acoustic conditions frequently occur; in spoken language or speaker recognition, channel, language, and environmental variations all contribute to domain shift.
	
	\subsubsection{Cross-domain adaptation}
\begin{figure}[tbp]
		\centering
		\includegraphics[width=8cm, height=3cm]{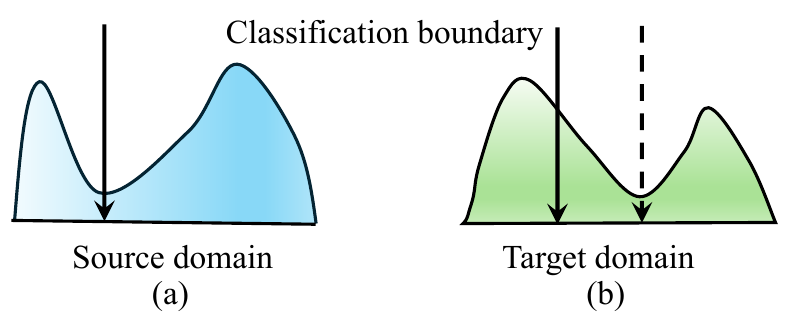}
		\caption{Domain adaptation: solid vertical line denotes the classification boundary trained on source domain data (a), while the dot vertical line represents the classification boundary on target domain data (b).}
		\label{fig:fig19}
	\end{figure}

	In domain adaptation, we assume that labeled data are available only in the source domain, whereas in the target domain only unlabeled data are observed. The model trained solely on the source domain is then expected to generalize to the target domain. Classical learning theory implicitly assumes that training (source) and test (target) samples are drawn from the same underlying distribution. However, as conceptually illustrated in Fig. \ref{fig:fig19}, the source and target distributions can differ significantly. Using a classifier trained on the source domain directly on the target domain inevitably leads to performance degradation.
	To mitigate this issue, unsupervised domain adaptation (UDA) seeks to reduce the discrepancy between source and target distributions. Let the joint distribution of features $\mathbf{x}$ and labels $\mathbf{y}$ be denoted as \( p(\mathbf{x},\mathbf{y}) \). For a joint distribution:
	\[
	p(\mathbf{x},\mathbf{y}) = p(\mathbf{x})p(\mathbf{y}|\mathbf{x}) = p(\mathbf{y})p(\mathbf{x}|\mathbf{y}).
	\]
	Depending on which part of the joint distribution changes across domains, several types of domain shift can arise (with $p^s(\cdot)$ and $p^s(\cdot)$ representing distributions in the source and target domains, respectively) \cite{Kouw2019}:
	
	\noindent\textbf{Label/target shift:}
	\[
	p^s(\mathbf{y}) \neq p^t(\mathbf{y}),\quad p^s(\mathbf{x}|\mathbf{y}) = p^t(\mathbf{x}|\mathbf{y}).
	\]
	
	\noindent\textbf{Covariate/feature shift:}
	\[
	p^s(\mathbf{x}) \neq p^t(\mathbf{x}),\quad p^s(\mathbf{y}|\mathbf{x}) = p^t(\mathbf{y}|\mathbf{x}).
	\]
	
	\noindent\textbf{Concept shift:}
	\[
	p^s(\mathbf{y}|\mathbf{x}) \neq p^t(\mathbf{y}|\mathbf{x}),\quad p^s(\mathbf{x}) = p^t(\mathbf{x}).
	\]
	
	\noindent\textbf{Conditional shift:}
	\[
	p^s(\mathbf{x}|\mathbf{y}) \neq p^t(\mathbf{x}|\mathbf{y}),\quad p^s(\mathbf{y}) = p^t(\mathbf{y}).
	\]
	
	\noindent\textbf{Conditional-target shift:}
	\[
	p^s(\mathbf{x}|\mathbf{y}) \neq p^t(\mathbf{x}|\mathbf{y}),\quad p^s(\mathbf{y}) \neq p^t(\mathbf{y}).
	\]
	
	OT provides a rigorous foundation for mitigating these types of shifts. Let the source and target domains be associated with probability measures \( \mu_s \) and \( \mu_t \), respectively, and for samples from the source and target domains $x_s  \in X_s$, $x_t  \in X_t$, let \( c(x_s,x_t) \) be a cost function measuring sample dissimilarity. OT seeks a transport plan \( \gamma \) minimizing the total transportation cost:
	\begin{equation}
		\inf_{\gamma \in \Pi(\mu_s,\mu_t)} \int_{X_s \times X_t} c(x_s,x_t)\, d\gamma(x_s,x_t),
	\end{equation}
	where \( \Pi(\mu_s,\mu_t) \) is the set of couplings with the prescribed marginals. The resulting transport plan or map can be used for:
	\begin{itemize}
		\item \textit{Feature alignment}: mapping features from the source to the target domain.
		\item \textit{Instance reweighting}: assigning importance weights to source samples to match target distribution.
		\item \textit{Model adaptation}: transformation of decision boundaries according to the aligned feature space.
	\end{itemize}
	OT offers several advantages for domain adaptation:  
	(1) it provides a mathematically principled discrepancy measure;  
	(2) it incorporates geometric structure of distributions;  
	(3) it accommodates distributions with partially overlapping support.  
	However, challenges remain, including computational cost, sensitivity to outliers, and the need to select an appropriate cost function.
	Among the most influential works is the OT-based domain adaptation pipeline proposed in \cite{Courty2014}. The method consists of three steps: (i) estimate the OT plan $T(\cdot)$ between the source and target distributions; (ii) transport the labeled source samples to the target domain; (iii) train a classifier in the transformed space with labels preserved as:
	\begin{equation}
		p^s(y|x^s) = p^t(y|T(x^s)).
	\end{equation}
	
	\subsubsection{Domain adaptation in the deep learning framework}
	In deep learning, a model can be viewed as the composition of a feature extractor and a classifier formulated as:
	\begin{equation}
		y\left( {\bf x} \right) = f\left( {{\bf x};\theta _{f_{\rm c} } ,\theta _{f_{{\rm fea}} } } \right) = f_{\rm c}  \circ f_{{\rm fea}} \left( {\bf x} \right),
	\end{equation}
	where \( f_{{\rm fea}}: \mathcal{X} \to \mathcal{Z} \) extracts latent features \( \mathbf{z}=f_{{\rm fea}}(\mathbf{x}) \), and \( f_{\rm c}: \mathcal{Z}\to\mathcal{Y} \) maps the features to the label probabilities, where $\theta_{f_{\rm c}}$ and $\theta_{f_{{\rm fea}}}$ are their corresponding model parameters. Domain adaptation typically aims to find a latent space where the joint distributions align:
	\begin{equation}
		p^s(\mathbf{z},\mathbf{y}) \approx p^t(\mathbf{z},\mathbf{y}), 
	\end{equation}
	where 
	\begin{equation}	
		p^m(\mathbf{z},\mathbf{y}) = p^m(\mathbf{y}|\mathbf{z})\, p^m(\mathbf{z}),
	\end{equation}
	with $m\in\{s,t\}$ belongs to the source or target domains. The expected risk in the target domain satisfies \cite{Chuang2020,LiuBound,Kouw2019}:
	\begin{equation}
		R^t(f) \le R^s(f) + L_{f_{{\rm fea}}}(p^s_{\mathbf{z}},p^t_{\mathbf{z}}) + R_{\mathbf{y}},
	\end{equation}
	where \( L_{f_{{\rm fea}}} \) measures the distribution discrepancy in the latent space, and \( R_{\mathbf{y}} \) reflects the mismatch of labeling functions. Most studies aim to reduce this upper bound by minimizing both the source-domain classification error and the distribution discrepancy.
	
	\subsection{Domain-Invariant Representation Learning}
	The goal of domain-invariant representation learning is to discover a latent space where feature distributions of source and target domains become similar. There are two families of approaches: discrepancy-based methods and adversarial learning-based methods.
	
	\subsubsection{Minimizing distribution discrepancy}
	In this framework, the model parameters are obtained by minimizing:
	\begin{equation}
		L(\theta_{f_{{\rm fea}}},\theta_{f_{\rm c}}) \mathop  = \limits^\Delta 
		\sum_i L_{\mathrm{CE}}^s(\mathbf{y}_i^s,\hat{\mathbf{y}}_i^s)
		+ \lambda_{\rm fea} L_{f_{{\rm fea}}}(p_{\mathbf{z}}^s,p_{\mathbf{z}}^t),
	\end{equation}
	where the second term aligns the feature distributions with a weighting coefficient $\lambda_{\rm fea}$. The methods based on Maximum Mean Discrepancy (MMD)-based and Central Moment Discrepancy (CMD) belong to this category \cite{PanMMD2009,LongMMD2013,MMD}.
	
	\subsubsection{Domain adversarial learning}
	\label{sub_DAL}
	Another approach uses adversarial training, where a domain discriminator \( f_{\rm d}(\cdot) \) (with model parameters $\theta_{f_{\rm d}}$) attempts to distinguish between the source and target features, while the feature extractor attempts to make them indistinguishable \cite{DANN2016}. The objective is as follows.
	\begin{equation}
		\begin{array}{l}
			\begin{aligned}		
				L(\theta_{f_{{\rm fea}}},\theta_{f_{\rm c}},\theta_{f_{\rm d}})
				&\mathop  = \limits^\Delta \sum_i L_{\mathrm{CE}}^s(\mathbf{y}_i^s,\hat{\mathbf{y}}_i^s) \\
				&- \lambda_{\rm d} \sum_{\mathbf{z}_i \in \{D^s \cup D^t\}} 
				L_{f_{\rm d}}({f_{\rm d}}(\mathbf{z}_i), d_i),\\
			\end{aligned}
		\end{array}
	\end{equation}
	with the standard min–max optimization:
	\begin{equation}
		(\theta_{f_{{\rm fea}}}^*,\theta_{f_{\rm c}}^*) \mathop  = \limits^\Delta 
		\arg\min_{\theta_{f_{{\rm fea}}},\theta_{f_{\rm c}}} L,
		\qquad
		\theta_{f_{\rm d}}^* \mathop  = \limits^\Delta \arg\max_{\theta_{f_{\rm d}}} L.
	\end{equation}
	
	\subsubsection{OT-guided learning in deep neural networks}
	Although moment-matching and adversarial approaches can align marginal feature distributions, they typically ignore the geometric structure of multimodal distributions, potentially degrading class discriminability. OT-based approaches explicitly model geometric relationships and have shown a strong performance in domain adaptation \cite{Courty2014}. Using the Kantorovich-Rubinstein duality, the Wasserstein distance becomes \cite{ShenAAAI2017}:
	\begin{equation}
		L_h(p_{\mathbf{z}}^s,p_{\mathbf{z}}^t)
		\mathop  = \limits^\Delta \sup_{h\in\mathcal{L}_{\rm {lip}}} 
		E_{\mathbf{z}\sim p_{\mathbf{z}}^s}[h(\mathbf{z})]
		- 
		E_{\mathbf{z}\sim p_{\mathbf{z}}^t}[h(\mathbf{z})],
	\end{equation}
	where $h \in \mathcal{L}_{\rm {lip}}$ means $h$ belongs to the set of 1-Lipschitz functions. In practice, a neural network approximates \( h \), and the empirical loss is:
	\begin{equation}
		L_{\rm wd}(\mathbf{z}^s,\mathbf{z}^t) \mathop  = \limits^\Delta
		\frac{1}{|D^s|} \sum_{\mathbf{z}_i\in D^s} h(\mathbf{z}_i)
		-
		\frac{1}{|D^t|} \sum_{\mathbf{z}_i\in D^t} h(\mathbf{z}_i),
	\end{equation}
	with a gradient penalty enforcing the Lipschitz constraint:
	\[
	L_{\rm grad}(\tilde{\mathbf{z}}) \mathop  = \limits^\Delta (\|\nabla_{\tilde{\mathbf{z}}}h(\tilde{\mathbf{z}})\|_2 - 1)^2.
	\]
	The resulting min–max optimization is:
	\begin{equation}
		L_h \mathop  = \limits^\Delta
		\min_{\theta_{f_{{\rm fea}}}}
		\max_{\theta_h}
		L_{\rm wd}(\mathbf{z}^s,\mathbf{z}^t)
		- \alpha_{\rm lip} L_{\rm grad}(\tilde{\mathbf{z}}),
	\end{equation}
	where $\alpha_{\rm lip}$ is the weighting coefficient, $|D^s|$ and $|D^t|$ denote the number of samples from source and target domains. To additionally incorporate label information, Joint Distribution OT (JDOT) framework has been proposed which aligns both features and labels \cite{CourtyNIPS2017,DamodaranECCV2018}. Inspired by JDOT, an OT-based domain adaptation model has been proposed for cross-domain speaker/language recognition, building upon pre-trained x-vectors, achieving strong performance \cite{LuICASSP2021} (will be introduced in more details in section \ref{sect:otLID}). 
	
\begin{figure}[tbp]
		\centering
		\includegraphics[width=8cm, height=4cm]{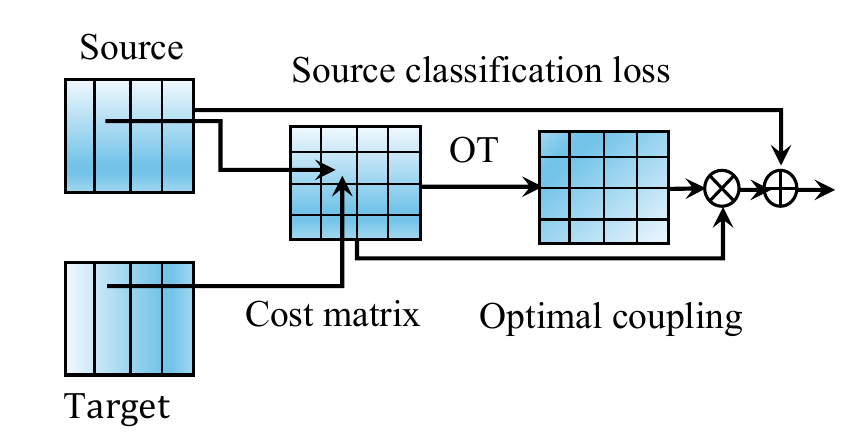}
		\caption{General computational framework with an OT embedding layer.}
		\label{fig:fig21}
	\end{figure}

	A summary of the overall computational pipeline is illustrated in Fig. \ref{fig:fig21}. 
	As shown in this figure, there are two loss objectives to be estimated, one is the source classification loss, and the other is domain feature distribution discrepancy loss. The OT part is a layer in the forward estimation. The Sinkhorn layer is used to estimate the optimal coupling or plan, and then multiply with Cost matrix to obtain the domain discrepancy loss. In the back propagation step, there is no parameter updated in this OT layer, that is, the back propagation jump to the front layers for feature representation learning.
	
	\subsection{OT for Speech enhancement}
	Speech enhancement (SE) aims to transform distorted or noisy speech to clean ones. Deep learning-based model frameworks for SE try to estimate the transform function with a large quantity of noisy-clean speech pairs, and use the learned transform function for SE in various test conditions. In real application, there is a mismatch between training and testing conditions which results in degraded performance. Cross-domain adaptation techniques are usually applied to mitigate domain-gaps. Moreover, conventional learning frameworks only try to learn the SE transform function from acoustic signal aspect, while linguistic or semantic information, another modality of speech besides acoustic aspect, should play an important role in speech intelligibility or understanding in adversarial environments. Therefore, cross-modal knowledge transfer during the SE transform function learning should be considered, particularly when many pretrained large language models encoding linguistic or semantic knowledge have already been available. In both cases, OT takes a suitable position in designing loss function or as a measure of domain/modal discrepancy measurement.             
	\subsubsection{Wasserstein distance-based Phone-Fortified Perceptual Loss for SE}  
\begin{figure}[tbp]
		\centering
		\includegraphics[width=8cm, height=4cm]{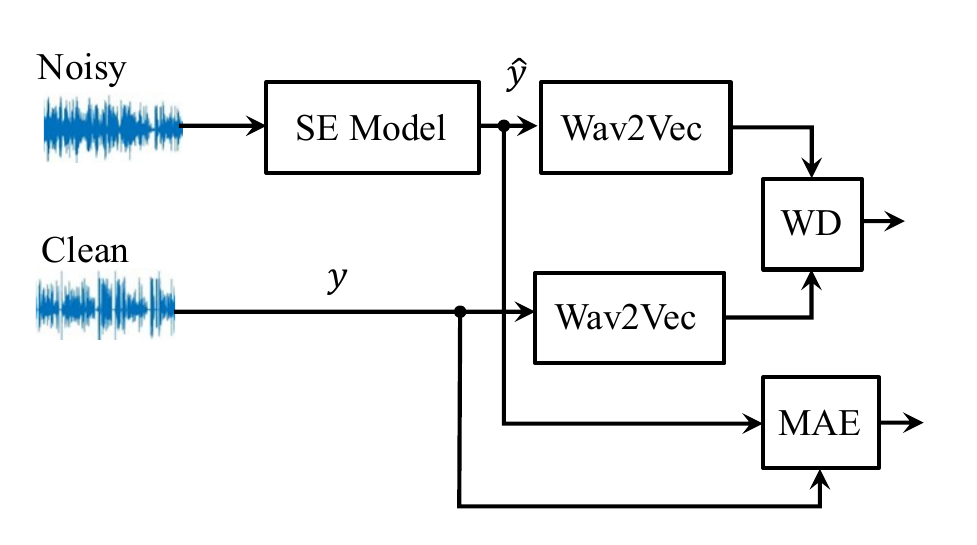}
		\caption{SE model framework enhanced with Phone-Fortified Perceptual Loss (PFPL) in a transformed latent space, WD: Wasserstein distance, MAE: mean absolute error.}
		\label{fig:fig22}
	\end{figure}

	Traditionally, deep learning-based SE is formulated as a sample-to-sample mapping regression task, i.e., transforming a noisy sample into a clean instance. On the other hand, OT formulates the SE as a distribution mapping task, i.e., transforming the distribution of noisy speech signal to a clean one. We may combine these two types of mapping process to achieve better enhancement performance. Based on this consideration, a learning framework has been proposed and explained in Fig. \ref{fig:fig22} \cite{HsiehIS2021}.
	In this figure, `wav2vec' is  a pretrained model block \cite{wav2vec2.0}. This block is used to explore latent space which is supposed to encode rich phonetic and semantic information for speech intelligibility. A loss function shown in this figure `WD', Kantorovich-Rubinstein dual form of W-distance is specially designed as:
	\begin{equation}
		L_{{\rm WD}}  \mathop  = \limits^\Delta \mathop {\sup }\limits_{\left\| h \right\|_L  \le 1} E_{z \sim \mu } \left[ {h(z)} \right] - E_{\hat z \sim \nu } \left[ {h(\hat z)} \right],
	\end{equation}  
	where $z = \phi _{{\rm wav2vec}} (y)$, $\hat z = \phi _{{\rm wav2vec}} (\hat y)$, and $\phi _{{\rm wav2vec}}(\cdot)$ is a transform function of the wav2vec block, function $h(\cdot)$ belongs to a set of all 1-Lipschitz functions. The conventional MAE loss is defined as:  
	\begin{equation}
		L_{{\rm MAE}}  \mathop  = \limits^\Delta ||y - \hat y||_1.
	\end{equation}
	
	The proposed Phone-Fortified Perceptual Loss (PFPL) for SE is defined as:
	\begin{equation}
		L_{{\rm PFPL}} (y,\hat y)\mathop  = \limits^\Delta  L_{{\rm MAE}}  + L_{{\rm WD}}.
	\end{equation} 
	In this definition, the OT loss (distribution-based mapping) and MAE loss (sample-based mapping) are composed in the estimation of the SE model. The OT-based loss defined on wav2vec explored latent feature could help to capture rich discriminative power of different phones for improving speech intelligibility. By implementing the SE model with U-net, SE experiments were carried out and showed that the proposed PFPL loss could enhance the SE performance, and has high correlations with the evaluation metrics PESQ and STOI scores \cite{HsiehIS2021}. 
	\subsubsection{Discriminator-constrained OT network for cross-domain SE}
\begin{figure*}[tbp]
		\centering
		\includegraphics[width=10cm, height=5cm]{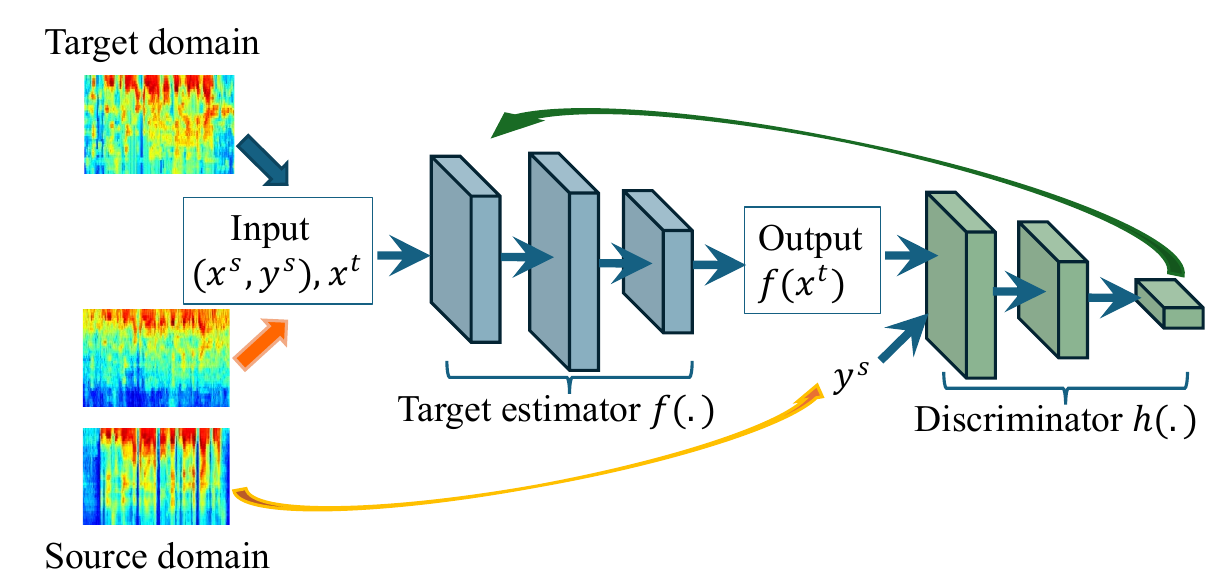}
		\caption{Discriminator-constrained OT network for SE.}
		\label{fig:fig23}
	\end{figure*}

	For mitigating domain mismatch, adversarial training is widely applied in machine learning field. In adversarial training, adaptive distance objective functions based on OT could be designed in adaptation. However, most studies based on OT for domain adaptation are for classification tasks, while adaptation in SE is a regression which is a difficult task since we need to maintain accurate speech structures. In the study \cite{LinNeurIPS2021}, an OT-based learning framework for SE with an discriminative training was designed, i.e., discriminator-constrained optimal transport network (DOTN), for improving the speech quality. The proposed model framework is showed in Fig. \ref{fig:fig23}.
	As shown in this figure, the model framework consists of an OT-based domain adaptation module and a discriminator. The adaptation module is designed based on the following minimization process, where an OT-based alignment objective function is involved in:
	\begin{equation}
		\mathop {\min }\limits_{\gamma ,f} (L_{{\rm Rec}}  + L_{{\rm Align}} ),
	\end{equation} 
	where target estimation loss and alignment losses are designed as: 
	\begin{equation}
		L_{{\rm Rec}}  \mathop  = \limits^\Delta \frac{1}{{N^s }}\sum\limits_i {||y_i^s  - f(x_i^s )||^2 },
	\end{equation}
	and
	\begin{equation}
		L_{{\rm Align}}  \mathop  = \limits^\Delta \sum\limits_{i,j} {\gamma _{i,j} (\alpha_x ||x_i^s  - x_j^t ||^2  + \beta_y ||y_i^s  - f(x_j^t )||^2 )},
	\end{equation}
	where $f(\cdot)$ is a target estimator, $\gamma _{i,j}$ is OT coupling, $\alpha_x$ and $\beta_y$ are the weighting coefficient for feature and label. Minimization of $L_{{\rm Rec}}$ tries to remind the source domain knowledge to avoid the catastrophic-forgetting issue, while $L_{{\rm Align}}$ is used for domain alignment. The domain discriminator is designed by following the algorithm of WGAN as \cite{ArjovskyICML2017}:
	\begin{equation}
		\mathop {\min }\limits_f \mathop {\max }\limits_{h \in \mathcal{L}_{\rm lip}} \left\{ {\mathbb{E}_{y \sim P_{Y^s } } (h(y)) - \mathbb{E}_{x \sim P_{X^t } } (h(f(x))} \right\},
	\end{equation}    
	where $h \in \mathcal{L}_{\rm lip}$ means $h$ belongs to the set of 1-Lipschitz functions. This proposed DOTN model combines OT and GAN is the first work that performs purely unsupervised SE domain adaptation, and obtained better performance than existing weakly-supervised domain adaptation approaches \cite{LinNeurIPS2021}.
	
	\subsubsection{Cross-modal knowledge transfer for SE}
\begin{figure}[tbp]
		\centering
		\includegraphics[width=8cm, height=5.5cm]{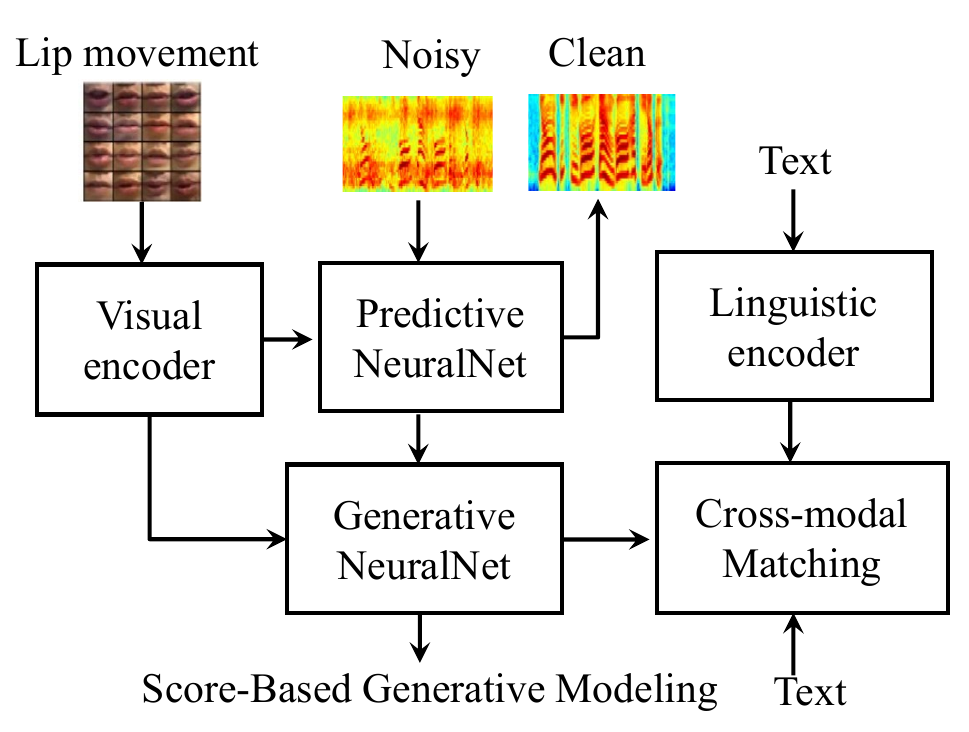}
		\caption{Multi-modal knowledge transfer learning for SE.}
		\label{fig:fig24}
	\end{figure}

	During speech communications, besides acoustic speech, visual cues, body gestures, and linguistic knowledge are also involved to remove confusion in real communications, particularly in adversarial environments or communication scenarios. Therefore, SE with multi-modality information, i.e., visual, acoustic speech, and text for designing SE algorithms have been intensively investigated in recent years. Conventional SE methods try to estimate the mapping function based on speech signal only, it is possible to encode the linguistic information in estimating the mapping function for SE, particularly a lot of pretrained large language models (LLMs) have been available. Here shows a multi-modal knowledge transfer learning framework for SE \cite{LinMP2025} in Fig. \ref{fig:fig24}. In this multi-modal SE framework, besides integrating video information and audio from a camera and microphone, linguistic information is provided during enhancement model training stage. And the linguistic information, audio and video information are integrated via a cross-model matching process where cross-modal matching could be adopted \cite{LinMP2025}. Since the enhancement model has already encoded the linguistic knowledge, text information is not required in inference stage \cite{HungKH2025}. 
	With this cross-modal learning paradigm, the performance of SE has improved significantly \cite{LinMP2025, HungKH2025}.
	\subsection{Cross-modal knowledge transfer for ASR}
\begin{figure}[tbp]
		\centering
		\includegraphics[width=8cm, height=3.5cm]{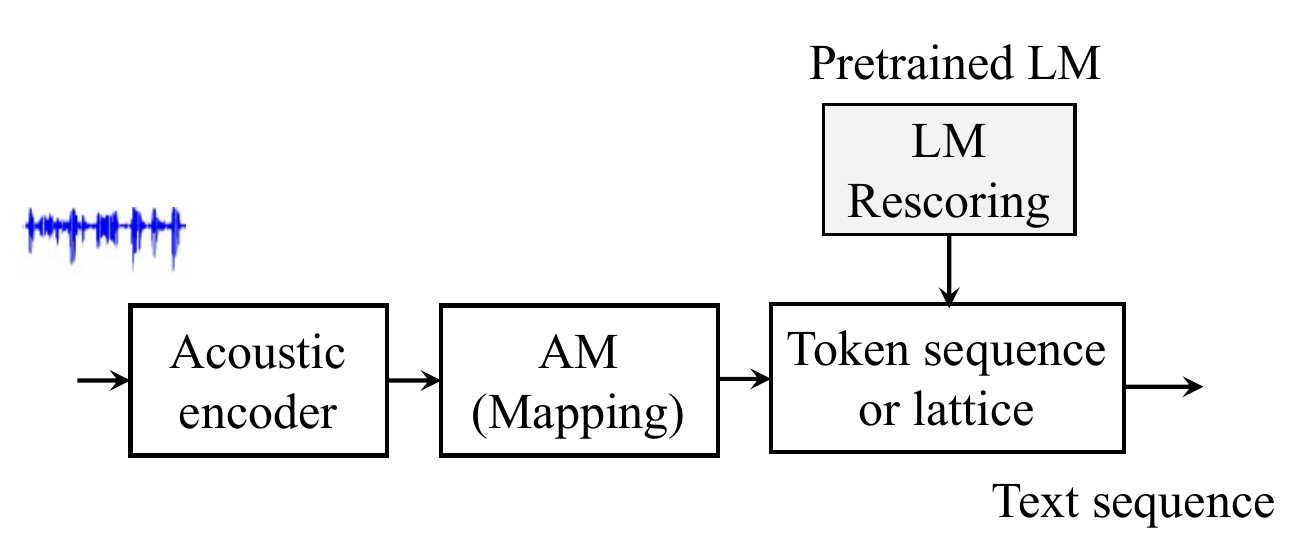}
		\caption{Conventional pipeline for ASR.}
		\label{fig:fig26}
	\end{figure}

	Linguistic knowledge information is one of the most important information in ASR. Fig. \ref{fig:fig26} shows a conventional two-stage pipeline for ASR where a language model (LM) encoding linguistic knowledge is involved.
	In this pipeline, LM is usually used as a rescoring process after acoustic decoding as the first stage. This two-stage processing slows recognition, and acoustic encoding could not take linguistic information in feature exploration learning in the first stage. With rapid advance and development of pretrained LLMs and end to end ASR models \cite{Li2022}, it is necessary to integrate or transfer knowledge of LLMs into ASR while not increasing the complexity of decoding either with knowledge distillation \cite{KD2015} or attention co-training \cite{Kim2017, Hori2017}, particularly the way with transferring the linguistic knowledge from LLMs during acoustic encoding \cite{FNAR-BERT,NARBERT,KuboICASSP2022, Choi2022, Futami2022, Higuchi2023, CIFBERT1, CIFBERT2, CTCBERT1, CTCBERT2, wav2vecBERTSLT2022, Cross2021}. In this way, when performing speech recognition, we do not need any later usage of language model since the linguistic knowledge has been distilled to acoustic encoding. This could be helpful for fast and parallel decoding in ASR since there is no need for a second-pass rescoring process. 
	
	\subsubsection{Cross-modal linguistic knowledge transfer for ASR}
\begin{figure}[tbp]
		\centering
		\includegraphics[width=8cm, height=5.5cm]{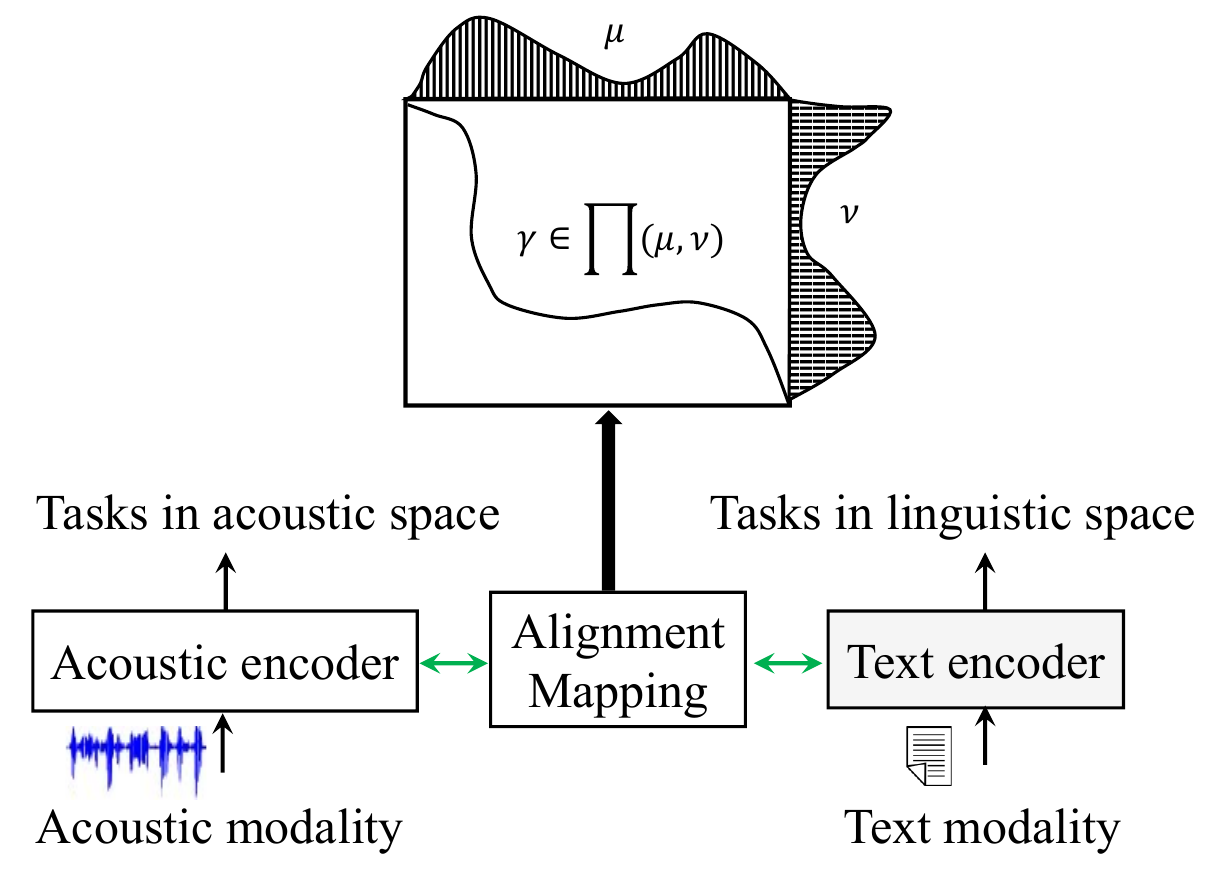}
		\caption{Cross-modal alignment and mapping for knowledge transfer learning based on OT coupling (acoustic-linguistic modalities).}
		\label{fig:fig20}
	\end{figure}

\begin{figure}[tbp]
		\centering
		\includegraphics[width=8cm, height=5cm]{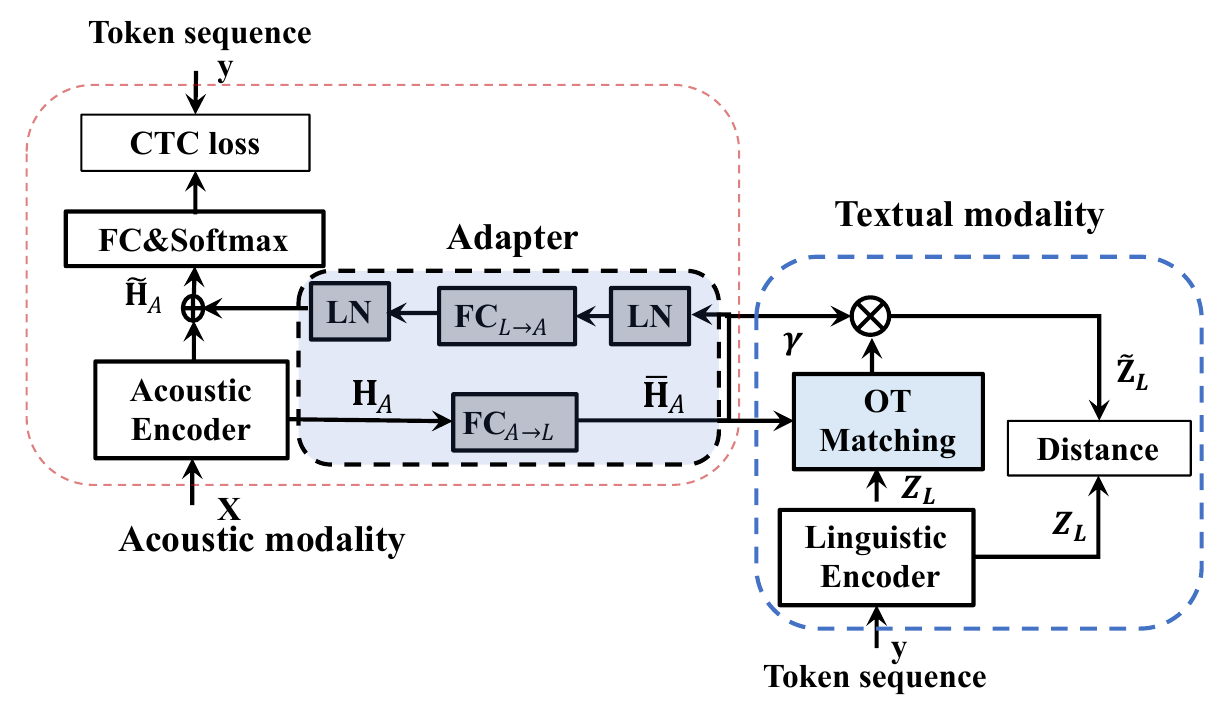}
		\caption{Cross-modal knowledge transfer for ASR based on OT-matching.}
		\label{fig:fig27}
	\end{figure}

	During speech communication or perception, usually multi-modalities are involved in sensing and decision making, for example, in speech communication, modalities corresponding to audio, video and text are integrated. For knowledge transfer learning, multi-modal information could be transferred among different modalities for improving performance on tasks in a specific uni-modality. An example of acoustic-linguistic knowledge transfer learning is shown in Fig. \ref{fig:fig20}.
	In this figure, there are two modalities, one is the acoustic modality, the other is text modality, and they have shared or co-occurred information since audio speech always has corresponding transcriptions. In this sense, these two modalities can be regarded as two sides of a coin during speech communication. However, the distributions of the feature in these two modalities are different (or heterogeneous). For efficient knowledge transfer learning, feature alignment and mapping between acoustic and text modalities are required. If we regard the two modalities follow two different probability distributions $\mu$ and $\nu $, the alignment between the two modalities is naturally recast to finding OT coupling $\gamma \in \Pi(\mu,\nu)$ as illustrated in Fig. \ref{fig:fig20}. Based on this idea, an OT-based knowledge transfer framework (conformer-based acoustic model \cite{conformer2020} and pretrained BERT model \cite{BERT}) has been proposed as detailed in Fig. \ref{fig:fig27} \cite{LuASRU2023}. 
	In this framework, two uni-modal encoders, i.e., acoustic and linguistic encoders, are used for exploring acoustic and linguistic feature representations. Between these encoders, there is an `Adapter' module for feature transforms during knowledge transfer learning. In Fig. \ref{fig:fig27}, an `OT matching' module is designed which serves as the key in cross-modal knowledge transfer learning. The function of this OT matching module is to align and match acoustic and linguistic representations for efficient transfer learning. Further details are provided in the following sections.
	
	\subsubsection{Acoustic and linguistic feature representations}
	As illustrated in Fig. \ref{fig:fig27}, the acoustic and linguistic representations ${\bf H}_A$ and ${\bf Z}_L$ are extracted by: 
	\begin{equation}
			{\bf H}_A = {\rm Encoder}_{A} ({\bf X}); \quad {\bf Z}_L = {\rm Encoder}_{L} ({\bf y}),	
	\end{equation}
	where ${\bf X}$ and ${\bf y}$ are acoustic and linguistic inputs, ${\rm Encoder}_{A}(\cdot)$ and ${\rm Encoder}_{L}(\cdot)$ denote the transforms of the two encoders, respectively. For paired speech-text sequences, they are represented as ${\bf H}_A = [ \mathbf{h}_1, \dots, \mathbf{h}_{l_a} ]\in \mathbb{R}^{l_a \times d_a}$ and ${\bf Z}_L = [ \mathbf{z}_1, \dots, \mathbf{z}_{l_t} ] \in \mathbb{R}^{l_t \times d_t} $, where $d_a$ and $d_l$ are dimensions of acoustic and linguistic features, $l_a$ and $l_t$ are sequence lengths of acoustic and linguistic features. In Fig. \ref{fig:fig27}, the dimension matching transform from acoustic to linguistic spaces ${\rm FC}_{A \to L}(\cdot)$ is obtained as  ${\bf \bar H}_A  = \left[ {{\bf \bar h}_1 ,...,{\bf \bar h}_{l_a } } \right] \in R^{l_a  \times d_t }$ with element estimated from ${\bf \bar h}_i  = {\rm FC}_{A \to L} \left( {{\bf h}_i } \right)$.
	\subsubsection{Sinkhorn-based OT for cross-modal alignment}
	Given acoustic and linguistic feature sequences ${\bf \bar H}_{A}$ and ${\bf Z}_L$ (obtained from text encoder if no further process is required), respectively, as: 
	\begin{equation}
		\begin{array}{l}
			{\bf \bar H}_{A}  = \left[ {{\bf \bar h}_1 ,{\bf \bar h}_2 ,...,{\bf \bar h}_i ,...,{\bf \bar h}_{l_a } } \right], \\ 
			{\bf Z}_L  = \left[ {{\bf z}_1 ,{\bf z}_2 ,...,{\bf z}_j ,...,{\bf z}_{l_t } } \right]. \\ 
		\end{array}
		\label{eq:twoseqs}
		\vspace{-1mm}
	\end{equation}
	Suppose that the two sequences in Eq. (\ref{eq:twoseqs}) are sampled from two probability distributions with weight vectors ${\bf a}  = \left[ {a_1 ,a_2 ,...,a_i ,...,a_{l_a } } \right]$ and 
	${\bf b}  = \left[ {b_1 ,b_2 ,...,b_j ,...,b_{l_t } } \right]$. (if no prior information is available, set $a_i  = {1 \mathord{\left/
			{\vphantom {1 {l_a }}} \right.
			\kern-\nulldelimiterspace} {l_a }}$ and $b_j  = {1 \mathord{\left/
			{\vphantom {1 {l_t }}} \right.
			\kern-\nulldelimiterspace} {l_t }}$ as uniform distributions). The OT distance between the two sequences is defined as:       
	\begin{equation}
		L_{\rm OT} \mathop  = \limits^\Delta  \mathop {\min }\limits_{\gamma  \in \prod {\left( {{{\bf \bar H}_{A}} ,{{\bf Z}_L} } \right)} } \left\langle {\gamma ,{\bf C}} \right\rangle, 
		\label{eq:ot}
	\end{equation}
	where ${\gamma}$ is a transport coupling set defined as:
	\begin{equation}
		\prod {\left( {{{\bf \bar H}_{A}} ,{{\bf Z}_L} } \right)} \mathop  = \limits^\Delta  \left\{ {\gamma  \in R_ + ^{l_a  \times l_t } \left| {\gamma {\bf 1}_{l_t }  = {\bf a},\gamma ^T {\bf 1}_{l_a }  = {\bf b}} \right.} \right\}.
		\label{eq:coupling}
	\end{equation}
	In Eq. (\ref{eq:coupling}), ${\bf 1}_{l_a }$ and ${\bf 1}_{l_t }$ are vectors of ones with dimensions $l_a$ and $l_t$, respectively. In Eq. (\ref{eq:ot}), $\bf C$ is a distance matrix (or ground cost metric) with element ${c_{i,j} }$ defined as pairwise cosine distance:
	\begin{equation}
		c_{i,j}  = {\bf C} \left( {{\bf \bar h}_i ,{\bf z}_j } \right)\mathop  = \limits^\Delta  1 - \cos \left( {{\bf \bar h}_i ,{\bf z}_j } \right).
		\label{eq:Cost}
		\vspace{-1mm}
	\end{equation}
	A fast estimation of OT has been introduced through the celebrated EOT \cite{Cuturi2013} (See Section \ref{sect:EROTSH}), where the EOT loss is defined as:
	\begin{equation}
		L_{\rm EOT} \left( {{\bf \bar H}_{ A} ,{\bf Z}_L } \right)\mathop  = \limits^\Delta  \mathop {\min }\limits_{\gamma  \in \prod {\left( {{{\bf \bar H}_{ A}} ,{{\bf Z}_L} } \right)} } \left\langle {\gamma ,{\bf C}} \right\rangle  - \lambda_1   H\left( \gamma  \right), 
		\label{eq:EOTloss}  
	\end{equation}
	where $\lambda _1 $ is a regularization coefficient, and $H\left( \gamma  \right) =  - \sum\limits_{i,j} {\gamma _{i,j} \log \gamma _{i,j} } $ is the entropy of coupling matrix. The solution of Eq. (\ref{eq:EOTloss}) can be implemented with the Sinkhorn algorithm as \cite{Cuturi2013}:
	\begin{equation}
		\gamma _{\lambda _1 }  = diag\left( {{\bf u} } \right)*{\bf K}*diag\left( {{\bf v} } \right),
		\label{eq:gamma}
	\end{equation}
	where ${\bf K} = \exp \left( { - \frac{{\bf C}}{{\lambda_1 }}} \right)$, ${{\bf u} }$ and ${{\bf v} }$ are two scaling (or re-normalization) vectors.
	\subsubsection{Temporal order preserved OT}
	In the original estimation of OT in Eq. (\ref{eq:EOTloss}), the two sequences in Eq. (\ref{eq:twoseqs}) are treated as two sets without considering their temporal order relationship. In speech, temporal order information is crucial in OT coupling during cross-modal alignment, meaning that neighboring frames in an acoustic sequence should be progressively coupled with neighboring tokens in a linguistic sequence which has been investigated in \cite{LuSLT2024}. For the sake of clarity, the two sequences in Eq. (\ref{eq:twoseqs}) can be further represented with temporal order information as:
	\begin{equation}
		\begin{array}{l}
			{\bf \bar H}_{A}  = \left[ {\left( {{\bf \bar h}_1 ,1} \right),\left( {{\bf \bar h}_2 ,2} \right),...,\left( {{\bf \bar h}_i ,i} \right),...,\left( {{\bf \bar h}_{l_a } ,l_a } \right)} \right], \\ 
			{\bf Z}_L  = \left[ {\left( {{\bf z}_1 ,1} \right),\left( {{\bf z}_2 ,2} \right),...,\left( {{\bf z}_j ,j} \right),...,\left( {{\bf z}_{l_t } ,l_t } \right)} \right]. \\ 
		\end{array}	
		\label{eq:inputseq}
	\end{equation} 
	During the alignment of the two sequences for knowledge transfer, it is crucial to consider that elements with significant cross temporal distances might not be likely to be coupled. In other words, the coupling pairs with high probabilities between the two sequences should be distributed along the diagonal line of the temporal coherence positions. Based on this consideration, the temporal coupling prior could be defined as a two dimensional Gaussian distribution \cite{Su2017}. The fundamental concept is that the coupled pairs should not deviate significantly from the diagonal line of temporal coherence positions between the two sequences, which can be defined as:    
	\begin{equation}
		p_{i,j} \mathop  = \limits^\Delta  \frac{1}{{\sigma \sqrt {2\pi } }}\exp \left( { - \frac{{d_{i,j}^2 }}{{2\sigma ^2 }}} \right),
		\label{eq:P}
	\end{equation}
	where $\sigma$ is a variation variable controlling the impact of the cross-temporal distance $d_{i,j}$ as defined in Eq. (\ref{eq:dij}).
	\begin{equation}
		d_{i,j}  = \frac{{\left| {{\raise0.5ex\hbox{$\scriptstyle i$}
						\kern-0.1em/\kern-0.15em
						\lower0.25ex\hbox{$\scriptstyle {l_a }$}} - {\raise0.5ex\hbox{$\scriptstyle j$}
						\kern-0.1em/\kern-0.15em
						\lower0.25ex\hbox{$\scriptstyle {l_t }$}}} \right|}}{{\sqrt {{\raise0.5ex\hbox{$\scriptstyle 1$}
						\kern-0.1em/\kern-0.15em
						\lower0.25ex\hbox{$\scriptstyle {l_a^2 }$}} + {\raise0.5ex\hbox{$\scriptstyle 1$}
						\kern-0.1em/\kern-0.15em
						\lower0.25ex\hbox{$\scriptstyle {l_t^2 }$}}} }}.
		\label{eq:dij}
	\end{equation}
	In Eq. (\ref{eq:dij}), the cross-temporal distance is defined on the normalized sequence lengths in acoustic and linguistic spaces. In this definition, it is evident that the farther the distance between a paired position and the temporal diagonal line, the lower the possibility of their correspondence in transport coupling. By incorporating this temporal coherence prior as regularization, the new OT is defined as:
	\begin{equation}
		L_{\rm{TOT}} ({\bf \bar H}_{A} ,{\bf Z}_L )\mathop  = \limits^\Delta  \mathop {\min }\limits_{\mathclap{\gamma  \in \prod {({{\bf \bar H}_{A}} ,{{\bf Z}_L} )} }}  < \gamma ,{\bf C} >  - \lambda _1   H(\gamma ) + \lambda _2   \mathrm{KL}(\gamma ||P),
		\label{eq:TOP}
	\end{equation}
	where $\lambda _1$ and $\lambda _2$ are two trade off parameters. In Eq. (\ref{eq:TOP}), $\mathrm{KL}(\gamma||P)$ is the KL-divergence between the transport coupling matrix $\gamma$ and the temporal prior correspondence matrix $P$ with elements defined in Eq. (\ref{eq:P}). Building upon the definitions of KL-divergence and entropy, Eq. (\ref{eq:TOP}) can be further expressed to:
	\begin{equation}
		L_{\rm{TOT}} ({\bf \bar H}_{A} ,{\bf Z}_L )\mathop  = \limits^\Delta  \mathop {\min }\limits_{\gamma  \in \prod {({{\bf \bar H}_{A}} ,{{\bf Z}_L} )} }  < \gamma ,{\bf \tilde C} >  - \tilde \lambda H(\gamma ),
		\label{eq:TOPC}
	\end{equation}
	where $\tilde \lambda  = \lambda _1    + \lambda _2$, and combined ground cost matrix as
	\begin{equation}
		{\bf \tilde C} = {\bf C} - \lambda _2  \log P,
		\label{eq:newC}
	\end{equation}
	where elements in $P$ are defined in Eq. (\ref{eq:P}). Following the procedures outlined in \cite{Cuturi2013}, the solution of Eq. (\ref{eq:TOPC}) is obtained using the Sinkhorn algorithm as:
	\begin{equation}
		\gamma _{\tilde \lambda }  = diag\left( {{\bf u} } \right)*{\bf \tilde K}*diag\left( {{\bf v} } \right),
	\end{equation}
	where ${\bf \tilde K} = \exp \left( { - \frac{{{\bf \tilde C}}}{{\tilde \lambda }}} \right)$. Substituting variables in Eq. (\ref{eq:newC}) into ${\bf \tilde K}$, we can obtain the following:
	\begin{equation}
		{\bf \tilde K} = P^{\frac{{\lambda _2}}{{\lambda_1    + \lambda_2  }}} \exp ( - \frac{{\bf C}}{{\lambda_1  + \lambda_2 }}).	
		\label{eq:newG}	
	\end{equation}
	From this equation, we can see that the transport coupling between the two sequences is further constrained by their temporal order correspondence. Temporal order preserved OT involves several hyper-parameters that can be challenging to control. For the sake of simplification, their effects can be consolidated into a reduced number of hyper-parameters. For example, considering Eq. (\ref{eq:newC}), the impact of variation $\sigma$ in Eq. (\ref{eq:P}) and $\lambda_2$ in Eq. (\ref{eq:TOP}) can be combined into a single control parameter $\beta_{\rm dist}$, defined as:
	\begin{equation}
		{\bf \tilde C} = {\bf C} + \beta_{\rm dist} d_{i,j}^2,
		\label{eq:combined}
	\end{equation}
	and the Sinkhorn algorithm is applied to this consolidated cost function ${\bf \tilde C}$ for OT in real implementations.  
	\subsubsection{Loss function}
	In the transfer learning model as illustrated in Fig. \ref{fig:fig27}, two loss functions are involved: the cross-modal alignment and matching loss (in the right branch of Fig. \ref{fig:fig27}) and the recognition loss (either CTC-based loss \cite{Graves2012, CTCASR} or CE-based classification loss). For cross-modal alignment, the acoustic feature is first projected to the linguistic space using OT transform as:
	\begin{equation}
		\begin{array}{l}
			\begin{aligned}
				{\bf \tilde Z}_L &\mathop  = \limits^\Delta  {\rm OT}\left( {{\bf \bar H}_{A}  \to {\bf Z}_{L} } \right), \\ 
				&= \left( {\gamma ^* } \right)^T  \times {\bf \bar H}_{A}  \in R^{l_t  \times d_t },  \\ 
			\end{aligned}
		\end{array}	
	\end{equation}
	where $\left({\gamma ^* } \right)^T$ is the transpose of the OT coupling obtained from the solution of Sinkhorn-based OT. Subsequently, the alignment loss is directly defined on the projected space as
	\begin{equation}
		L_{{\rm align}}  = \sum\limits_{j = 1}^{l_t} {1 - \cos \left( {{\bf \tilde z}_L^j ,{\bf z}_L^j } \right)}, 
		\label{eq:Align}
	\end{equation}
	where ${\bf \tilde z}_L^j$ and ${\bf z}_L^j$ are row vectors of feature matrices ${\bf \tilde Z}_{L}$ and ${\bf Z}_{L}$ (according in temporal dimensions), respectively. For efficient transfer of linguistic knowledge to acoustic encoding, the following transforms are designed as indicated in Fig. \ref{fig:fig27}:
	\begin{equation}			
		{\bf \tilde H}_{A}   = {\bf H}_A  + {\rm LN(FC}_{L \to A} {\rm (LN(}{{\bf \bar H}_{A}} {\rm )))} \in \mathbb{R}^{l_a  \times d_a }. 	
		\label{eq:adapter}
	\end{equation}
	Based on this new representation ${\bf \tilde H}_{A}$, the probability prediction for ASR is formulated as ${\bf \tilde P} = {\rm Softmax}( {{\rm FC}({\bf \tilde H}_{A})} )$,
	where `FC' is a fully connected linear transform with output size the same as that of vocabulary tokens. Finally, the total loss in model training is defined as
	\begin{equation}
		L\mathop  = \limits^\Delta  \eta L_{{\rm C}} ({\bf \tilde P},{\bf y} ) + (1 - \eta ){L_{{\rm align}}},   
		\label{eq:totalloss} 	
	\end{equation}
	where $L_{{\rm C}} ({\bf \tilde P},{\bf y} )$ is a classification loss, either CE or CTC loss (as shown in Fig. \ref{fig:fig27}). After training the model, only the branch of the acoustic modality of Fig. (\ref{fig:fig27}) is retained for the inference of ASR. The alignment and matching are performed on the final representations from encoders of the two modalities \cite{LuASRU2023, LuSLT2024}, as an extension, the alignment and matching also has been designed on different layers of acoustic encoders in order to explore rich acoustic information with consideration of linguistic knowledge \cite{LuICASSP2024}. Moreover, the study in \cite{LuICASSP2024} also indicated the connection between conventional transformer-based cross-attention and OT where OT-based transport can be regarded as a double-side (row and column) normalization process; therefore, the matching and alignment are also named as Sinkhorn attention \cite{SinkhornAtt2020,MialonICLR2021,SanderAISTATS2022}.
	
	\subsubsection{Adding structure to OT in Cross-modal linguistic knowledge transfer for ASR}
	\label{sec:gmotASR}
\begin{figure}[tbp]
		\centering
		\includegraphics[width=7cm, height=5cm]{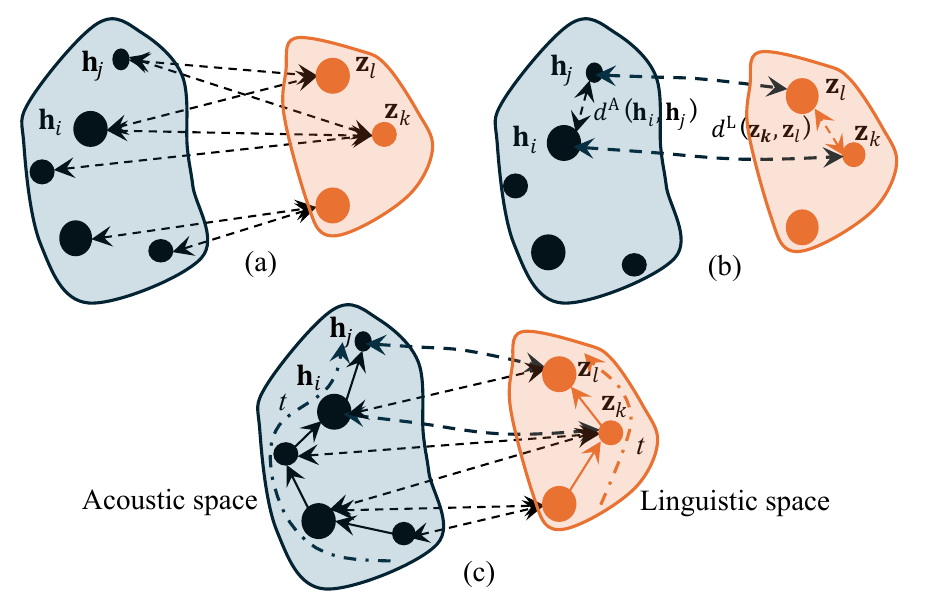}
		\caption{Graph matching OT in Cross-modal knowledge transfer learning for ASR.}
		\label{fig:fig28}
	\end{figure}

	The original concept of OT is defined on sample points between two sets as illustrated in Fig. \ref{fig:fig28}-(a), either acoustic or linguistic space, $\left[ {{\bf h}_1 ,{\bf h}_2 ,...,{\bf h}_i ,...,{\bf h}_j ,...} \right]$ or $\left[ {{\bf z}_1 ,{\bf z}_2 ,...,{\bf z}_l ,...,{\bf z}_k ,...} \right]$, can be regarded as samples and can be independently associated or coupled during transportation between the two spaces, if we add structure (e.g., topological structure) on points in a space, for example, by defining the distance metric between two points in each space as illustrated in Fig. \ref{fig:fig28}-(b), $d^{\rm A} \left( {{\bf h}_i ,{\bf h}_j } \right)$ and $d^{\rm L} \left( {{\bf z}_l ,{\bf z}_k } \right)$ in acoustic and linguistic spaces, the association or coupling can be designed on their distance relationship between two points. Moreover, as it is known, information in acoustic speech and text are sequentially encoded, that is the temporal order information of points should be considered during matching between the two sets. For example, as shown in Fig. \ref{fig:fig28}-(c), temporal order information is attached to each node. From these explanations, we can see that during OT, we could add these kinds of structured information. This idea has been addressed and named as graph matching OT in cross-modal knowledge transfer for ASR in \cite{LuIS2025}.
	
\begin{figure}[tbp]
		\centering
		\includegraphics[width=7cm, height=4cm]{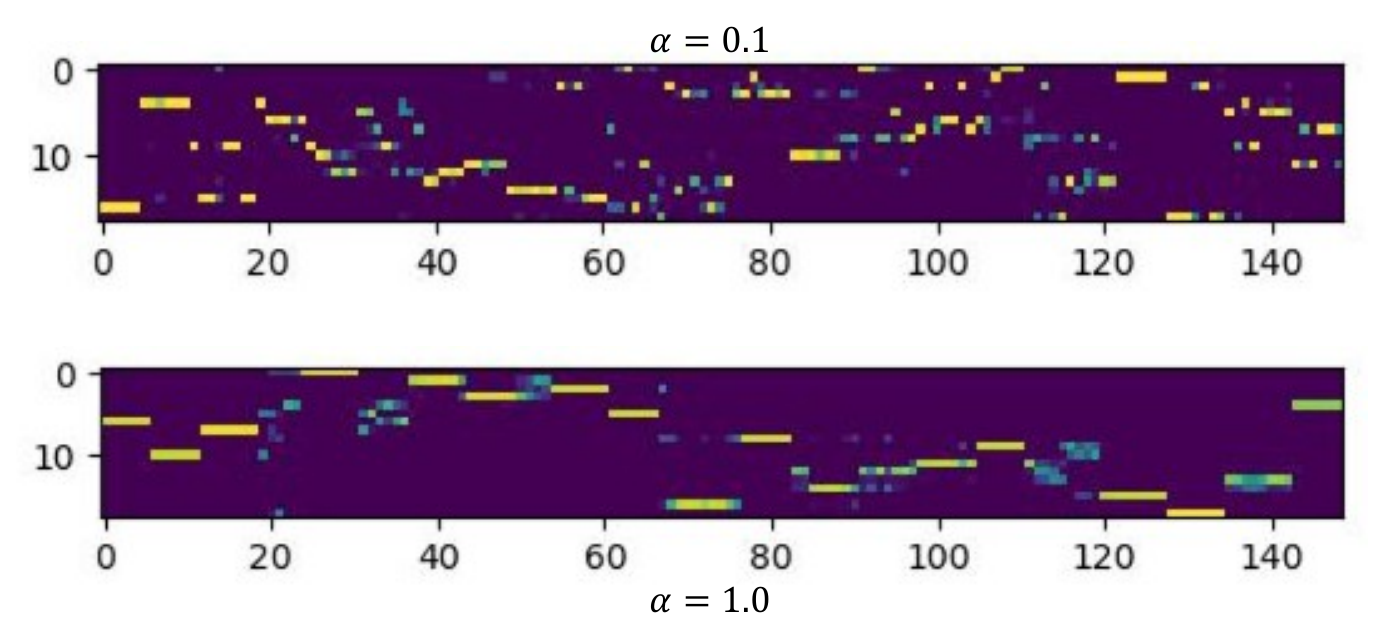}
		\caption{Segmental structure of speech in matching between acoustic and linguistic features based on OT.}
		\label{fig:fig29}
	\end{figure}

\begin{figure}[tbp]
		\centering
		\includegraphics[width=7cm, height=4cm]{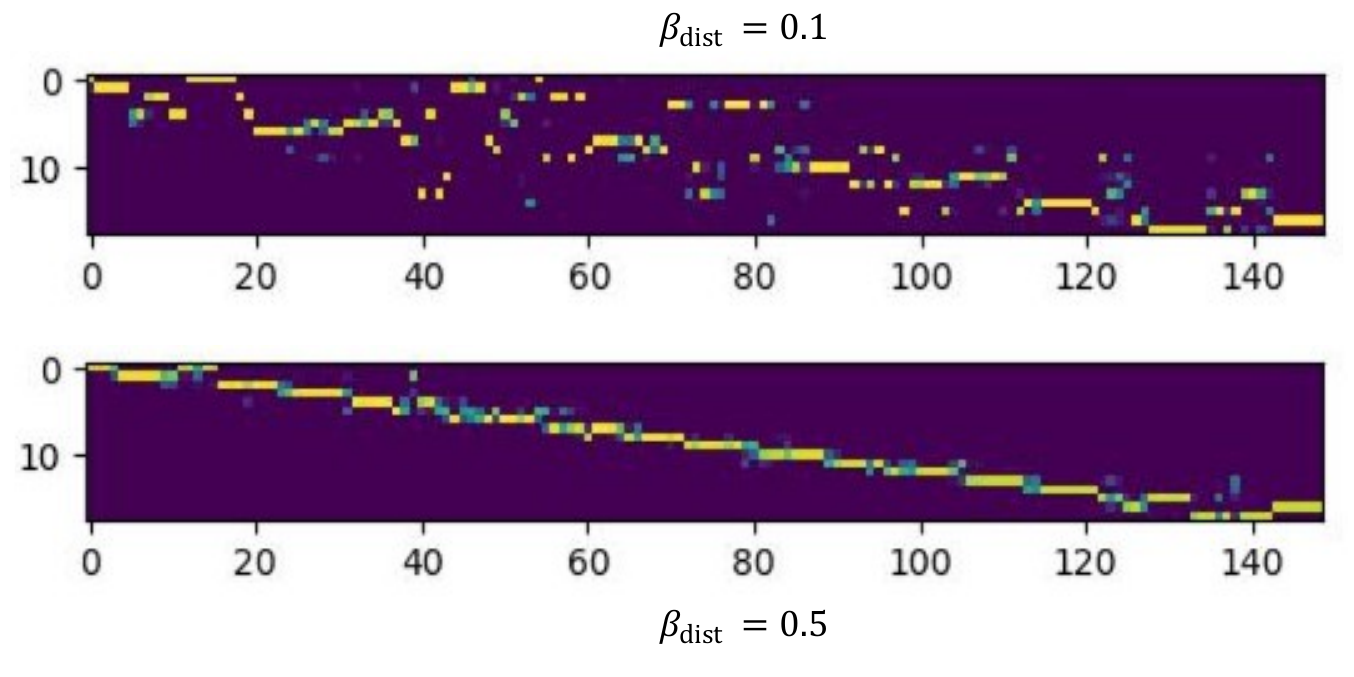}
		\caption{Temporal order structure of speech in matching between acoustic and linguistic features based on OT.}
		\label{fig:fig30}
	\end{figure}

\begin{figure}[tbp]
		\centering
		\includegraphics[width=7cm, height=4cm]{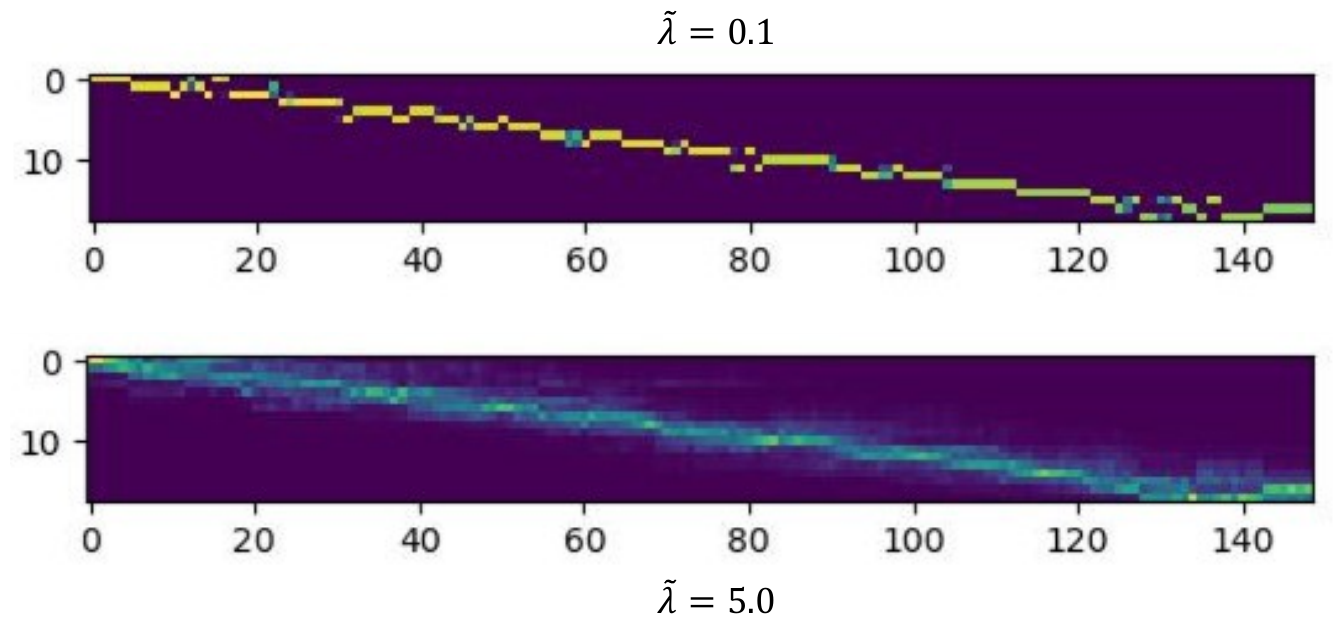}
		\caption{Entropy regularization of coupling in matching between acoustic and linguistic features based on OT.}
		\label{fig:fig31}
	\end{figure}

	The behavior of adding different structures in OT for acoustic and linguistic matching and alignment has been well analyzed with different hyper-parameters. For example, as shown in Fig. \ref{fig:fig29}, with adjusting weighting importance between matching on nodes and edges (the trade-off weighting coefficient $\alpha$ in Eq. (\ref{eq:fgwot}) to control the importance between WD and GWD). In this figure, horizontal and vertical are indexes of acoustic embedding and its corresponding linguistic token embedding sequences. With small $\alpha$, we can see that there are many small segments of the matching or coupling, but when $\alpha$ increases, we can see a clear structure of the matching or coupling segments between the embedding of acoustic and linguistic feature embeddings. 
	Changing the coefficient $\beta_{\rm dist}$ in Eq. (\ref{eq:combined}), the importance of the temporal order information can be controlled in alignment and matching based on OT. The behavior is shown in Fig. \ref{fig:fig30}. we can see that the alignment between acoustic and linguistic representations follows a strong monotonic corresponding relationship. 
	Moreover, in the solution of the OT based on the Sinkhorn algorithm, the regularization of the OT coupling is adopted. In the definition Eq. (\ref{eq:TOPC}), with controlling of the regularization coefficient $\hat \lambda$, the expansion and smooth behavior can be observed as shown in Fig. \ref{fig:fig31}. 
	With properly setting these hyper-parameters, state of the art ASR performance could be obtained which showed the advantages of cross-modal alignment and matching \cite{LuASRU2023, LuSLT2024}.  
	\subsubsection{Unbalanced OT in Cross-modal linguistic knowledge transfer for ASR}
\begin{figure}[tbp]
		\centering
		\includegraphics[width=6cm, height=4.5cm]{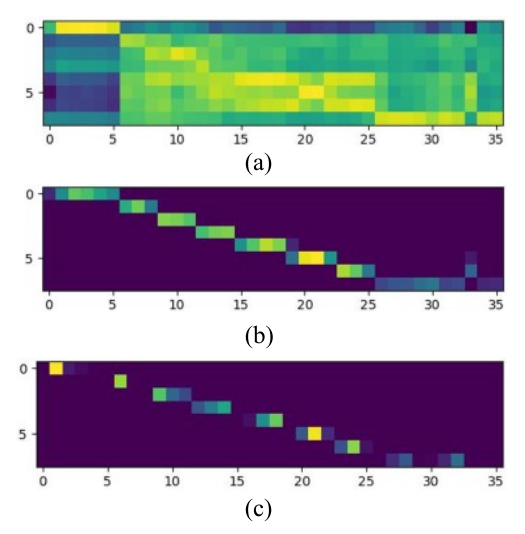}
		\caption{Alignment between acoustic and linguistic tokens (indexes of acoustic (Horizontal) and linguistic (Vertical) embedding sequences): (a) cosine similarity matrix; (b) $\alpha_1=1.0$, $\alpha_2=1.0$; (c) $\alpha_1=0.01$, $\alpha_2=1.0$.}
		\label{fig:fig38}
	\end{figure}

	With different setting of $\alpha_1$ and $\alpha_2$ of UOT in Eq. (\ref{eq:uot}), the alignment behaviors between acoustic and linguistic features can be well controlled: (a) {\bf Acoustic-to-Linguistic (A2L) alignment}: To ensure that every linguistic token is aligned, we set \( \alpha_2 > \alpha_1 \). This forces mass coverage over linguistic units while permitting selective skipping or discarding of noisy and outlier acoustic frames, including NULL acoustic matching; (b) {\bf Linguistic-to-Acoustic (L2A) alignment}: To account for as much of the acoustic input as possible, we set \( \alpha_1 > \alpha_2 \), ensuring that the acoustic frames are matched even if some linguistic tokens are less activated.	This matching strategy in model training tries to encourage bidirectional consistency, and the resulting optimal transport plan \( \gamma^\ast \) represents a soft alignment matrix that assigns probabilistic mass between acoustic and linguistic elements, effectively grounding linguistic tokens in observed speech while avoiding overfitting to background noisy frames in transfer learning \cite{LuICASSP2026}. The effect is shown in Fig. \ref{fig:fig38}. From this figure, we can see that the alignment between the acoustic and linguistic embeddings is well controlled by properly setting the regularization parameters $\alpha_1$ and $\alpha_2$ to meet the requirement of unbalanced matching between them. 	
	\subsection{Cross-domain adaptation in various identification and detection tasks}
	\label{sect:otLID}
\begin{figure}[tbp]
		\centering
		\includegraphics[width=7cm, height=6cm]{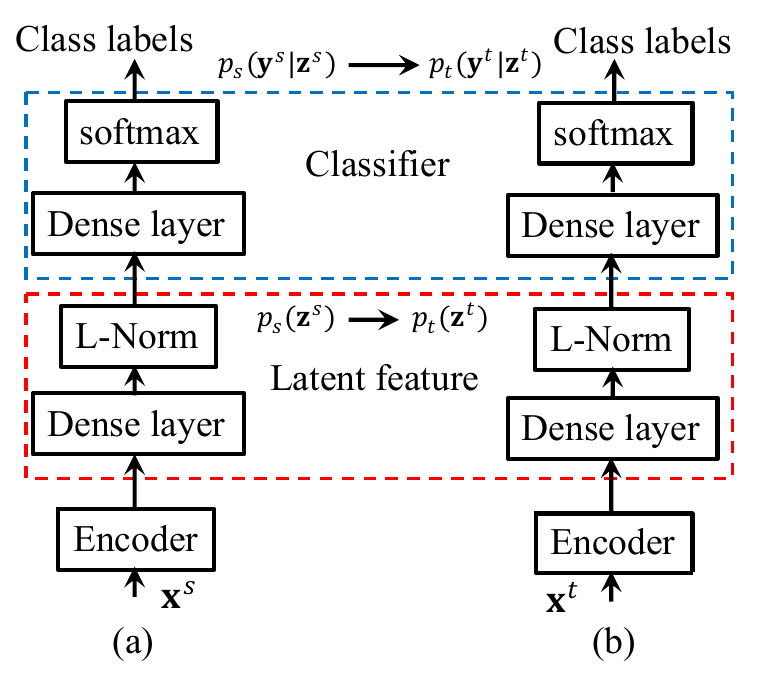}
		\caption{Cross-domain adaptation for LID based on a Siamese neural network architecture for source (left) and target (right) domain samples.}
		\label{fig:fig25}
	\end{figure}

\begin{figure}[tbp]
		\centering
		\includegraphics[width=8cm, height=4cm]{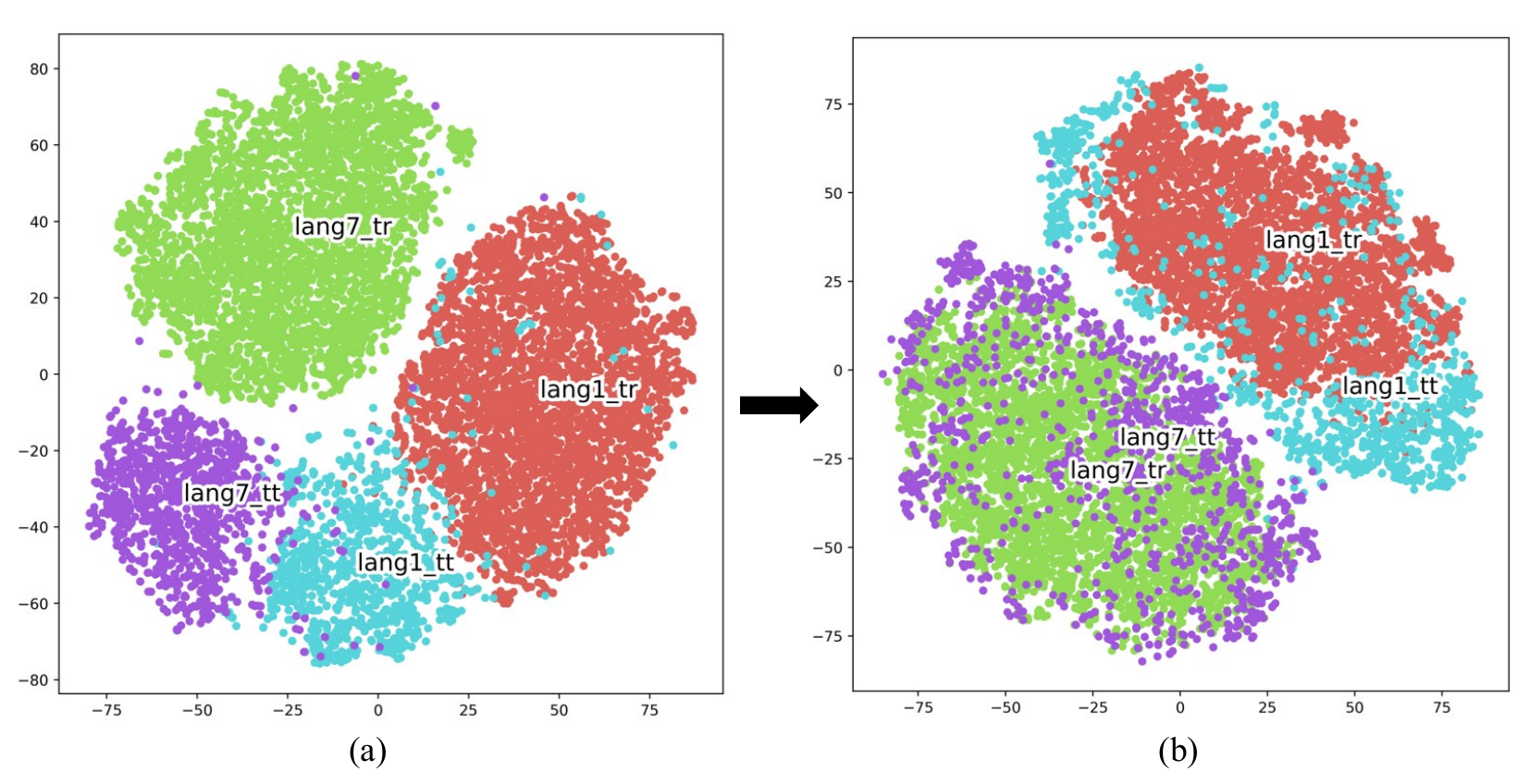}
		\caption{Visualization of feature representations via t-SNE in cross-domain LID before (a) and after (b) adaptation, lang\{1,7\}\_tr, lang\{1,7\}\_tt denote language IDs 1 and 7 for training and test.}
		\label{fig:fig33}
	\end{figure}

\begin{figure}[tbp]
		\centering
		\includegraphics[width=8.5cm, height=3.5cm]{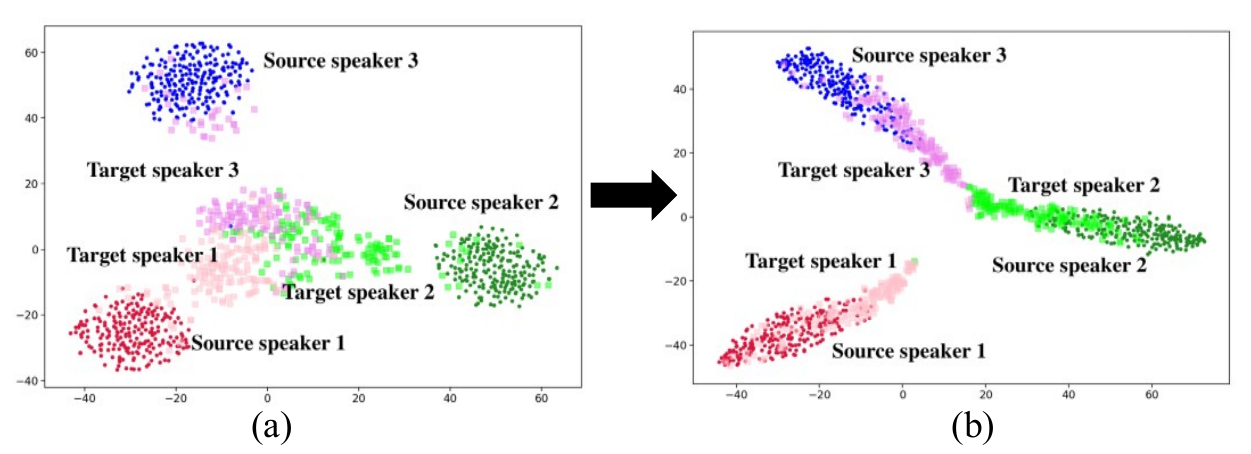}
		\caption{Feature representation in cross-domain speaker recognition before (a) and after (b) adaptation.}
		\label{fig:fig34}
	\end{figure}

	In various speech tasks, for example LID and SR, the domain mismatch problem is very common and challenging. Reducing domain mismatch for improving domain generalization performance is necessary where the distance metric between domain distributions is involved \cite{SpeakerACM2018}. OT is quite suitable for these tasks, as OT naturally defines the distance metric between probability distributions. For the cross-domain LID task, an OT-based adaptation neural model framework has been proposed \cite{LuICASSP2021}. The model framework is illustrated in Fig. \ref{fig:fig25} where both training and test networks share the same set of weight parameters in a Siamese network architecture. And the x-vector is adopted as input feature \cite{XVector2018} with fixed pretrained model parameters. The adaptation is designed at both the feature and the classifier levels to reduce the mismatch between the distributions of training and testing sets. A similar idea has been proposed for SR \cite{ZhangTASLP2024, YangTIFS2026}. 
	The effect of adaptation for LID and SR is shown in Figs. \ref{fig:fig33} and \ref{fig:fig34}. In these figures, the t-SNE visualizations \cite{TSNE2008} of the features of the language and speaker clusters are plotted. From these figures, we can see that there exist obvious gaps between the training/source and test/target domains (two given language IDs marked lang1\_\{tr, tt\} and lang7\_\{tr, tt\} in Fig. \ref{fig:fig33}-a; Speaker IDs marked with \{Source, Target\} speaker \{1, 2, 3\} in Fig. \ref{fig:fig34}-a). After adaptation, distributions of features in training (source) and test (target) domains are pushed to neighboring regions while keeping their class discrimination structure.   
	OT-based domain alignment and adaptation have also been applied in speech emotion recognition tasks \cite{ZhangIS2023,ZhangICASSP2024}, where speech emotion samples come from different languages and recording environments. 
	
	In recent years, due to the requirement of trust-worthy speech communications, audio spoof detection (ASD) has been increasing its importance in various applications. ASD, also known as anti-spoofing refers to techniques designed to distinguish between bona fide speech (genuine/real human speech) and artificially generated or manipulated speech. Most ASD models are designed and trained with a large number of training data samples. It is difficult to use the same models to deal with all spoof speech generated with various algorithms and from versatile domains \cite{DeepFakeSv2025}. Cross-domain ASD and OT-based adaptation techniques have been developed under the same line of cross-domain tasks in LID and SVR, and significant performance gains were obtained \cite{ZhangTASLP2025, ZhangTIFS2025}. 
	
	
	\section{Future directions}
	The application of OT to speech processing has shown promising results across various tasks and applications, either for multi-modal representation transfer learning, or cross-modal domain adaptation and integration. However, several important directions remain open for further exploration. First, scalability and efficiency remain critical challenges. Classical OT formulations often incur high computational cost, especially when applied to long speech sequences or high-dimensional feature spaces. Developing more efficient computational algorithms is essential in engineering implementations, such as stochastic OT, mini-batch OT, and low-rank or sparsity-constrained transport plans, to enable real-time and large-scale deployment in speech applications systems. A particularly promising direction is the integration of OT with SDE-based generative modeling frameworks. Recent advances in diffusion models and flow matching methods have demonstrated strong capabilities in modeling complex data distributions. OT provides a natural perspective for describing the evolution of probability distributions over time, which aligns closely with the dynamics defined by SDEs. Moreover, with lifting the states of the dynamics in a functional space, composition operator theory (e.g., Koopman operator) may be applied to investigate the nonlinear dynamic properties with linear operator analysis. Future work may explore the theoretical connections between OT and diffusion bridges/processes, transfer operator learning, as well as their practical implications for domain or modality alignment and matching. 
	
	Multi-modal learning also remains a fertile area for OT-based methods. Speech is often accompanied by complementary modalities, such as visual information (e.g., lip movements), articulatory signals, and textual context. Speech processing requires methods and algorithms for multimodal feature learning or multimodal information fusion \cite{MultimodalSv2024, MultimodalSv2025}. OT offers a natural framework for aligning heterogeneous feature distributions across these modalities. In particular, domain adaptation and modality gap minimization should receive increased attention, especially in scenarios involving mismatched distributions across audio, visual, and textual streams. Future research may focus on conditional and dynamic transport formulations that explicitly reduce cross-modal discrepancies, handle temporal asynchrony, and improve robustness under missing or noisy modalities. In addition, theoretical advances are needed to better understand the role of OT in speech modeling. This includes studying the properties of transport-based alignments in sequential data, their relationship to probabilistic formulations such as SDEs and diffusion models, and their implications for representation learning and generalization. Such insights could guide the design of more principled and effective OT-based loss or objective functions. 
	
	In summary, future research on OT for speech lies at the intersection of efficiency, generative modeling, multi-modality, and robustness. In particular, a deeper exploration of OT in conjunction with SDE-based generative frameworks and transfer operator theory (as well as composition operator theory), along with a stronger focus on domain adaptation and modality gap minimization, will be crucial for advancing next-generation speech technologies. 
	
	
	
	\bibliographystyle{IEEEtran}

\begin{thebibliography}{1}
		
		\bibitem{AmbrosioBook2008} 
		Luigi Ambrosio, Nicola Gigli, Giuseppe Savar{\'e}, Gradient Flows In Metric Spaces and in the Space of Probability Measures, Springer, 2008.
		
		\bibitem{AmbrosioBook2021}
		L. Ambrosio, E. Bru{\'e} and D. Semola, ``Lectures on Optimal Transport," Springer, 2021.
		
		\bibitem{AmosICML2017}
		B. Amos, L. Xu, and J. Kolter, ``Input convex neural networks," the Proceedings of the 34th International Conference on Machine Learning, Vol. 70, pp. 146-155, 2017.
		
		\bibitem{Anderson1982}
		B. Anderson, ``Reverse-time diffusion equation models," Stochastic Processes and their Applications, 12(3):313-326, 1982.

        \bibitem{Arase2023}
        Yuki Arase, Han Bao, and Sho Yokoi, ``Unbalanced Optimal Transport for Unbalanced Word Alignment," In Proceedings of the 61st Annual Meeting of the Association for Computational Linguistics (Volume 1: Long Papers), pp. 3966-3986, 2023.
        
		\bibitem{ArjovskyICML2017}
		M. Arjovsky, S. Chintala, L. Bottou, ``Wasserstein Generative Adversarial Networks," International Conference on Machine Learning, PMLR: 214–223, 2017.
		
		\bibitem{wav2vec2.0}
		A. Baevski, Y. Zhou, A. Mohamed, and M. Auli, ``Wav2vec 2.0: A framework for self-supervised learning of speech representations," in \emph{Proc. of NeurIPS}, 2020. 
		
		\bibitem{FNAR-BERT}
		Y. Bai, J. Yi, J. Tao, Z. Tian, Z. Wen and S. Zhang, "Fast End-to-End Speech Recognition Via Non-Autoregressive Models and Cross-Modal Knowledge Transferring From BERT," \emph{IEEE/ACM Transactions on Audio, Speech, and Language Processing}, vol. 29, pp. 1897-1911, 2021.
		
		\bibitem{Benamou2000}
		J. Benamou, Y. Brenier, ``A computational fluid mechanics solution to the Monge-Kantorovich mass transfer problem," Numerische Mathematik, vol. 84, pp. 375-393, 2000.
		
		\bibitem{Benamou2014}
		J. Benamou, G. Carlier, M. Cuturi, Marco, L. Nenna, Luca, G. Peyr\'{e}, ``Iterative Bregman Projections for Regularized Transportation Problems," SIAM Journal on Scientific Computing, Vol.37, No.2, pp. A1111-AA1138, 2015.
		
		\bibitem{Bhardwaj2022}
		R. Bhardwaj,T. Vaidya, S. Poria, ``Towards solving NLP tasks with optimal transport loss," Journal of King Saud University-Computer and Information Sciences, Vol. 34, No. 10, pp. 9441-9452, 2022.
		
		\bibitem{Blondel2018}
		M. Blondel, V. Seguy, A. Rolet, ``Smooth and Sparse Optimal Transport," Proceedings of the Twenty-First International Conference on Artificial Intelligence and Statistics, pp. 880-889, 2018.
		
		\bibitem{Bonneel2023}
		N. Bonneel, J. Digne, ``A survey of optimal transport for computer graphics and computer vision," Computer Graphics Forum (Eurographics STAR 2023), Vol. 42, No. 2, pp. 439-460, 2023.
		
		\bibitem{BortoliNIPS2021}
		V. De Bortoli, J. Thornton, J. Heng, A. Doucet, ``Diffusion schr\"{o}dinger bridge with applications to score-based generative modeling," International Conference on Neural Information Processing Systems, 2021.
		
		\bibitem{Brenier1991}
		Yann Brenier, ``Polar Factorization and Monotone Rearrangement of Vector-Valued Functions," Communications on Pure and Applied Mathematics,vol. 44, pp. 375-417,1991.
		
		\bibitem{Brunton2022}
		S. Brunton, M. Budi\v{s}i\'{c}, E. Kaiser, J. Kutz, ``Modern Koopman Theory for Dynamical Systems," SIAM Review, Vol. 64, No. 2, pp. 229-340, 2022.
		
		\bibitem{Bunne2023}
		C. Bunne, ``Neural Optimal Transport for Dynamical Systems: Methods and Applications in Biomedicine," ETH Zurich thesis, Zürich, Switzerland, 2023.
		
		\bibitem{ChapelPOT2020}
		L. Chapel, M. Alaya, G. Gasso, ``Partial optimal transport with applications on positive-unlabeled learning," Advances in Neural Information Processing Systems, Vol. 33, pp. 2903-2913, 2020.
		
		\bibitem{ChenNLP2019}
		Liqun Chen, Yizhe Zhang, Ruiyi Zhang, Chenyang Tao, Zhe Gan, Haichao Zhang, Bai Li, Dinghan Shen, Changyou Chen, and Lawrence Carin, ``Improving sequence-to-sequence learning via	optimal transport," arXiv preprint arXiv:1901.06283, 2019.
		
		\bibitem{Chen2014}
		Y. Chen, T. Georgiou and M. Pavon, ``On the Relation Between Optimal Transport and Schr{\"o}dinger Bridges: A Stochastic Control Viewpoint," Journal of Optimization Theory and Applications, Vol. 169, pp. 671-691, 2014.  
		
		\bibitem{NODE2018}
		R. Chen, Y. Rubanova, J. Bettencourt, D. Duvenaud, ``Neural ordinary differential equations," arXiv preprint arXiv:1806.07366, 2018.
		
		\bibitem{Chen2021}
		Y. Chen, T. Georgiou and M. Pavon, ``Stochastic Control Liaisons: Richard Sinkhorn Meets Gaspard Monge on a Schr{\"o}dinger Bridge," SIAM Review, Vol. 2, pp. 249-313, 2021.  
		
		\bibitem{ChizatUOT2016}
		L. Chizat, G. Peyré, B. Schmitzer, F. Vialard, ``Scaling algorithms for unbalanced transport problems," arXiv preprin arXiv:1607.05816, 2016.
		
		\bibitem{Choi2022}
		K. Choi, H. Park, ``Distilling a Pretrained Language Model to a Multilingual ASR Model," in \emph{Proc. of INTERSPEECH}, pp. 2203-2207, 2022.
		
		\bibitem{Chuang2020}
		C. Chuang, A. Torralba, and S. Jegelka,``Estimating generalization under distribution shifts via domain-invariant representations," in \emph{Proc. of International Conference on Machine Learning (ICML)}, pp. 1984-1994, 2020.
		
	
		\bibitem{Courty2014}
		N. Courty, R. Flamary, D. Tuia and A. Rakotomamonjy, ``Optimal Transport for Domain Adaptation," In IEEE Transactions on Pattern Analysis and Machine Intelligence, Vol. 39 (9), pp. 1853–1865, 2014.
		
		\bibitem{CourtyNIPS2017}
		N. Courty, R. Flamary, A. Habrard, A. Rakotomamonjy, ``Joint distribution optimal transportation for domain adaptation," Neural Information Processing Systems, 2017.
	
		\bibitem{Cuturi2013}
		Marco Cuturi, ``Sinkhorn Distances: Lightspeed Computation of Optimal Transport," In Advances in Neural Information Processing Systems, pp. 2292-2300, 2013.
		
		\bibitem{DamodaranECCV2018}
		B. Damodaran, B. Kellenberger, R Flamary, D Tuia, N Courty, DeepJDOT:Deep Joint Distribution Optimal Transport for Unsupervised Domain Adaptation, ECCV, 2018.   
		
		\bibitem{CTCBERT1}
		K. Deng, S. Cao, Y. Zhang, L. Ma, G. Cheng, J. Xu, P. Zhang, ``Improving CTC-Based Speech Recognition Via Knowledge Transferring from Pre-Trained Language Models," in \emph{Proc. of ICASSP}, pp. 8517-8521, 2022.
		
		\bibitem{CTCBERT2}
		K. Deng, Z. Yang, S. Watanabe, Y. Higuchi, G. Cheng, P. Zhang, ``Improving Non-Autoregressive End-to-End Speech Recognition with Pre-Trained Acoustic and Language Models," in \emph{Proc. of ICASSP}, pp. 8522-8526, 2022.
		
		\bibitem{BERT}
		J. Devlin, M. Chang, K. Lee, and K. Toutanova, ``Bert: Pretraining of deep bidirectional transformers for language understanding," \emph{arXiv preprint}, arXiv:1810.04805, 2018.	
		
		\bibitem{LTFlow2025}
		K. Do, D. Coeurjolly, P. Memari, N. Bonneel, ``Linear-Time Transport with Rectified Flows," ACM Transactions on Graphics (TOG), Vol. 44 (4), No. 118, pp. 1-13, 2025.
		
		\bibitem{SpeakerACM2018}
		M. Farr\'{u}s, ``Voice Disguise in Automatic Speaker Recognition," ACM Computing Surveys (CSUR), Vol. 51 (4), No. 68, pp. 1-22, 2018.
		
		\bibitem{FigalliBook2023}
		Alessio Figalli, Federico Glaudo, An Invitation to Optimal Transport, Wasserstein Distances, and Gradient Flows, EMS society, 2023.
		
		\bibitem{FlamaryNIPS2014}
		R. Flamary, N. Courty, D. Tuia, A. Rakotomamonjy, ``Optimal transport with Laplacian regularization: Applications to domain adaptation and shape matching," NIPS Workshop on Optimal Transport and Machine Learning, 2014.
		
		\bibitem{Futami2022}
		H. Futami, H. Inaguma, M. Mimura, S. Sakai, T. Kawahara, ``Distilling the Knowledge of BERT for CTC-based ASR," CoRR abs/2209.02030, 2022.
		
		\bibitem{Galichon2016}
		A. Galichon, Optimal Transport Methods in Economics. Princeton University Press, 2016.
		
		\bibitem{DANN2016}
		Y. Ganin, E. Ustinova, H. Ajakan, P. Germain, H. Larochelle, F. Laviolette, M. Marchand, and V. Lempitsky, ``Domain-adversarial training of neural networks," \emph{Journal of Machine Learning Research},  vol. 17, no. 1, pp. 1-35, 2016.
		
		\bibitem{Gangbo1996}
		Wilfrid Gangbo, Robert J. McCann, ``The geometry of optimal transportation," Acta Math.,  177(2), pp. 113-161, 1996.
		
		\bibitem{GAN2014}
		I. Goodfellow, J. Pouget-Abadie, M. Mirza, B. Xu, D. Warde-Farley, S. Ozair, A. Courville, Y. Bengio, ``Generative Adversarial Nets," Proceedings of the International Conference on Neural Information Processing Systems, pp. 2672-2680, 2014.
		
		\bibitem{conformer2020}
		A. Gulati, J. Qin, C. Chiu, N. Parmar, Y. Zhang, J. Yu, W. Han, S. Wang, Z. Zhang, Y. Wu, R. Pang, ``Conformer: Convolution augmented transformer for speech recognition," \emph{arXiv preprint}, arXiv:2005.08100, 2020.
		
		\bibitem{Graves2012}
		A. Graves, ``Sequence transduction with recurrent neural networks," \emph{arXiv preprint}, arXiv:1211.3711, 2012.
		
		\bibitem{CTCASR}
		A. Graves, and N. Jaitly, ``Towards end to-end speech recognition with recurrent neural networks," in \emph{Proc. ICML}, pp. 1764–1772, 2014.
		
		\bibitem{CIFBERT1}
		M. Han, F. Chen, J. Shi, S. Xu, B. Xu, ``Knowledge Transfer from Pre-trained Language Models to Cif-based Speech Recognizers via Hierarchical Distillation," \emph{arXiv preprint}, arXiv:2301.13003, 2023.
		
		\bibitem{CIFBERT2}
		M. Han, L. Dong, Z. Liang, M. Cai, S. Zhou, Z. Ma, B. Xu, ``Improving End-to-End Contextual Speech Recognition with Fine-Grained Contextual Knowledge Selection," in \emph{Proc. of ICASSP}, pp. 8532-8536, 2022.
		
		
		\bibitem{Higuchi2023}
		Y. Higuchi, T. Ogawa, T. Kobayashi, S. Watanabe, ``BECTRA: Transducer-Based End-To-End ASR with Bert-Enhanced Encoder," in \emph{Proc. of ICASSP}, pp. 1-5, 2023.
		
		\bibitem{KD2015}
		G. Hinton, O. Vinyals, and J. Dean, ``Distilling the knowledge in a neural network," \emph{arXiv preprint}, arXiv:1503.02531, 2015.
		
		\bibitem{DDPM2020}
		J. Ho, A. Jain, and P. Abbeel, ``Denoising diffusion probabilistic models," Advances in neural information processing systems, Vol 33, pp. 6840-6851, 2020.
		
		\bibitem{Holderrieth2026}
		P. Holderrieth, Peter, E. Erives, ``An Introduction to Flow Matching and Diffusion Models," arXiv: 2506.02070, 2026.
		
		\bibitem{Hori2017}
		T. Hori, S. Watanabe, and J. R. Hershey, ``Joint ctc/attention decoding for end-to-end speech recognition," in \emph{Proc. of ACL}, vol. 1, pp. 518–529, 2017.
		
		\bibitem{HsiehIS2021}
		T. A. Hsieh, C. Yu, S. W. Fu, X. Lu, and Y. Tsao, ``Improving Perceptual Quality by Phone-Fortified Perceptual Loss using Wasserstein Distance for Speech Enhancement," ISCA-Interspeech, 2021.
		
		\bibitem{HungKH2025}
		K. Hung, X. Lu, S. Fu, H. Tseng, H. Lin, C. Lin, and Y. Tsao, ``Linguistic knowledge transfer learning for speech enhancement," arXiv preprint arXiv:2503.07078, 2025.
		
		\bibitem{Hyvarinen2005}
		A. Hyv{\"a}rinen, ``Estimation of non-normalized statistical models by score matching," Journal of Machine Learning Research, Vol.6, pp. 695-709, 2005.
		
		\bibitem{Hyvarinen2007}
		A. Hyv{\"a}rinen, ``Some extensions of score matching.," Computational statistics \& data analysis, 51(5), pp.2499-2512, 2007.
		
		\bibitem{MatrixScale2016}
		M. Idel, ``A review of matrix scaling and Sinkhorn's normal form for matrices and positive maps," arXiv: Rings and Algebras, 2016.
		
			
		\bibitem{Jordan1998}
		R. Jordan, D. Kinderlehrer, F. Otto, ``The Variational Formulation of the Fokker-Planck Equation," SIAM Journal on Mathematical Analysis Vol. 29, Iss. 1, 1998.
		
		\bibitem{KhamisPAMI2024}
		A. Khamis, R. Tsuchida, M. Tarek, V. Rolland, L. Petersson, ``Scalable Optimal Transport Methods in Machine Learning: A Contemporary Survey," IEEE Transactions on Pattern Analysis and Machine Intelligence, doi:10.1109/TPAMI.2024.3379571, 2024.
		
		\bibitem{Kim2017}
		S. Kim, T. Hori, and S. Watanabe, ``Joint CTC-attention based end-to-end speech recognition using multi-task learning," in \emph{Proc. of ICASSP}, pp. 4835–4839, 2017.
		
		\bibitem{VAE2013}
		D. Kingma, M. Welling, ``Auto-Encoding Variational Bayes," arXiv preprint arXiv:1312.6114, 2013.
		
		\bibitem{VAE2019}
		D. Kingma,M. Welling, ``An Introduction to Variational Autoencoders," Foundations and Trends in Machine Learning, Vol. 12, No. 4, pp. 307-392, 2019.
		
		\bibitem{CNF2021}
		I. Kobyzev, S. Prince, and M. Brubaker, ``Normalizing flows: An introduction and review of current methods," IEEE Transactions on Pattern Analysis and Machine Intelligence, 2021.
		
		\bibitem{Kolouri2016}
		S. Kolouri, S. Park, M. Thorpe, D. Slep\v{c}ev, and G. Rohde, ``Transport-based analysis, modeling, and learning from signal and data distributions," arXiv preprint arXiv:1609.04767, 2016.
		
		\bibitem{KolouriSPM2017}
		S. Kolouri, S. Park, M. Thorpe, D. Slepcev, G. K. Rohde, ``Optimal Mass Transport: Signal Processing and Machine-learning Applications," IEEE Signal Processing Magazine, Vol. 34 (4), pp. 43-59, 2017.
		
		\bibitem{Kornilov2024}
		N. Kornilov, P. Mokrov, A. Gasnikov, and A. Korotin, ``Optimal Flow Matching: Learning Straight Trajectories in Just One Step," The Thirty-eighth Annual Conference on Neural Information Processing Systems, 2024.
		
		\bibitem{NOPT2021}
		A. Korotin, L. Li, A. Genevay, J. Solomon, A. Filippov, E. Burnaev, ``Do Neural Optimal Transport Solvers Work? A Continuous Wasserstein-2 Benchmark," Advances in Neural Information Processing Systems, Vol. 34, pp. 14593-14605, 2021.
		
		\bibitem{NOPT2022}
		A. Korotin, V. Egiazarian, A. Asadulaev, R. Safin, Ruslan, E. Burnaev, ``Neural Optimal Transport," arXiv preprint arXiv:2201.12220, 2022.
		
		\bibitem{Kouw2019}
		W. M. Kouw and M. Loog, ``A Review of Domain Adaptation without Target Labels," \emph{IEEE transactions on pattern analysis and machine intelligence}, 43(3), pp. 766-785, 2019.
		
		\bibitem{KuboICASSP2022}
		Y. Kubo, S. Karita, M. Bacchiani, ``Knowledge Transfer from Large-Scale Pretrained Language Models to End-To-End Speech Recognizers," in \emph{Proc. of ICASSP}, pp. 8512-8516, 2022.
		
		\bibitem{KusnerICML2015}
		M. Kusner, Y. Sun, N. Kolkin, K. Weinberge, ``From Word Embeddings To Document Distances," Proceedings of the 32nd International Conference on Machine Learning (ICML), Vol. 37, pp. 957-965, 2015.
		
		\bibitem{Lasota1994}
		A. Lasota, and M. Mackey, Chaos, Fractals and Noise. Stochastic Aspects of Dynamics, Applied Mathematical Sciences vol. 97, New York Springer, 1994.
		
		\bibitem{LeICML2023}
		P. Le, H. Gong, C. Wang, J. Pino, B. Lecouteux, D. Schwab, ``Pre-training for Speech Translation: CTC Meets Optimal Transport," \emph{arXiv preprint}, CoRR abs/2301.11716, 2023.
		
		\bibitem{Leonard2014}
		C. L{\'e}onard, ``A survey of the Schr{\"o}dinger problem and some of its connections with optimal transport," Discrete \& Continuous Dynamical Systems-A, Vol. 34, No. 4, pp. 1533-1574, 2014.
		
		\bibitem{DeepFakeSv2025}
		M. Li, A. Yasaman, X. Zhang, ``A Survey on Speech Deepfake Detection," ACM Computing Surveys, Vol. 57 (7), No. 165, pp. 1-38, 2025.
		
		\bibitem{LiBook2015}
		J. Li, L. Deng, R. Haeb-Umbach, Y. Gong, Robust Automatic Speech Recognition: A Bridge to Practical Applications, Academic Press, 2015.
		
		\bibitem{Li2022}
		J. Li, ``Recent advances in end-to-end automatic speech recognition," \emph{APSIPA Transactions on Signal and Information Processing}, DOI 10.1561/116.00000050, 2022.
		
		\bibitem{LinNeurIPS2021}
		H. Lin, H. Tseng, X. Lu, Y. Tsao, ``Unsupervised Noise Adaptive Speech Enhancement by Discriminator-Constrained Optimal Transport," in \emph{Proc. of NeurIPS}, pp. 19935-19946, 2021.
		
		\bibitem{LinMP2025}
		M. Lin, J. Hou, C. Chen, S. Chien, J. Chen, X. Lu, and Y. Tsao, ``Bridging the multi-modality gaps of audio, visual and linguistic for
		speech enhancement," arXiv preprint arXiv:2501.13375, 2025.
		
		\bibitem{Lipman2022}
		Y. Lipman, Y., R. Chen, H. Ben-Hamu, M. Nickel, and M. Le, ``Flow matching for generative modeling," arXiv preprint arXiv:2210.02747, 2022.
		
		\bibitem{Lipman2024}
		Y. Lipman,  M. Havasi, P. Holderrieth, N. Shaul, M. Le, B. Karrer, R. Chen, D. Lopez-Paz, H. Ben-Hamu, I. Gat, ``Flow Matching Guide and Code," arXiv:2412.06264, 2024.
		
		\bibitem{LiuBound}
		X. Liu, Z. Guo, S. Li, F. Xing, J. You, C. Kuo, G. Fakhri, J. Woo, ``Adversarial Unsupervised Domain Adaptation with Conditional and Label Shift: Infer, Align and Iterate," in \emph{Proc. of IEEE/CVF International Conference on Computer Vision (ICCV)}, pp. 10347-10356, 2021.
		
		\bibitem{LiuX2023}
		X. Liu, C. Gong, and Q. Liu, ``Flow straight and fast: Learning to generate and transfer data with rectified flow," International Conference on Learning Representations, 2023.
		
		\bibitem{LiuICLR2023}
		T. Liu, J. Puigcerver, M. Blondel, ``Sparsity-constrained optimal transport," Proceedings of the Eleventh International Conference on Learning Representations (ICLR), 2023.
		
		\bibitem{LongMMD2013}
		M. Long, J. Wang, G. Ding, J. Sun, and P. S. Yu, ``Transfer feature learning with joint distribution adaptation," in \emph{Proc. of the IEEE International Conference on Computer Vision}, pp. 2200-2207, 2013.
		
		\bibitem{wav2vecBERTSLT2022}
		K. Lu and K. Chen, ``A Context-aware Knowledge Transferring Strategy for CTC-based ASR," in \emph{Proc. of SLT}, pp. 60-67, 2022.
		
		\bibitem{LuICASSP2024}
		X. Lu, S. Shen, Y. Tsao, H. Kawai, ``Hierarchical Cross-Modality Knowledge Transfer with Sinkhorn Attention for CTC-based ASR," IEEE-ICASSP, 2024.
		
		\bibitem{LuICASSP2021}
		X. Lu, P. Shen, Y. Tsao, H. Kawai, ``Unsupervised Neural Adaptation Model Based on Optimal Transport for Spoken Language Identification," in \emph{Proc. of ICASSP}, pp. 7213-7217, 2021.
		
		\bibitem{LuASRU2023}
		X. Lu, S. Shen, Y. Tsao, H. Kawai, ``Cross-modal Alignment with Optimal Transport for CTC-based ASR," IEEE-ASRU, 2023.
		
		\bibitem{LuSLT2024}
		X. Lu, P. Shen, Y. Tsao, H. Kawai, ``Temporal Order Preserved Optimal Transport-based Cross-modal Knowledge Transfer Learning for ASR," IEEE-SLT, 2024.
		
		\bibitem{LuIS2025}
		X. Lu, P. Shen, Y. Tsao, H. Kawai, ``Cross-modal Knowledge Transfer Learning as Graph Matching Based on Optimal Transport for ASR," ISCA-Interspeech, 2025.
		
		\bibitem{LuICASSP2026}
		X. Lu, P. Shen, Y. Tsao, H. Kawai, ``Learning to Align with Unbalanced Optimal Transport in Linguistic Knowledge Transfer for ASR," IEEE-ICASSP, 2026. 
		
		\bibitem{TSNE2008}
		L. Maaten and G. Hinton, ``Visualizing data using t-sne," Journal of machine learning research, pp. 2579-2605, 2008.
		
		\bibitem{MaggiBook2023}
		Francesco Maggi, Optimal Mass Transport on Euclidean Spaces, Cambridge University Press, 2023.
		
		\bibitem{Makkuva2020}
		A. Makkuva, A. Taghvaei, S. Oh, J. Lee, ``Optimal Transport Mapping via Input Convex Neural Networks," Proceedings of the 37th International Conference on Machine Learning, Vol. 119, pp. 6672-6681, 2020.
		
		\bibitem{MialonICLR2021}
		G. Mialon, D. Chen, A. Aspremont, J. Mairal, ``A Trainable Optimal Transport Embedding for Feature Aggregation and its Relationship to Attention," ICLR 2021.
		
		\bibitem{MontesumaPAMI2025}
		E. Montesuma, F. Mboula and A. Souloumiac, ``Recent Advances in Optimal Transport for Machine Learning," IEEE Trans. Pattern Anal. Mach. Intell., vol.47, no. 22, pp. 1161–1180, 2025.
		
		\bibitem{NguyenUOT2023}
		Q. Nguyen, H. Nguyen,Y. Zhou, L. Nguyen, ``On unbalanced optimal transport: gradient methods, sparsity and approximation error," J. Mach. Learn. Res., Vol. 24, No.1, 2023.
		
		\bibitem{Otto2001}
		F. Otto, ``The geometry of dissipative evolution equations: the porous medium equation," Communications in Partial Differential Equations, 26(1–2), pp. 101–174, 2001.
		
		\bibitem{PanMMD2009}
		S. Pan, I. Tsang, J. Kwok and Q. Yang, ``Domain Adaptation via Transfer Component Analysis," in \emph{IEEE Transactions on Neural Networks}, vol. 22, no. 2, pp. 199-210, Feb. 2011.
		
		\bibitem{Papamakarios2021}
		G. Papamakarios, E. Nalisnick, D. Rezende, S. Mohamed, B. Lakshminarayanan, ``Normalizing flows for probabilistic modeling and inference," JMLR, Vol. 22, No. 1, pp. 1532-4435, 2021.
		
		\bibitem{PeyreBook2019}
		G. Peyr{\'e} and M. Cuturi, ``Computational Optimal Transport: With Applications to Data Science," Foundations and Trends® in Machine Learning, Vol. 11 (5-6), pp 355-607, 2019.
		
		\bibitem{Peyre2016}
		G. Peyr{\'e}, M. Cuturi, J. Solomon, ``Gromov-Wasserstein Averaging of Kernel and Distance Matrices," \emph{Proc. of ICML}, vol. 48, pp. 2664-2672, 2016.
		
		\bibitem{Gabriel2025}
		G. Peyr{\'e}, ``Optimal Transport for Machine Learners," arXiv:2505.06589, 2025.
		
		\bibitem{SanderAISTATS2022}
		M. Sander, P. Ablin, M. Blondel, G. Peyre, ``Sinkformers: Transformers with Doubly Stochastic Attention," in \emph{Proc. of AISTATS}, pp. 3515-3530, 2022.
		
		\bibitem{SantambrogioBook2015} 
		Filippo Santambrogio, Optimal Transport for Applied Mathematicians: Calculus of Variations, PDEs and Modeling, Springer, 2015.
		
		\bibitem{SDEBook2019}
		S. S{\"a}rkk{\"a}, and A. Solin, Applied Stochastic Differential Equations, Cambridge University Press, IMS Textbooks series, 2019. 
		
		\bibitem{Scetbon2021}
		M. Scetbon, M. Cuturi, and G. Peyr{\'e}, ``Low-Rank Sinkhorn Factorization," In International Conference on Machine Learning, 2021.
		
		\bibitem{MMD}
		B. Scholkopf, J. Platt, T. Thomas, ``A Kernel Method for the Two-Sample-Problem," in \emph{Proc. of the International Conference on Neural Information Processing Systems (NIPS)}, vol. 19, pp. 513-520, 2006.
		
		\bibitem{UOT2023}
		T. S{\'e}journ{\'e}, G. Peyr{\'e}, F. Vialard, ``Unbalanced Optimal Transport, from theory to numerics," in Handbook of Numerical Analysis,Vol. 24, pp. 407-471, 2023.
		
		\bibitem{ShiNIPS2023}
		Y. Shi, V. De Bortoli, A. Campbell, A. Doucet, ``Diffusion schr\"{o}dinger bridge matching," International Conference on Neural Information Processing Systems, 2023.
		
		\bibitem{ShenAAAI2017}
		J. Shen, Y. Qu, W. Zhang, Y Yu, ``Wasserstein Distance Guided Representation Learning for Domain Adaptation," AAAI Conference on Artificial Intelligence, 2017. 
		
		\bibitem{Sinkhorn1967}
		R. Sinkhorn and P. Knopp, ``Concerning nonnegative matrices and doubly stochastic matrices," Pacific Journal of Mathematics, 21(2):343–348, 1967.
		
		\bibitem{XVector2018}
		D. Snyder, D. Garcia-Romero, G. Sell, D. Povey and S. Khudanpur, ``X-Vectors: Robust DNN Embeddings for Speaker Recognition," IEEE International Conference on Acoustics, Speech and Signal Processing (ICASSP), pp. 5329-5333, 2018.
		
		\bibitem{Song2019}
		Y. Song, S. Ermon, ``Generative Modeling by Estimating Gradients of the Data Distribution," CoRR abs/1907.05600, 2019.
		
		\bibitem{SongICLR2021}
		Y. Song, J. Sohl-Dickstein, D. Kingma, A. Kumar, S. Ermon, B. Poole, ``Score-Based Generative Modeling through Stochastic Differential Equations," ICLR 2021.
		
		\bibitem{Su2017}
		B. Su, G. Hua, ``Order-Preserving Wasserstein Distance for Sequence Matching," Proceedings of the IEEE Conference on Computer Vision and Pattern Recognition (CVPR), 2017.
		
		\bibitem{SinkhornAtt2020}
		Y. Tay, D. Bahri, L. Yang, D. Metzler, D. Juan, ``Sparse Sinkhorn Attention," in \emph{Proc. of ICML}, pp. 9438-9447, 2020.
		
		\bibitem{Tong2024}
		A. Tong, K. Fatras, N. Malkin, G. Huguet, Y. Zhang, J. Brooks, G. Wolf, Y. Bengio, ``Improving and generalizing flow-based generative models with minibatch optimal transport," Transactions on Machine Learning Research, pp. 1-34, Mar., 2024.
		
		\bibitem{Vayer2020}
		T. Vayer, L. Chapel, R. Flamary, R. Tavenard, N. Courty, ``Fused Gromov-Wasserstein Distance for Structured Objects," \emph{Algorithms}, vol. 13, no. 9, 212, https://doi.org/10.3390/a13090212, 2020.
		
		\bibitem{VillaniBook2003} 
		Cédric Villani, Topics on optimal transportation, AMS society, 2003. 
		
		\bibitem{VillaniBook2008} 
		Cédric Villani, Optimal Transport: Old and New, Springer, 2008.
				
		\bibitem{YangTIFS2026}
		W. Yang, J. Wei, W. Lu, L. Li, X. Lu, ``Domain Adaptation for Speaker Verification Using Optimal Transport with Pseudo Label," IEEE Transactions on Information Forensics \& Security, 2026.
		
		\bibitem{Yokoi2020}
		Sho Yokoi, Ryo Takahashi, Reina Akama, Jun Suzuki, and Kentaro Inui, ``Word rotator’s distance," In Proceedings of the 2020 Conference on Empirical Methods in Natural Language Processing (EMNLP), pp. 2944–2960, 2020.
		
		\bibitem{YuBook2014}
		D. Yu, L. Deng, Automatic Speech Recognition: A Deep Learning Approach, Springer Publishing Company, 2014.
		
		\bibitem{NARBERT}
		F. Yu, K. Chen, and K. Lu, ``Non-autoregressive ASR Modeling using Pre-trained Language Models for Chinese Speech Recognition," \emph{IEEE/ACM Transactions on Audio, Speech, and Language Processing}, vol. 30, pp. 1474-1482, 2022
		
		\bibitem{MultimodalSv2025}
		Y. Yuan, Z. Li, B. Zhao, ``A Survey of Multimodal Learning: Methods, Applications, and Future," ACM Computing Surveys, Vol. 57 (7), No. 167, pp. 1-34, 2025.
		
		\bibitem{WangNIPS2014}
		H. Wang, A.  Banerjee, Arindam, ``Bregman alternating direction method of multipliers," Proceedings of the 28th International Conference on Neural Information Processing Systems, pp. 2816-2824, 2014.
				
		\bibitem{Cross2021}
		W. Wang, S. Ren, Y. Qian, S. Liu, Y. Shi, Y. Qian, M. Zeng, ``Optimizing Alignment of Speech and Language Latent Spaces for End-To-End Speech Recognition and Understanding," in \emph{Proc. of ICASSP}, pp. 7802-7806, 2021.	

        \bibitem{EDMD2015}
        M. O. Williams, I.G. Kevrekidis, C.W. Rowley, ``A Data–Driven Approximation of the Koopman Operator: Extending Dynamic Mode Decomposition, " Journal of Nonlinear Science 25, pp. 1307-1346, 2015.

        \bibitem{XingTOG2022}
        Jiankai Xing, Fujun Luan, Ling-Qi Yan, Xuejun Hu, Houde Qian, and Kun Xu. Differentiable rendering using rgbxy derivatives and optimal transport. ACM Transactions on Graphics (TOG), 41(6):1–13, 2022.

		\bibitem{Xie2018}
		Yujia Xie, Xiangfeng Wang, Ruijia Wang, and Hongyuan Zha, ``A fast proximal point method for computing exact wasserstein distance" In Uncertainty in artificial intelligence, pp. 433–453, PMLR, 2020.
		
		\bibitem{ZhangEMD2020}		
		Chi Zhang, Yujun Cai, Guosheng Lin, and Chunhua Shen, ``Deepemd: Few-shot image classification with differentiable earth mover’s distance and structured classifiers," In Proceedings of the IEEE/CVF Conference on Computer Vision and Pattern Recognition (CVPR), June 2020.
				
		\bibitem{ZhangICASSP2023}
		R. Zhang, J. Wei, X. Lu, W. Lu, D. Jin, L. Zhang, J. Xu, ``Optimal Transport with a Diversified Memory Bank for Cross-Domain Speaker Verification," IEEE-ICASSP, 2023.
		
		\bibitem{ZhangIS2023}
		R. Zhang, J. Wei, X. Lu, Y. Li, J. Xu, D. Jin, J. Tao, ``SOT: Self-supervised Learning-Assisted Optimal Transport for Unsupervised Adaptive Speech Emotion Recognition," ISCA-Interspeech, 2023.
		
		\bibitem{ZhangICASSP2024}
		R. Zhang, J. Wei, X. Lu, Y. Li, W. Lu, D. Jin, J. Xu, ``Self-supervised Domain Exploration with an Optimal Transport Regularization for Open Set Cross-domain Speech Emotion Recognition," IEEE-ICASSP, 2024.
		
		\bibitem{ZhangTASLP2024}
		R. Zhang, J. Wei, X. Lu, W. Lu, D. Jin, L. Zhang, J. Xu., ``Unsupervised Adaptive Speaker Recognition by Coupling-Regularized Optimal Transport," IEEE-TASLP, 2024
		
		\bibitem{ZhangTIFS2025}
		R. Zhang, J. Wei, X. Lu, L. Zhang, D. Jin, J. Xu, W. Lu, ``SHDA: Sinkhorn Domain Attention for Cross-Domain Audio Anti-Spoofing," IEEE Transactions on Information Forensics \& Security, 2025.
		
		\bibitem{ZhangTASLP2025}
		R. Zhang, J. Wei, X. Lu, L. Zhang, D. Jin, W. Lu, J. Xu, ``Multi-Sinkhorn Teacher Knowledge Aggregation Framework for Adaptive Audio Anti-Spoofing," IEEE Transactions on Audio, Speech, and Language Processing, 2025.
		
		\bibitem{DeepRobustASR2018}
		Z. Zhang, J. Geiger, J. Pohjalainen, A. Mousa, W. Jin, B. Schuller, ``Deep Learning for Environmentally Robust Speech Recognition: An Overview of Recent Developments," ACM Transactions on Intelligent Systems and Technology (TIST), Vol. 9, no. 5, pp. 1-28, 2018.
		
		\bibitem{MultimodalSv2024}
		F. Zhao, C. Zhang, B. Geng, ``Deep Multimodal Data Fusion," ACM Computing Surveys, Vol. 56 (9), No. 216, pp. 1-36, 2024.
		
		\bibitem{ZhouACL2023}
		Y. Zhou, Q. Fang, Y. Feng, ``CMOT: Cross-modal Mixup via Optimal Transport for Speech Translation," \emph{arXiv preprint},  arXiv:2305.14635, 2023.
		
		
		
		
		
				
				
		
		
		
			
	\end{thebibliography}
	
\end{document}